\documentclass[12pt]{article}
\usepackage{amsmath,mathtools,amsfonts}
\usepackage{amssymb}
\usepackage{bm}
\usepackage{graphicx}
\usepackage[all]{xy}
\usepackage{mathrsfs}
\usepackage{cite}
\usepackage{physics}
\usepackage{comment}
\usepackage{color}
\usepackage{ulem}

\newcommand{\sectiono}[1]{\section{#1}\setcounter{equation}{0}}

\newcommand{\bd}[1]{\boldsymbol{#1}}
\newcommand{\id}{\mathbb{I}}

\newcommand{\idSH}{\mathbb{I}_{\mathcal{SH}}}

\newcommand{\coder}[6]{\bm{#1}_{#2,#3}^{(#4)}|^{#5}_{#6}}
\newcommand{\os}{\mathfrak{o}_1(s)}
\newcommand{\ot}{\mathfrak{o}_2(t)}
\newcommand{\pdvt}{\partial_t}
\newcommand{\pdvs}{\partial_s}

\begin{document}

\baselineskip=17pt

\begin{titlepage}
\rightline{\tt KUNS-3138}
\rightline{\tt YITP-26-152}
\begin{center}
\vskip 2.5cm
{\Large \bf {Quantum action for heterotic string field theory\\ with quantum $L_\infty$ structure }}
\vskip 1.0cm
{\large {Yasu Gen$^a$, Hiroshi Kunitomo$^b$, and Jojiro Totsuka$^a$}}
\vskip 1.0cm
$^a${\it  Department of Physics, Kyoto University,}\\
{\it {Kitashirakawa Oiwakecho, Sakyo-ku, Kyoto 606-8502, Japan}}\\
$^b${\it {Center for Gravitational Physics and Quantum Information}}, \\
{\it {Yukawa Institute for Theoretical Physics,}}
{\it {Kyoto University}},\\
{\it {Kitashirakawa Oiwakecho, Sakyo-ku, Kyoto 606-8502, Japan}}\\
E-mail: yasugen@gauge.scphys.kyoto-u.ac.jp, kunitomo@yukawa.kyoto-u.ac.jp, george.yoshinaka@gauge.scphys.kyoto-u.ac.jp

\vskip 2.0cm

{\bf Abstract}
\end{center}
\noindent

We construct a quantum action for heterotic string field theory that satisfies the quantum Batalin-Vilkovisky equation. Assuming a triangulation of the bosonic moduli space, the prescription for inserting picture-changing operators is determined based on a quantum $L_\infty$ structure. After concretely constructing some of the lower-order interactions, we propose a general prescription, which generalizes that for constructing the classical interactions governed by a classical $L_\infty$ structure. Assuming it does not suffer from spurious poles, all quantum interactions are determined by iteratively solving differential equations and transforming the solutions via $IBL_{\infty}$-transfer. To demonstrate the consistency of our prescription, we compute the one-loop two-point functions (modulo the spurious pole issue) and show that they agree with those in the first-quantized formulation.

\end{titlepage}

\tableofcontents

\newpage

\sectiono{Introduction}

String field theory (SFT) is a candidate for a complete formulation of string theory 
and possesses significant theoretical and practical utility. Theoretically, it may help 
elucidate the physical principles underlying string theory, which is essential for establishing 
it as a fundamental theory of nature. Practically, SFT allows for the application of various 
techniques from conventional quantum field theory, both perturbative and non-perturbative, through 
appropriate generalizations. In particular, closed string field theory remains a compelling 
subject of research as a promising framework for a consistent theory of quantum gravity.

A gauge-invariant bosonic closed string field theory was first constructed as a classical theory
[\citen{Saadi:1989tb,Kugo:1989aa,Kugo:1989tk}] and subsequently completed as a consistent quantum theory
[\citen{Zwiebach:1992ie}].
Even at the classical level, the action takes a non-polynomial form with an infinite 
number of interaction terms, which is complicated but characterized by a sophisticated 
structure governed by the $L_\infty$ algebra. 
Through the Batalin-Vilkovisky (BV) quantization method \cite{Batalin:1981jr}, the classical action is extended
to a quantum master action that satisfies the quantum BV equation. The $L_\infty$ structure
characterizing the classical action is also extended to the quantum $L_\infty$ structure\cite{Zwiebach:1992ie},
also known as the loop homotopy structure \cite{Markl:1997bj}.
The $L_\infty$ structure guarantees the gauge invariance of the action, which is essentially 
equivalent to guaranteeing that contributions from each cell decomposed region, which is given
as those from several Feynman graphs in the string field theory, are smoothly connected 
at their boundaries and provide an amplitude as the integral of the whole moduli space.

Classical closed superstring field theories can also be constructed using the $L_\infty$ structure 
as a guiding principle \cite{Erler:2014eba,Kunitomo:2019glq,Kunitomo:2019kwk,Kunitomo:2021wiz,Jurco:2013qra}. In these theories, string interactions must be equipped with an appropriate number of picture-changing operators (PCOs) \cite{Friedan:1985ge}. The $L_\infty$ structure in this context also governs the configuration of PCOs at each interaction vertex. This simultaneously provides a prescription\footnote{
A systematic prescription, known as vertical integration, was first introduced in \cite{Sen:2014pia,Sen:2015hia}; an alternative approach was recently proposed in \cite{Erler:2017dgr}.} for resolving discrepancies in PCO placement at the boundaries of the cell regions in the moduli space. The purpose of this paper is to quantize heterotic string field theory via the BV method. This requires adding quantum correction terms to the action at each loop order, which are governed by the quantum $L_\infty$ structure. We extend the (classical) $L_\infty$ structure to the quantum $L_\infty$ structure and construct these higher-loop string vertices incorporating the requisite PCOs. Furthermore, we explicitly calculate the one-loop two-point amplitudes and demonstrate their consistency with results obtained from the first-quantized method. This provides robust confirmation of the validity of the constructed action.

The rest of the paper is organized as follows. In Section~\ref{sec:classical action}, we review classical string field theories with an $L_\infty$ structure for both bosonic strings and heterotic strings. We outline the basic ingredients and explain how the $L_\infty$ structure governs the classical action and its gauge invariance. This framework is extended to the quantum action in Section~\ref{sec:Quantum action}, where the quantum action satisfies the quantum BV master equation 
as a consequence of its quantum $L_\infty$ structure.

In Section~\ref{explicit}, we construct quantum heterotic string products that satisfy a quantum $L_\infty$ algebra. We first explicitly solve several quantum $L_\infty$ relations for simple cases, 
and encounter a contribution that could be a potential obstruction. A closer examination, however, shows that this term contains a loop in which two equivalent representations of states with opposite statistics propagate; their contributions cancel each other, causing the apparent obstruction to vanish. This cancellation property persists throughout the general construction and plays a crucial role in this paper. 

We then extend the method used for classical products to establish a general prescription for constructing quantum heterotic products. First, we extend the setup to a large Hilbert space. However, because it is difficult to impose constraints on states in loops at a later stage, we restrict the states propagating in loops to the small Hilbert space, which realizes the $\eta$ zero-mode insertion required by the index theorem for the $\eta\xi$ system in the large Hilbert space. This mismatch creates a technical problem: while a quantum $L_\infty$-algebra requires both the quantum $L_\infty$ relations and cyclicity, maintaining cyclicity becomes difficult under these conditions. Therefore, we temporarily require only the quantum $L_\infty$ relations, deferring the proof of cyclicity.

Within this relaxed condition, the problem reduces to a straightforward extension of the classical case, allowing us to construct heterotic string products in the large Hilbert space that satisfy the quantum $L_\infty$ relations. As the next step, we apply an $IBL_\infty$-transfer to transform these products into those in the small Hilbert space required for the quantum action. This construction is carried out up to terms that do not affect the quantum $L_\infty$ relations. 

Section~\ref{amplitudes} presents the calculation of the one-loop two-point amplitude for both the NS-NS and R-R cases, confirming the validity of our general construction. This verification is conducted while setting aside the issue of spurious poles; if spurious poles affect these results, a re-examination and refinement of the construction will be required. Section~\ref{Discussion} is devoted to discussion. 

 Finally, four appendices provide technical details omitted from the text. Appendix~\ref{prop of basis} summarizes the basic properties of the basis for string states needed to define the order-two coderivation used in quantum $L_\infty$ algebras. This order-two coderivation is essentially identical to what was previously termed a reflector. Its operator form, which is useful for various analyses, is presented in Appendix~\ref{sec:reflectors}. In Appendix~\ref{sec:cyclicity}, we provide the relations satisfied by string products in the large Hilbert space, in place of the cyclicity. Appendix~\ref{sec:IBL} contains detailed calculations for the $IBL_\infty$-transfer.

\sectiono{Closed string field theory with $L_\infty$ structure}\label{sec:classical action}

In this section, we summarize how the $L_\infty$ structure characterizes 
classical bosonic closed string field theory, and then show how it is extended to heterotic string field theory.

\subsection{Classical action for bosonic string field theory}
The classical action of closed string field theory can be constructed based on the cyclic $L_\infty$ 
structure \cite{Zwiebach:1992ie}. The kinetic term is written as
\begin{align}
    S_\mathrm{kin}=\frac{1}{2}\omega_{\textrm{B}}(\Phi,Q\Phi),
\end{align}
where 
$Q$ is the BRST operator and $\omega_{\textrm{B}}$ is a degree odd\footnote{
In this paper, we define the degree as the number modulo 2:
$0$ (even) or $1$ (odd).} symplectic form defined by 
the BPZ inner product:
\begin{align}
    \omega_{\textrm{B}}(A,B)=(-1)^A{}_{\textrm{BPZ}}\!\bra{A}c_0^-\ket{B}\ =\ -(-1)^{AB}\omega_{\textrm{B}}(B,A)\,,
\label{symp b}
\end{align}
where ${}_{\textrm{BPZ}}\bra{A}$ is the BPZ conjugate of $\ket{A}$\,.

The classical string field $\Phi$ is Grassmann even\footnote{
For closed string, the degree of the string field is equal to 
its Grassmann parity deg$(\Phi)=\epsilon(\Phi)$. We often denote it simply as $\Phi$ in the exponent.} 
and has restricted in the Hilbert space
$\mathcal{H}$ with the world-sheet ghost number 2:
%
%
\begin{align}
    N_{\textrm{gh}}(\Phi)=2,\qquad \Phi\in \mathcal{H}.
\label{ghost number}
\end{align}
In addition, 
it is restricted by the closed string constraints:
\begin{equation}
    b_0^-\Phi=0,\qquad
    L_0^-\Phi=0.
\end{equation}

The interaction of $n+1$ closed strings is written as
\begin{align}
    \frac{1}{(n+1)!}\omega_{\textrm{B}}(\Phi_1,L_n(\Phi_2,\dots,\Phi_{n+1})),
\end{align}
with the string product
\begin{align}
     L_{n}:\qquad\qquad &\mathcal{H}^{\wedge n}\qquad\qquad \longrightarrow \qquad\quad \mathcal{H},\nonumber\\
            \Phi_1 \wedge &\dots \wedge \Phi_n \qquad\longmapsto \qquad L_{n}(\Phi_1, \dots, \Phi_n),\nonumber
\end{align}
where a wedge $\wedge$ denotes the symmetrized tensor product defined by
\begin{align}
    \Phi_1 \wedge \dots \wedge \Phi_n = \sum_\sigma (-1)^{\epsilon(\sigma)} 
\Phi_{\sigma(1)}\otimes \dots \otimes \Phi_{\sigma(n)}.
\end{align}
The sign factor $(-1)^{\epsilon(\sigma)}$ is the one obtained by permutation
from $\{\Phi_1,\dots\Phi_n\}$ to $\{\Phi_{\sigma(1)},\dots\Phi_{\sigma(n)}\}$.
%
The total action is then written as
\begin{align}
    S=\sum_{n=1}^\infty\frac{1}{(n+1)!}\omega_{\textrm{B}}(\Phi,L_{n}(\underbrace{\Phi,\dots,\Phi}_n)),
\end{align}
with the identification $L_1=Q$.
The action $S$ is invariant under the gauge transformation
\begin{align}
    \delta\Phi=\sum_{n=1}^\infty\frac{1}{(n-1)!}L_{n}(\underbrace{\Phi,\dots,\Phi}_{n-1},\Lambda),
\end{align}
if the string product $L_n$ satisfy the $L_\infty$ relations,
\begin{align}\label{eq:Linfty relations}
    0&={\sum_\sigma}\sum_{\substack{k,l=1\\k+l=n+1}}^\infty(-1)^{\epsilon(\sigma)}\frac{1}{(k-1)!\,l!}
L_{k}(L_{l}(\Phi_{\sigma(1)},\dots,\Phi_{\sigma(l)}),\Phi_{\sigma(l+1)},\dots,\Phi_{\sigma(n)}),
\end{align}
and the cyclicity condition
\begin{align}\label{eq:cyclicity condition}
    \omega_{\textrm{B}}(\Phi_1,L_n(\Phi_2,\dots,\Phi_{n+1}))\ =\
(-1)^{\epsilon}\omega_{\textrm{B}}(\Phi_{n+1}, L_n(\Phi_1,\dots,\Phi_n)),
\end{align}
with the sign factor $\epsilon=\Phi_1+(\sum_{i=1}^n\Phi_i+1)\Phi_{n+1}$.
The string products satisfying these conditions is called the cyclic $L_\infty$ algebra
which we denote $(\mathcal{H},\omega_{\textrm{B}},\{L_n\})$.

The $L_\infty$ relations \eqref{eq:Linfty relations} are more compactly expressed by 
using the coalgebra representation.
From the string product $L_n$,
we can define a degree odd coderivations $\bd{L}_n$ acting on the symmetrized tensor 
algebra $\mathcal{SH}=\sum_{n=0}^\infty \mathcal{H}^{\wedge n}$ as
\begin{align}
    \bd{L}_n\ =\ \sum_{m=0}^\infty (L_n\wedge \id_m)\pi_{m+n},
\end{align}
where $\pi_l$ is the projection operator onto $\mathcal{H}^{\wedge l}$.
Then the $L_\infty$ relations \eqref{eq:Linfty relations} are concisely written as
\begin{align}
    [\bd{L},\bd{L}]=0,
\end{align}
with $\bd{L}\ =\ \sum_{n=1}^{\infty}\bd{L}_n$.

\subsection{Classical action for heterotic string field theory}

In the case of superstring field theory, the action  can also be constructed based on the cyclic $L_\infty$ algebra.
The ingredients, such as the string field, symplectic form, and string products, must be equipped with
the additional structure arising from the specific quantum number in RNS superstring theory: 
the picture number.
There are two sectors in the heterotic string: the NS sector describing space-time bosons and the R sector describing the space-time fermions. Their (first-quantized) Hilbert spaces have infinitely many equivalent representations, $\mathcal{H}_{\textrm{NS}}^{(p)}$ and $\mathcal{H}_{\textrm{R}}^{(p)}$, respectively. Here, $p$ is the picture number, taking integer values in the NS sector and half-integer values in the R sector. We assume that each Hilbert space is GSO-projected.
The classical heterotic string field $\Phi$ is Grassmann even and has 
two components with the world-sheet ghost number 2:
\begin{align}
    \Phi=\Phi_{\textrm{NS}}+\Phi_{\textrm{R}} \in \mathcal{H}=\mathcal{H}_{\textrm{NS}}^{(-1)}+\overline{\mathcal{H}}_{\textrm{R}}^{(-1/2)},\qquad
N_{\textrm{gh}}(\Phi)\ =\ 2,
\end{align}
where $\overline{\mathcal{H}}_{\textrm{R}}^{(-1/2)}$ denotes the space obtained by imposing an additional constraint on $\mathcal{H}_{\textrm{R}}^{(-1/2)}$, which will be specified shortly.
The string field $\Phi$ can also be embedded in the large Hilbert space as the one with the picture number $-1$ or $-1/2$
and satisfying the condition\footnote{In this paper, we denote the ghost zero-modes
$(\eta_0,\xi_0)$ as $(\eta,\xi)$ for notational simplicity.}
\begin{align}
    \eta\Phi\ =\ 0.
\end{align}
Since the heterotic string is a closed string, it satisfies the closed string constraints
\begin{equation}
 b_0^-\Phi\ =\ L_0^-\Phi\ =\ 0.
\end{equation}
The kinetic term of the action in the NS sector is written as
\begin{align}\label{eq:NS kin term}
    S_\mathrm{kin}^{\textrm{NS}}=\frac{1}{2}\omega(\Phi_{\textrm{NS}},Q\Phi_{\textrm{NS}}),
\end{align}
similarly to the bosonic string field theory.
The symplectic form $\omega$ is graded anti-symmetric and
defined in the same way as the bosonic case using the BPZ
inner-product. 
The picture number of $\Phi_{NS}$ has been chosen to be consistent
with the fact that $\omega$ does not vanish if and only if the total
picture number is equal to $-2$.

The kinetic term in the R sector, on the other hand, cannot be written in the same form
due to the picture number mismatch.
%
We introduce the PCOs $X$ and $Y$ defined by
\begin{align}
     X&=G_0\delta(\beta_0)+b_0\delta'(\beta_0),\\
    Y&=-2c_0^+\delta'(\gamma_0).
\end{align}
The operators $X$ and $Y$ act the states with picture number $-3/2$ and $-1/2$, 
and changes them to the corresponding states with $-1/2$ and $-3/2$, respectively. 
They are BPZ-even,
\begin{alignat}{2}
    \omega(A, XB)\ =&\ \omega(XA,B),\qquad &\ A,B &\in \mathcal{H}_{\textrm{R}}^{(-3/2)}\\
    \omega(A,YB)=&\ \omega(YA,B),\qquad &\ A,B &\in \mathcal{H}_{\textrm{R}}^{(-1/2)},
\end{alignat}
and are inverse of each other 
in the sense that they satisfy the relations
\begin{alignat}{2}\label{eq:XY relations}
    XYX\ =&\ X,\qquad& \textrm{on}\ &\mathcal{H}_{\textrm{R}}^{(-3/2)},\\
 YXY\ =&\ Y,\qquad&\ \textrm{on}\ &\mathcal{H}_{\textrm{R}}^{(-1/2)}.
\end{alignat}
We can define projection operators as
\begin{alignat}{2}
    \mathcal{P}_{\textrm{XY}}\ =&\ XY\,,\qquad \textrm{on}\ &\mathcal{H}_{\textrm{R}}^{(-1/2)},\\
     \mathcal{P}_{\textrm{YX}}\ =&\ YX\,,\qquad \textrm{on}\ &\mathcal{H}_{\textrm{R}}^{(-3/2)}.
\end{alignat}
The PCO $X$ is BRST invariant and is written as a BRST exact form
\begin{equation}
[Q,X]\ =\ 0,\qquad X\ =\ [Q,\textrm{H}(\beta_0)],
\label{X exact}
\end{equation}
using the Heaviside step function H$(x)$ in the large Hilbert space.
For later use, we also define the operator well-defined in the large Hilbert space,
\begin{equation}
 \widetilde{X}\ =\ [Q, \Xi],
\end{equation}
using
\begin{align}\label{eq:Xi}
    \Xi\ =\ \xi + \Big(\textrm{H}(\beta_0)\eta\xi - \xi\Big) P_{-3/2} + \Big(\xi\eta\textrm{H}(\beta_0)- \xi\Big)P_{-1/2}\,,,
\end{align}
where we denoted the projector onto the states with the picture number $n$ as $P_n$.
This $\widetilde{X}$ is identical to $X$ when it acts on states in the small
Hilbert space with the picture number $-3/2$ \cite{Erler:2016ybs}. 
We can define the symplectic form $\Omega$ in the R sector,
compensating for the mismatch in the picture number by inserting the operator $Y$:
\begin{equation}
\Omega(A,B)\ \equiv\ \omega(YA, B),\qquad A,B \in \mathcal{H}_{\textrm{R}}^{(-1/2)},.
\end{equation}
This is the natural symplectic form in the subspace 
$\overline{{\mathcal{H}}}^{(-1/2)}_{\textrm{R}}$ restricted by the constraint
\begin{equation}
\mathcal{P}_{\textrm{XY}} A\ =\ A,\qquad A\in\mathcal{H}_{\textrm{R}}^{(-1/2)}.
\end{equation}
Using the relations (\ref{X exact}), 
$\mathcal{P}_{\textrm{XY}}$ is shown to be compatible 
with the BRST cohomology:
\begin{equation}
Q\mathcal{P}_{\textrm{XY}}\ =\ \mathcal{P}_{\textrm{XY}}Q\mathcal{P}_{\textrm{XY}}.
\end{equation}
Therefore, we impose the constraint
\begin{equation}
 \mathcal{P}_{\textrm{XY}}\Phi_{\textrm{R}}\ =\ \Phi_{\textrm{R}},
\label{XY constraint}
\end{equation}
on the R string field\footnote{
This constraint can be seen to naturally arise from the supermoduli integration
\cite{Kunitomo:2015usa}.} without spoiling the $L_\infty$ structure and define the kinetic term in the R sector by
\begin{align}\label{eq:R kin term}
    S_\mathrm{kin}^{\textrm{R}}=\frac{1}{2}\Omega(\Phi_{\textrm{R}}, Q\Phi_{\textrm{R}}).
\end{align}

In order to denote the NS and R sectors collectively,
we introduce the operators $\mathcal{G}$ and $\mathcal{G}^{-1}$ as
\begin{equation}
    \mathcal{G}\ =\ \pi^0+X\pi^1,\qquad
    \mathcal{G}^{-1}\ =\ \pi^0+Y\pi^1,
\end{equation}
with the projector onto the NS (R) component,
$\pi^0 \Phi=\Phi_{\textrm{NS}}$ ($\pi^1\Phi=\Phi_{\textrm{R}}$) and denote
the relations (\ref{eq:XY relations}) as
\begin{align}\label{eq:G projector}
    \mathcal{G}\mathcal{G}^{-1}\mathcal{G}=\mathcal{G},\qquad\mathcal{G}^{-1}
\mathcal{G}\mathcal{G}^{-1}=\mathcal{G}^{-1},\qquad[Q,\mathcal{G}]=0.
\end{align}
Then, the constraint \eqref{XY constraint} is denoted as
\begin{equation}
 \mathcal{P}_{\mathcal{G}\mathcal{G}^{-1}}\Phi\ =\ \Phi,
\end{equation}
with  $\mathcal{P}_{\mathcal{G}\mathcal{G}^{-1}}\ =\ \mathcal{G}\mathcal{G}^{-1}$,
and the kinetic terms \eqref{eq:NS kin term} and \eqref{eq:R kin term} are simply written as
\begin{align}
    S_\mathrm{kin}=\frac{1}{2}\Omega(\Phi, Q\Phi)=\frac{1}{2}\omega(\Phi,\mathcal{G}^{-1}Q\Phi),
\end{align}
with $\Omega(\cdot,\cdot)=\omega(\cdot,\cdot)$ for the NS sector.
For later use, we also introduce the symplectic form $\omega_{\textrm{l}}$ in the large Hilbert space $\mathcal{H}_{\textrm{l}}$ 
defined by the BPZ inner product in $\mathcal{H}_l$.
The symplectic form $\omega_{\textrm{l}}$ is related to $\omega$ as
\begin{align}
    \omega_{\textrm{l}}(\xi A, B)=\omega(A,B),
\end{align}
if $A,B\in \mathcal{H}$.

Similar to the the bosonic string field theory, the interaction of heterotic strings is described by using 
the string products, 
\begin{equation}
L_n\ :\ (\mathcal{H})^{\wedge n}\rightarrow \mathcal{H},
\end{equation}
satisfying the cyclic
$L_\infty$ algebra $(\mathcal{H}^{\textrm{res}},\Omega, \{L_n\})$. The action and gauge transformation have
the same form of the bosonic string field theory:
\begin{align}
    S\ =\ \sum_{n=1}^\infty \frac{1}{(n+1)!}\Omega(\Phi,L_{n}(\underbrace{\Phi,\dots,\Phi}_{n})),
\end{align}
and 
\begin{align}
    \delta\Phi=\sum_{n=1}^\infty\frac{1}{(n-1)!}L_{n}(\underbrace{\Phi,\dots,\Phi}_{n-1},\Lambda).
\end{align}
The heterotic string products, however, must have a proper picture number depending on 
the number of R string field in the input so that the interaction term does not vanish.
Suppose first that the number of R-string inputs is even and let it be $2r$.
Such an interaction term should be given, except for a combinatorial factor, by
\begin{align}
&\Omega(\Phi_{\textrm{NS}}, L_{n}(\underbrace{\Phi_{\textrm{NS}},\dots,\Phi_{\textrm{NS}}}_{n-2r},
\underbrace{\Phi_{\textrm{R}},\dots,\Phi_{\textrm{R}}}_{2r}))
\nonumber\\
&\hspace{5mm}
=\ \omega(\Phi_{\textrm{NS}},L_{n}(\underbrace{\Phi_{\textrm{NS}},\dots,\Phi_{\textrm{NS}}}_{n-2r},
\underbrace{\Phi_{\textrm{R}},\dots,\Phi_{\textrm{R}}}_{2r})).
\end{align}
Here, we used the fact that the R string field describes the space-time fermion  and thus
the output is in the NS sector. Since $\Phi_{\textrm{NS}}$ has picture number $-1$ and $\Phi_{\textrm{R}}$ 
has picture number $-1/2$, 
$L_{n}$ must have the picture number $n-r-1$ so that the total picture number is equal to $-2$
\cite{Friedan:1985ge}.
Next, suppose that the number of R-string inputs is odd, $2r+1$. 
The interaction term is given, except for a combinatorial factor, as
\begin{align}
    &\Omega(\Phi_{\textrm{R}},L_{n}(\underbrace{\Phi_{\textrm{NS}},\dots,
\Phi_{\textrm{NS}}}_{n-2r-1},\underbrace{\Phi_{\textrm{R}},\dots,\Phi_{\textrm{R}}}_{2r+1}))
\nonumber\\
&\hspace{5mm}
=\ \omega(\Phi_{\textrm{R}},YL_{n}(\underbrace{\Phi_{\textrm{NS}},\dots,
\Phi_{\textrm{NS}}}_{n-2r-1},\underbrace{\Phi_{\textrm{R}},\dots,\Phi_{\textrm{R}}}_{2r+1})).
\end{align}
The picture number that $L_{n}$ should have is $n-r-1$ also in this case.
In order to characterize these two case in common, it is convenient to
introduce the Ramond number:
\begin{align}
    \textrm{Ramond number}\ =\ \textrm{number of Ramond inputs}\ -\ \textrm{number of Ramond outputs}\,,
    \nonumber
\end{align}
which is equal to $2r$ in the above two cases.
Using the Ramond number, we can denote the above two cases as two components of string product $L_{n}^{(n-r-1)}\mid_{2r}$:\footnote{Note that if we fix the number of 
inputs the Ramond number 
has an upper bound by definition and $n-r$ cannot be negative. 
Hereafter, we consider the products with the Ramond number that is out of range to be zero.}
\begin{align}
    L_{n}^{(n-r-1)}|_{2r}=\pi^0L_{n}^{(n-r-1)}\mid_{2r}
(\underbrace{\pi^0,\dots,\pi^0}_{n-2r},\underbrace{\pi^1,
\dots,\pi^1}_{2r})+\pi^1L_{n}^{(n-r-1)}\mid_{2r}(\underbrace{\pi^0,\dots,\pi^0}_{n-2r-1},
\underbrace{\pi^1,\dots,\pi^1}_{2r+1}).
\end{align}
%
However, the Ramond number is not convenient to consider the cyclicity \eqref{eq:cyclicity condition} since it connects the products with different Ramond numbers:
\begin{align}
    \Omega\big(\Phi_{\textrm{NS}},&L_{n}^{(n-r-1)}\mid_{2r}(\underbrace{\Phi_{\textrm{NS}},\dots,\Phi_{\textrm{NS}}}_{n-2r},
\underbrace{\Phi_{\textrm{R}},\cdots,\Phi_{\textrm{R}}}_{2r})\big)\nonumber\\
 &=\ 
\Omega(\Phi_{\textrm{R}}, L_{n}^{(n-r)}\mid_{2r-2}(\underbrace{\Phi_{\textrm{NS}},\dots,\Phi_{\textrm{NS}}}_{n-2r+1},
\underbrace{\Phi_{\textrm{R}},\cdots,\Phi_{\textrm{R}}}_{2r-1})).
\label{cyc comp}
\end{align}
So we also introduce the cyclic Ramond number:
\begin{align}
    \textrm{cyclic Ramond number}\ =\ \textrm{number of Ramond inputs}\ +\ \textrm{number of Ramond outputs}\,,\nonumber
\end{align}
which is invariant under cyclic permutations.
We denote it as a superscript after the slash, as in $L_{n}\mid^{2r}$. 
The two components of $L_n\mid^{2r}$ have different picture numbers
 \begin{align}
    L_{n}|^{2r}\ =\ 
L_{n}^{(n-r-1)}\mid^{2r}_{2r}
(\underbrace{\pi^0,\dots,\pi^0}_{n-2r},\underbrace{\pi^1,\dots,\pi^1}_{2r})
+
L_{n}^{(n-r)}\mid^{2r}_{2r-2}(\underbrace{\pi^0,\dots,\pi^0}_{n-2r+1},
\underbrace{\pi^1,\dots,\pi^1}_{2r-1}).
\end{align}
and thus, $L_n\mid^{2r}$ does not have a definite picture number.

In addition, the R component of heterotic string products must satisfy the condition
 \begin{equation}
  \mathcal{P}_{XY}L_{n}^{(n-r)}\mid^{2r}_{2r-2}\ =\ L_{n}^{(n-r)}\mid^{2r}_{2r-2}
 \end{equation}
to be consistent with the constraint (\ref{XY constraint}).
This is automatically satisfied 
if $L_{n}\mid^{2r}$ has the form
\begin{equation}
 L_{n}\mid^{2r}=\mathcal{G}\tilde{L}_{n}^{(n-r-1)}\mid^{2r}.
\label{L to b}
\end{equation}
In this case, the cyclicity condition simply connects the two components of
\begin{equation}
\tilde{L}_{n}^{(n-r-1)}\mid^{2r}\ =\ \tilde{L}_{n}^{(n-r-1)}\mid^{2r}_{2r}
+\ \tilde{L}_{n}^{(n-r-1)}\mid^{2r}_{2r-2} 
\end{equation}
as
\begin{align}
    \omega\big(\Phi_{\textrm{NS}},&\tilde{L}_{n}^{(n-r-1)}\mid^{2r}_{2r}(\underbrace{\Phi_{\textrm{NS}},\dots,\Phi_{\textrm{NS}}}_{n-2r},
\underbrace{\Phi_{\textrm{R}},\cdots,\Phi_{\textrm{R}}}_{2r})\big)\nonumber\\
 &=\ 
\omega(\Phi_{\textrm{R}}, \tilde{L}_{n}^{(n-r-1)}\mid^{2r}_{2r-2}(\underbrace{\Phi_{\textrm{NS}},\dots,\Phi_{\textrm{NS}}}_{n-2r+1},
\underbrace{\Phi_{\textrm{R}},\cdots,\Phi_{\textrm{R}}}_{2r-1})).
\end{align}
A specific way to construct $L_{n}\mid^{2r}$ of the form (\ref{L to b})
was given in \cite{Kunitomo:2019glq},
which we extend, in section \ref{explicit}, to the interaction for the quantum action: the products satisfying the quantum $L_\infty$ algebra.

\sectiono{Quantum action with quantum $L_\infty$ structure}\label{sec:Quantum action}

Let us consider quantizing heterotic string field theory based on its $L_\infty$ structure.
For that, we need to apply the Batalin-Vilkovisky formalism \cite{Batalin:1981jr}, which involves the following steps. First, we introduce all necessary fields and anti-fields, including the space-time ghosts and anti-ghosts, by extending the classical string field to the quantum one\footnote{We use the same symbol for the quantum string field as for the classical one. Hereafter, the symbol $\Phi$ denotes the quantum string field unless specified.} while removing the ghost number restriction
(\ref{ghost number}). 
Under the anti-bracket defined by the symplectic structure $\Omega$, we can identify
the component $\phi_{\textrm{NS}}+\phi_{\textrm{R}}$ with the field and the component 
$\psi_{\textrm{NS}}+\psi_{\textrm{R}}$ with the anti-field in the so-called gauge-fixed basis, 
the basis in which the BV action reduces to the gauge-fixed action when
we remove all the anti-fields.
Then, if necessary, we incorporate the quantum corrections into the action so that
the quantum action $S$ satisfies the (quantum) BV master equation  \cite{Batalin:1983ggl}:
\begin{equation}
 \frac{1}{2}(S,S)+\hbar\Delta S\ =\ 0.
\label{quantum BV}
\end{equation}
For closed bosonic string field theory, it was shown that
the quantum action can be constructed by introducing the corrections 
so as to satisfy the quantum $L_\infty$ algebra \cite{Zwiebach:1992ie,Markl:1997bj}.

\subsection{Quantum $L_\infty$ algebra in bosonic string field theory}

Let us introduce the quantum $L_\infty$ algebra, which provides a quantum action satisfying the quantum BV equation (\ref{quantum BV}).
For clarity, we first consider the case of bosonic string field theory \cite{Zwiebach:1992ie}.
The quantum master action is given by
\begin{align}
    S=\sum_{g=0}^\infty \sum_{n=0}^\infty \frac{\hbar^g}{(n+1)!}\omega(\Phi,L_{g,n}(\underbrace{\Phi,\dots,\Phi}_n)),
\label{bosonic action}
\end{align}
where $L_{g,n}$ are the string products, 
\begin{equation}
L_{g,n}\ :\ \mathcal{H}^{\wedge n}\rightarrow \mathcal{H}, 
\end{equation}
which are extensions of the classical products $L_n=L_{0,n}$,
defined using the genus-$g$ Riemann surface with $(n+1)$ boundaries.
The quantum action (\ref{bosonic action}) satisfies the quantum BV master equation 
if the string products $\{L_{g,n}\}$ satisfy the quantum $L_\infty$ relations,
\begin{subequations}\label{eq:loop relations}
\begin{align}
    0\ =&\ {\sum_\sigma}\sum_{\substack{g_1+g_2=g\\k+l=n+1}}
\frac{(-1)^{\epsilon(\sigma)}}{l!\,(n-l)!}L_{g_1,k}(L_{g_2,l}
(\Phi_{\sigma(1)},\dots,\Phi_{\sigma(l)}),\Phi_{\sigma(l+1)},\dots,\Phi_{\sigma(n)})
\nonumber\\
 &\hspace{35mm}
+ \frac{1}{2}\sum_sL_{g-1,n+2}(\Tilde{e}_s,e_s,\Phi_1,\dots,\Phi_n),\label{eq:main id}
\end{align}
and cyclicity
\begin{align}
    \omega(\Phi_1,L_{g,n}(\Phi_2,\dots,\Phi_{n+1}))
=-(-1)^{\Phi_1}\omega(L_{g,n}(\Phi_1,\dots,\Phi_n),\Phi_{n+1}),\label{eq:Quantum cyclic}
\end{align}
\end{subequations}
where $\{e_s\}$ is a basis of $\mathcal{H}$ and $\{\Tilde{e}_s\}$ is its dual in the sense that
\begin{align}\label{eq: bosonic basis relation}
    \omega(\Tilde{e}_s,e_t)\ =\  (-1)^{e_s}\delta_{s,t}.
\end{align}
By the definition of $\omega$ in Eq.~(\ref{symp b}),
the degrees of $e_s$ and $\Tilde{e}_s$ are opposite:
\begin{align}
\textrm{deg}(\tilde{e}_s)\ =\ \textrm{deg}(e_s) + 1,\qquad \mod{2}.
\end{align}
If $(\mathcal{H},\omega,\{L_{g,n}\})$ satisfies the relations \eqref{eq:loop relations}, 
it is called a quantum $L_\infty$ algebra \cite{Munster:2011ij} (or a loop homotopy algebra \cite{Markl:1997bj}). It is an extension of the cyclic $L_\infty$ algebra\footnote{
These properties of the basis are summarized in Appendix~\ref{prop of basis} to make the paper self-contained.
}.

As in the classical case, it is convenient to use the coalgebra representation.
In preparation, we introduce an element $\theta_{\textrm{B}}$ as \cite{Markl:1997bj}:
\begin{equation}
\theta_{\textrm{B}}\ \equiv \sum_s e_s\otimes \tilde{e}_s.
\label{reflector B}
\end{equation}
It generates a (zero-length) handle by gluing the two inputs of $L_{g,n}$\,, thereby increasing its genus by one.
Due to the symmetric property 
\begin{equation}
 \theta_{\textrm{B}}\ 
=\ \sum_se_s\otimes \tilde{e}_s\ 
=\ \sum_s \tilde{e}_s\otimes e_s\ 
=\frac{1}{2}\sum_s e_s \wedge \tilde{e}_s\ \in\mathcal{H}^{\wedge2},
\label{theta sym}
\end{equation}
we can regard $\theta_B$ as a degree-odd 
map from $\mathcal{H}^{\wedge 0}$ to $\mathcal{H}^{\wedge 2}$.
The basic properties of the $\theta_{\textrm{B}}$ are summarized in Appendix~\ref{sec:reflectors}\,.
One of the important properties of $\theta_{\textrm{B}}$ is
\begin{equation}\label{Q theta}
 (Q\wedge\id)\theta_{\textrm{B}}\ =\ 0.
\end{equation}
%
%

From these maps, we can define the operators acting on $\mathcal{SH}$ as
\begin{align}
 \bd{L}_{g,n}\ =&\ \sum_{m=0}^\infty (L_{g,n}\wedge\id_m)\pi_{n+m},\\
 \bd{\theta}_{\textrm{B}}\ =&\ \sum_{m=0}^\infty (\theta_{\textrm{B}}\wedge\id_m)\pi_m.
\end{align}
as in the classical case. However, the map $\bd{\theta}_B$ is not a coderivation,
but rather its generalization known as a (strict) coderivation of order two \cite{Munster:2011ij}.
In general, an operator constructed similarly from a map with $r$ or fewer outputs is called a coderivation of order $r$\,\footnote{If the number of outputs of the map
is exactly equal to $r$, it is called a strict coderivation of order $r$.}.
In this sense, we refer to the conventional coderivation as a (strict) coderivation of order one, if necessary.
If we define
\begin{align}
    \bd{L}\ =\ \sum_{g,n}\hbar^g\bd{L}_{g,n},
\end{align}
then, the relations \eqref{eq:loop relations} can be combined into a single equation and written as
\begin{align}\label{eq:loop relation coderibative}
    [\bd{L}+\hbar\bd{\theta}_{\textrm{B}},\bd{L}+\hbar\bd{\theta}_{\textrm{B}}]\ =\ 0.
\end{align}
This is equivalent to the two equations
\begin{subequations} 
\begin{align}
 \pi_1[\bd{L}+\hbar\bd{\theta}_{\textrm{B}},\bd{L}+\hbar\bd{\theta}_{\textrm{B}}]\ =&\ 0,
\label{condition 1}\\
\pi_2[\bd{L}+\hbar\bd{\theta}_{\textrm{B}},\bd{L}+\hbar\bd{\theta}_{\textrm{B}}]\ =&\ 0,
\label{condition 2}
\end{align}
\end{subequations}
since $\bd{L}+\hbar\bd{\theta}_{\textrm{B}}$ is a coderivation of order two.
Using $[\bd{Q},\bd{\theta}_{\textrm{B}}]=0$ (which follows from the relation in Eq.~(\ref{Q theta})),
the first condition in Eq.~(\ref{condition 1}) is nothing other than the quantum $L_\infty$ relations
in Eq.~\eqref{eq:main id}. The second condition in Eq.~(\ref{condition 2}) is equivalent 
to the relation 
 \begin{equation}
\sum_s\,L_{g,n+1}(\id^{\wedge n}\wedge e_s)\wedge \tilde{e}_s\ =\ 0,
\label{eq:second relation}
\end{equation}
under the first condition, and holds if the cyclicity condition in Eq.~\eqref{eq:Quantum cyclic} holds as is shown in Appendix~\ref{sec:cyclicity}.
In the classical limit $\hbar=0$, the equation \eqref{eq:loop relation coderibative}
reduces to the $L_\infty$ relation
\begin{align}
    [\bd{L}_{\textrm{cl}}, \bd{L}_{\textrm{cl}}]\ =\ 0,\qquad \bd{L}_{\textrm{cl}}\ =\ \sum_{n=0}^\infty\bd{L}_{0,n+1},
\end{align}
and the quantum $L_\infty$ algebra $(\mathcal{H},\omega_{\textrm{B}},\bd{L}+\hbar\bd{\theta}_{\textrm{B}})$ reduces to 
the cyclic $L_\infty$ algebra $(\mathcal{H},\omega_{\textrm{B}},\bd{L}_{\textrm{cl}})$.

\subsection{Quantum action for heterotic string field theory}

The quantum action for heterotic string field theory has the same form as that
of the bosonic closed string field theory:
\begin{align}
    S=\sum_{g=0}^\infty \sum_{n=0}^\infty\frac{\hbar^g}{(n+1)!}\Omega(\Phi,L_{g,n}(\underbrace{\Phi,\dots,\Phi}_n)).
\label{q action het}
\end{align}
The heterotic string products are defined by the map
\begin{equation}
 L_{g,n}\ :\ (\mathcal{H}^{\textrm{res}})^{\wedge n}\ \longrightarrow\ \mathcal{H}^{\textrm{res}},
\end{equation}
using genus-$g$ Riemann surface with $(n+1)$ boundaries. Its \textit{spin structures} 
are realized by two sectors (NS and R) and GSO projection.
As it is determined only by the types of inputs,
the proper picture number for a given product $L_{g,n}$ is independent of $g$.
However, the quantum ($g\ne0$) products require $2g$ more PCO insertions than the classical ones
since they have picture number of $-2g$ without PCO insertions \cite{Alvarez-Gaume:1988lei}.
We indicate the number of inserted PCOs in the superscript for a quantum product, 
which is more convenient for construction than denoting the proper picture number:
\begin{align}
    L_{g,n}=\sum_rL_{g,n}^{(2g+n-r-1)}|_{2r}.
\end{align}

We introduce bases of $\mathcal{H}_{\textrm{NS}}^{(p)}$ with general picture number $p\in\mathbb{Z}$ and of $\mathcal{H}_{\textrm{R}}^{(p)}$ with $p\in\mathbb{Z}+\frac{1}{2}$ by $\{e^0_{(s,p)}\}$ and $\{e^1_{(s,p)}\}$, respectively. The bases $\{\tilde{e}^0_{(s,p)}\}$ and $\{\tilde{e}^1_{(s,p)}\}$
are their duals in the sense of the symplectic form $\omega$ in the small Hilbert space:
\begin{align}
    \omega(\tilde{e}_{(s,p)}^a,e_{(t,q)}^b)\ =\ (-1)^{\tilde{e}_s^a}\delta^{a,b}\delta_{s,t}\delta_{p+q+2,0}. 
\label{small basis}
\end{align}
Similar to the bosonic case, we can show that
\begin{align}
\sum_s e^a_{(s,p)}\otimes \tilde{e}^a_{(s,-p-2)}\ =\ \sum_s\tilde{e}^a_{(s,p)}\otimes e^a_{(s,-p-2)}\,.
\label{exchange e and e-tilde}
\end{align}

We define an element $\theta=\theta\mid^0+\theta\mid^2\in\mathcal{H}^{\wedge2}$ by
\begin{align}
 \theta\mid^0\ =&\ \sum_s e^0_{(s,-1)}\otimes \tilde{e}^0_{(s,-1)}\ =\
\frac{1}{2}\sum_s e^0_{(s,-1)}\wedge \tilde{e}^0_{(s,-1)},
\nonumber\\
\theta\mid^2\ =&\ \frac{1}{2}\sum_s\left(
e^1_{(s,-3/2)}\otimes\tilde{e}^1_{(s,-1/2)}
+ \tilde{e}^1_{(s,-1/2)}\otimes e^1_{(s,-3/2)} 
\right) \ =\
\frac{1}{2}\sum_s e^1_{(s,-3/2)}\wedge \tilde{e}^1_{(s,-1/2)}\,.
\label{reflector}
\end{align}
Each of them satisfies the relation
\begin{equation}
 (Q\wedge\id)\,\theta\mid^0\ =\ 0\,,\qquad
 (Q\wedge\id)\,\theta\mid^2\ =\ 0\,,
\end{equation}
independently.\footnote{The operator form and its basic properties are given in Appendix~\ref{sec:reflectors}.}
We also introduce an element $\Theta=\Theta\mid^0+\Theta\mid^2$ in the restricted space by
\begin{subequations}\label{def Theta}
    \begin{align}
\Theta\mid^0\ =&\ \theta\mid^0,\\
\Theta\mid^2\ =&\ \frac{1}{2}\sum_s Xe^1_{(s,-3/2)} \wedge \tilde{e}^1_{(s,-1/2)} 
\nonumber\\
=&\ \frac{1}{2}\sum_s X\tilde{e}^1_{(s,-3/2)} \wedge e^1_{(s,-1/2)}\,.
\end{align}
\end{subequations}
Note that $\{Xe^1_{(s,-3/2)}\}$ is a basis of $\overline{\mathcal{H}}_{\textrm{R}}^{(-1/2)}$.
The R component, $\Theta\mid^2$, is BRST exact in the large Hilbert space:
\begin{subequations}
    \begin{align}
    \Theta\mid^2\ =&\ [Q,\Xi\theta\mid^2],\\
    \Xi\theta\mid^2\ =&\ \frac{1}{2}\sum_s \Xi e^1_{(s,-3/2)} \wedge \tilde{e}^1_{(s,-1/2)}
    \nonumber\\ 
    =&\ \frac{1}{2}\sum_s \Xi \tilde{e}^1_{(s,-3/2)} \wedge e^1_{(s,-1/2)}\,.
\end{align}
\end{subequations}
Similar to the bosonic string field theory, we define the (strict) coderivations of 
order two  $\bd{\theta}$, $\bd{\Theta}$, and $\bd{\Xi\theta}$ as
\begin{equation}
 \bd{\theta}\ =\ \sum_{m=0}^\infty(\theta\wedge\id_m)\pi_m,\qquad
 \bd{\Theta}\ =\ \sum_{m=0}^\infty(\Theta\wedge\id_m)\pi_m,\qquad
 \bd{\Xi\theta}\mid^2\ =\ \sum_{m=0}^\infty(\Xi\theta\mid^2\wedge\id_m)\pi_m.
\label{Theta def}
\end{equation}
If we consider the coderivation of order two $\bd{L}+\hbar\bd{\Theta}$
the action (\ref{q action het}) is a solution to the quantum BV equation (\ref{quantum BV})
if $(\mathcal{H}^{\textrm{res}},\Omega,\bd{L}+\hbar\bd{\Theta})$ is a quantum $L_\infty$ algebra:
\begin{align}
    [\bd{L}+\hbar\bd{\Theta}, \bd{L}+\hbar\bd{\Theta}]\ =\ 0\,,
\end{align}
or equivalently
\begin{subequations}
\begin{align}
    \pi_1[\bd{L}+\hbar\bd{\Theta}, \bd{L}+\hbar\bd{\Theta}]\ =\ 0\,,\label{q het main}\\ 
    \pi_2[\bd{L}+\hbar\bd{\Theta}, \bd{L}+\hbar\bd{\Theta}]\ =\ 0\,.\label{q het cyc}
\end{align}
\end{subequations}
The former constitutes the quantum $L_\infty$ relations, whereas the latter reduces, upon using the quantum $L_\infty$ relations, to the conditions 
\begin{subequations}\label{cyc hetero}
\begin{align}
&\   \sum_s  L_{g,n+1}^{(2g+n-r)}\mid^{2r}_{2r}(\id^{\wedge n}\wedge e^0_{(s,-1)})
    \wedge \tilde{e}^0_{(s,-1)}\ =\ 0\,,\label{q het cyc 1}\\
&\   \sum_s L_{g,n+1}^{(2g+n-r+1)}\mid^{2r}_{2r-2}(\id^{\wedge n}\wedge e^1_{(s,-1/2)})
    \wedge X\tilde{e}^1_{(s,-3/2)}\ =\ 0\,,\label{q het cyc 2}\\
&\   \sum_s \Big(L_{g,n+1}^{(2g+n-r)}\mid^{2r}_{2r}(\id^{\wedge n}\wedge  e^1_{(s,-1/2)})
    \wedge X\tilde{e}^1_{(s,-3/2)} \nonumber\\
&\qquad\qquad  + L_{g,n+1}^{(2g+n-r+1)}\mid^{2r}_{2r-2}(\id^{\wedge n}
    \wedge e^0_{(s,-1)})\wedge \tilde{e}^0_{(s,-1)} \Big)\ =\ 0\,,\label{q het cyc 3}
\end{align}
\end{subequations}
which hold if $L$ is cyclic with respect to $\Omega$ (analogous to the bosonic case).\footnote{
We have used the relation $\sum_sXe^1_{(s,-3/2)}\otimes\tilde{e}^1_{(s,-1/2)}=\sum_se^1_{(s,-1/2)}\otimes X\tilde{e}^1_{(s,-3/2)}$\,, which is derived in Appendix~\ref{sec:reflectors}.
}
We give a specific way to construct $\{L_{g,n}^{(2g+n-r-1)}\mid_{2r}\}$ in the next section.

\sectiono{
Construction of quantum heterotic string products}\label{explicit}

Now, let us construct quantum heterotic string products that constitute the quantum action of heterotic string field theory. Before considering a way to construct general products, we first solve the quantum $L_\infty$ relations in a few simple cases to see how to proceed. Next, we propose a prescription for constructing the general string products to obtain the quantum action. It must be noted that the following arguments do not take into account the existence of spurious poles. Some modifications will be necessary if this causes any problems.

\subsection{Explicit constructions for simple cases}

Following the prescription given in Refs.~\cite{Erler:2013xta,Erler:2014eba}, we individually examine the leading-order quantum correction, that is, the one-loop correction, for a few simple cases.

\subsubsection{Tadpole product}

The simplest correction is the tadpole product $L_{1,0}^{(1)}\mid^0_0$ (no input and one output), which was discussed and explicitly constructed in Ref.~\cite{Erler:2017pgf}. Let us start by examining how we can construct it.
The equation of the quantum $L_{\infty}$ algebra that $L_{1,0}^{(1)}$ must satisfy is
\begin{align}
[\bd{Q}, \bd{L}_{1,0}^{(1)}\mid^0_0\,]+[\bd{L}_{0,2}^{(1)}\mid^0_0,\bd{\Theta}\mid^0]
    +[\bd{L}_{0,2}^{(0)}\mid^2_2,\bd{\Theta}\mid^2]\ =\ 0\,,
\label{tad pole}
\end{align}
using the coalgebra representation. Since the left-hand side $\bd{I}_{1,0}$ is BRST-closed, we can rewrite it in a BRST-exact form by extending it in the large Hilbert space as an intermediate step.
Actually, substituting the expression for the classical product $\bd{L}_{0,2}^{(1)}\mid^0_0=[\bd{Q},\bd{\lambda}_{0,2}^{(1)}\mid^0_0\,]$\,,  
we can rewrite it as
\begin{align}
 [\bd{Q},\left(\bd{L}_{1,0}^{(1)}\mid^0_0 + [\bd{\lambda}_{0,2}^{(1)}\mid^0_0,\bd{\Theta}\mid^0]
 -[\bd{L}_{0,2}^{(0)}\mid^2_2,\bd{\Xi\theta}\mid^2]\right)]\ =\ 0\,,
 \label{tad pole 2}
\end{align}
using  $\bd{\Theta}\mid^2=[\bd{Q},\bd{\Xi\theta}\mid^2]$\,. 
Then, $\bd{L}_{1,0}^{(1)}\mid^0_0$ can be solved as
\begin{align}
    \pi_1\bd{L}_{1,0}^{(1)}\mid^0_0\ =\ \pi_1\left([\bd{Q},\bd{\lambda}_{1,0}^{(1)}\mid^0_0]-[\bd{\lambda}_{0,2}^{(1)}\mid^0_0,\bd{\Theta}\mid^0]
 +[\bd{L}_{0,2}^{(0)}\mid^2_2,\bd{\Xi\theta}\mid^2]\right)\,.
\end{align}
The part where the outputs go into $\mathcal{H}\wedge\mathcal{H}$ holds
because the product $L_{0,2}$ is cyclic (with respect to $\Omega$)\,.
The even coderivation (gauge product) $\bd{\lambda}_{1,0}^{(1)}\mid^0_0$ contained in a trivially BRST-invariant term is determined so that $\bd{L}_{1,0}^{(1)}\mid^0_0$ is in the small Hilbert space, that is, commutes with $\bd{\eta}$:
\begin{align}
   0\ =&\  [\bd{\eta},\bd{L}_{1,0}^{(1)}\mid^0_0\,]\nonumber\\
   =&\ -\left([\bd{Q},[\bd{\eta},\bd{\lambda}_{1,0}^{(1)}\mid^0_0\,]] 
   +[\bd{L}_{0,2}^{(0)}\mid^0_0,\bd{\Theta}\mid^0]+[\bd{L}_{0,2}^{(0)}\mid^2_2,\bd{\theta}\mid^2]\right)\,,
   \label{tad pole 3}
\end{align}
where we used
\begin{align}
    [\bd{\eta},\bd{\lambda}_{0,2}^{(1)}\mid^0_0]\ =\ \bd{L}_{0,2}^{(0)}\mid^0_0,\qquad
    [\bd{\eta},\bd{\Xi\theta}\mid^2]\ =\ \bd{\theta}\mid^2\,.
\end{align}
If we set 
\begin{align}
    [\bd{\eta},\bd{\lambda}_{1,0}^{(1)}\mid^0_0]=\bd{L}_{1,0}^{(0)}\mid^0_0\,,
    \label{lambda10}
    \end{align}
Eq.~(\ref{tad pole 3}) becomes an equation of the quantum $L_{\infty}$ algebra for the tadpole product without PCO-insertion, 
\begin{align}
&\    [\bd{Q},\bd{L}_{1,0}^{(0)}\mid^0_0] 
   +[\bd{L}_{0,2}^{(0)}\mid^0_0,\bd{\theta}\mid^0]+[\bd{L}_{0,2}^{(0)}\mid^2_2,\bd{\theta}\mid^2]\ =\ 0\,.
\end{align}
The tadpole product $L_{1,0}^{(0)}\mid^0_0$ satisfying this equation can be constructed 
in a similar manner to that of the bosonic closed string field theory. For example, we can take
\begin{align}
    \lambda_{1,0}^{(1)}\mid^0_0\ =\ \xi L_{1,0}^{(0)}\mid^0_0\,,
\end{align}
as the simplest solution of (\ref{lambda10}). 
The tadpole product $\bd{L}_{1,0}^{(1)}\mid^0_0$ has thus been constructed.

\subsubsection{R-R two-point product}

The second simplest correction is the two-point product $\bd{L}_{1,1}^{(2)}$. It has one input and one output, both in the NS sector or the R sector. 
We consider the R-R two-point product $\bd{L}_{1,1}^{(2)}\mid^2_0$ first since the construction is simpler. The equation that $\bd{L}_{1,1}^{(2)}\mid^2_0$ must satisfy is given by
\begin{align}
[\bd{Q},\bd{L}_{1,1}^{(2)}\mid^2_0]+[\bd{L}_{0,2}^{(1)}\mid^2_0,\bd{L}_{1,0}^{(1)}\mid^0_0]
    +[\bd{L}_{0,3}^{(2)}\mid^2_0,\bd{\Theta}\mid^0]+[\bd{L}_{0,3}^{(1)}\mid^4_2,\bd{\Theta}\mid^2]\ =\ 0\,.
    \label{RR}
\end{align}
Again, one can show that the left-hand side is BRST-closed using a quantum $L_\infty$ relation for the tadpole product (\ref{tad pole}).
However, unlike the tadpole case, even if we rewrite it using the expressions for tree-level products,
\begin{align}
    \pi_1\bd{L}_{0,2}^{(1)}\mid^2_0\ =&\ X\pi_1\bd{L}_{0,2}^{(0)}\mid^2_0\,,\nonumber\\
    \pi_1\bd{L}_{0,3}^{(2)}\mid^2_0\ =&\ X\pi_1\left([\bd{Q},\bd{\lambda}_{0,3}^{(1)}\mid^2_0]
    +[\bd{L}_{0,2}^{(0)}\mid^2_0,\bd{\lambda}_{0,2}^{(1)}\mid^0_0]+\bd{L}_{0,2}^{(0)}\mid^2_0 \bd{\Xi L}_{0,2}^{(0)}\mid^2_0\right)\,,
    \nonumber\\
    \pi_1\bd{L}_{0,3}^{(1)}\mid^4_2\ =&\ X\pi_1\bd{L}_{0,3}^{(0)}\mid^4_2\,,
\end{align}
it appears that there are extra terms that at first glance cannot be written in a BRST-exact form:
\begin{align}
 &\ \pi_1[\bd{Q},X\left(\bd{\overline{L}}_{1,1}^{(1)}\mid^2_0-[\bd{L}_{0,2}^{(0)}\mid^2_0,\bd{\lambda}_{1,0}^{(1)}\mid^0_0]+[\bd{\lambda}_{0,3}^{(1)}\mid^2_0,\bd{\Theta}\mid^0]-[\bd{L}_{0,3}^{(0)}\mid^4_2,
    \bd{\Xi\theta}\mid^2]\right)] + X\mathcal{E}\ =\ 0\,,\\
 &\ \mathcal{E}\ =\ -\frac{1}{2}\sum_s\Bigg(L_{0,2}^{(0)}\mid^2_0\left(L_{0,2}^{(0)}\mid^2_2(\id\wedge\Xi  e^1_{(s,-3/2)})\wedge \tilde{e}^1_{(s,-1/2)}\right)
 \nonumber\\
   &\hspace{50mm} 
- (-1)^{e^1_s}L_{0,2}^{(0)}\mid^2_0\left(L_{0,2}^{(0)}\mid^2_2
(\id\wedge e^1_{(s,-1/2)})\wedge \Xi \tilde{e}^1_{(s,-3/2)}\right)\Bigg)\,,
    \label{RR 2}
\end{align}
where $\bd{\overline{L}}_{1,1}^{(1)}$ is defined by $\pi_1\bd{L}_{1,1}^{(2)}\mid^2_0=X\pi_1\bd{\overline{L}}_{1,1}^{(1)}\mid^2_0$ so that
the output R-state is in the restricted space. In order to write $\mathcal{E}$ in this form,
we used the relation
\begin{align}\label{R reflector in large}
&\ \sum_s\Bigg(\Xi L_{0,2}^{(0)}\mid^2_0(\id\wedge e^0_{(s,-1)})\wedge \tilde{e}^0_{(s,-1)}
    \nonumber\\
&\hspace{30mm}  + (-1)^{e^1_s}L_{0,2}^{(0)}\mid^2_2(\id \wedge e^1_{(s,-1/2)})
    \wedge\Xi \tilde{e}^1_{(s,-3/2)}\Bigg)\ =\ 0\,,
\end{align}
which follows from the cyclicity of $L$ with respect to $\omega$\,.
At first glance, this $\mathcal{E}$ is nonzero and appears to give a problematic contribution.
However, as detailed in Appendix~\ref{sec:reflectors}, the factor
\begin{align}
&\    - \sum_s\left(\Xi e^1_{(s,-3/2)}\otimes \tilde{e}^1_{(s,-1/2)}-(-1)^{e^1_s}e^1_{(s,-1/2)}\otimes \Xi \tilde{e}^1_{(s,-3/2)} \right)\ 
\nonumber\\
& =\ 
    -(\Xi\otimes\id)\sum_se^1_{(s,-3/2)}\otimes \tilde{e}^1_{(s,-1/2)}+(\id\otimes\Xi)\sum_{s}e^1_{(s,-1/2)}\otimes \tilde{e}^1_{(s,-3/2)}\,,
    \label{vanish}
\end{align}
appearing in $\mathcal{E}$ represents the (picture-number-projected) reflector in the large Hilbert space. 
It sews the two inputs of the string products and effectively generates a handle \textit{in the large Hilbert space} with a picture-number projection inserted, the contribution of which vanishes due to the absence of an $\eta$ zero mode \cite{Carow-Watamura:1989vmd,Carow-Watamura:1989cxv,Lechtenfeld:1989wu,DiVecchia:1988tc}. Hence, $\mathcal{E}$ vanishes. As also demonstrated in the next subsection, this cancellation mechanism is required throughout the process and plays a crucial role in establishing the quantum $L_\infty$ algebra.

Now, we can solve Eq.~(\ref{RR}) and determine $\overline{\bd{L}}_{1,1}^{(1)}\mid^2_0$ as
\begin{align}
    \pi_1\bd{\overline{L}}_{1,1}^{(1)}\mid^2_0\ =\ \pi_1\left([\bd{Q},\bd{\lambda}_{1,1}^{(1)}\mid^2_0]+[\bd{L}_{0,2}^{(0)},\bd{\lambda}_{1,0}^{(1)}\mid^0_0]-[\bd{\lambda}_{0,3}^{(1)}\mid^2_0,\bd{\Theta}\mid^0]+[\bd{L}_{0,3}^{(0)}\mid^4_2,\bd{\Xi\theta}\mid^2]
    \right)\,.
\end{align}
The gauge product $\bd{\lambda}_{1,1}^{(1)}\mid^2_0$ is determined by requiring
\begin{align}
    0\ =&\ - [\bd{\eta},\bd{\overline{L}}_{1,1}^{(1)}\mid^2_0]
    \nonumber\\
    =&\ [\bd{Q},[\bd{\eta},\bd{\lambda}_{1,1}^{(1)}\mid^2_0]]+[\bd{L}_{0,2}^{(0)},\bd{L}_{1,0}^{(0)}\mid^0_0]+[\bd{L}_{0,3}^{(0)}\mid^2_0,\bd{\Theta}\mid^0]+[\bd{L}_{0,3}^{(0)}\mid^4_2,\bd{\theta}\mid^2]\,.
\end{align}
If we set $[\bd{\eta},\bd{\lambda}_{1,1}^{(1)}\mid^2_0]=\bd{L}_{1,1}^{(0)}\mid^2_0$, this becomes
a quantum $L_\infty$-relation,
\begin{align}
  [\bd{Q},\bd{L}_{1,1}^{(0)}\mid^2_0]+[\bd{L}_{0,2}^{(0)},\bd{L}_{1,0}^{(0)}\mid^0_0]+[\bd{L}_{0,3}^{(0)}\mid^2_0,\bd{\theta}\mid^0]+[\bd{L}_{0,3}^{(0)}\mid^4_2,\bd{\theta}\mid^2]\ =\ 0\,,
\end{align}
as for the string products without PCO insertion, which can be obtained in a similar manner to the bosonic product.

\subsubsection{NS-NS two-point}

We can obtain the NS-NS two-point product $\bd{L}_{1,1}^{(2)}\mid^0_0$ in a similar manner, although it requires more steps. Using the expressions for the classical products,
\begin{align}
    2\bd{L}_{0,3}^{(2)}\mid^0_0\ =&\ [\bd{Q},\bd{\lambda}_{0,3}^{(2)}\mid^0_0]+[\bd{L}_{0,2}^{(1)}\mid^0_0,\bd{\lambda}_{0,2}^{(1)}\mid^0_0]\,,
    \nonumber\\
    \pi_1\bd{L}_{0,3}^{(1)}\mid^2_2\ =&\ \pi_1\left([\bd{Q},\bd{\lambda}_{0,3}^{(1)}\mid^2_2]+[\bd{L}_{0,2}^{(0)}\mid^2_2,\bd{\lambda}_{0,2}^{(1)}\mid^0_0] +\bd{L}_{0,2}^{(0)}\mid^2_2\bd{\Xi L}_{0,2}^{(0)}\mid^2_0
    \right)\,,
\end{align}
the quantum $L_{\infty}$ relation for $\bd{L}_{1,1}^{(2)}\mid^0_0$,
\begin{align}
  [\bd{Q},\bd{L}_{1,1}^{(2)}\mid^0_0]+[\bd{L}_{0,2}^{(1)}\mid^0_0,\bd{L}_{1,0}^{(1)}\mid^0_0]+[\bd{L}_{0,3}^{(2)}\mid^0_0,\bd{\Theta}\mid^0]+[\bd{L}_{0,3}^{(1)}\mid^2_2,\bd{\Theta}\mid^2]\ =\ 0\,,
\end{align}
can be written as
\begin{align}
    0\ =&\ \Big[\bd{Q},\Big(2\bd{L}_{1,1}^{(2)}\mid^0_0-[\bd{L}_{1,0}^{(1)}\mid^0_0,\bd{\lambda}_{0,2}^{(1)}\mid^0_0]-[\bd{L}_{0,2}^{(1)}\mid^0_0,\bd{\lambda}_{1,0}^{(1)}\mid^0_0]
\nonumber\\
&\qquad      +[\bd{\lambda}_{0,3}^{(2)}\mid^0_0,\bd{\Theta}\mid^0]+[\bd{\lambda}_{0,3}^{(1)}\mid^2_2,\bd{\Theta}\mid^2]-[\bd{L}_{0,3}^{(1)}\mid^2_2,\bd{\Xi\theta}\mid^2]
    \Big)\Big]  + \mathcal{E}\,.
    \label{NSNS 1}
\end{align}
The additional contribution $\mathcal{E}$ in this case involves more terms than the previous case, and more considerations are needed to show that they vanish.
Some of them are terms containing a handle generated by 
\begin{align}
    L_{0,2}^{(1)}\mid^0_0(\id \wedge e^0_{(s,-1)})\wedge \tilde{e}^0_{(s,-1)}\qquad\textrm{or}\qquad 
    \lambda_{0,2}^{(1)}\mid^0_0(\id \wedge e^0_{(s,-1)})\wedge \tilde{e}^0_{(s,-1)}\,.
\end{align}
The former vanishes due to the cyclicity of $L_{0,2}^{(1)}\mid^0_0$, whereas the latter does not since
$\lambda_{0,2}^{(1)}\mid^0_0$ is not closed in the small Hilbert space and therefore is not cyclic with respect to $\Omega$\,. 
Alternatively, it can be rewritten in a similar form to (\ref{R reflector in large}):
\begin{align}
&\    \sum_{s}\lambda_{0,2}^{(1)}\mid^0_0(\id \wedge e^0_{(s,-1)})\wedge \tilde{e}^0_{(s,-1)}\ 
\nonumber\\
&\ =\ 
    - \frac{1}{2}\sum_s\Big(L_{0,2}^{(0)}\mid^0_0(\id \wedge \xi e^0_{(s,-1)})\wedge \tilde{e}^0_{(s,-1)}
    -(-1)^{e^0_s}L_{0,2}^{(0)}\mid^0_0(\id \wedge e^0_{(s,0)})\wedge\xi \tilde{e}^0_{(s,-2)}
\Big)\,.
\end{align}
Using this relation and the relation following from the cyclicity,
\begin{align}
&\    \sum_s\Xi L_{0,2}^{(0)}\mid^2_0(\id \wedge e^1_{(s,-1/2)})\wedge X\tilde{e}^1_{(s,-3/2)}
 =\    \sum_s(-1)^{\tilde{e}_s^1}L_{0,2}^{(1)}\mid^2_0(\id \wedge e^1_{(s,-1/2)})\wedge\Xi \tilde{e}^1_{(s,-3/2)}\,,
\end{align}
the extra terms $\mathcal{E}$ are now written as
\begin{align}
\mathcal{E}  =&\ 
    -\frac{1}{2}\sum_s\Bigg(L_{0,2}^{(1)}\mid^0_0\left(L_{0,2}^{(0)}\mid^0_0(\id\wedge\xi e^0_{(s,-1)})\wedge \tilde{e}^0_{(s,-1)}\right)\nonumber\\
 &\hspace{30mm}   - (-1)^{e^0_s}L_{0,2}^{(1)}\mid^0_0\left(L_{0,2}^{(0)}\mid^0_0(\id\wedge e^0_{{(s,0)}})\wedge \xi \tilde{e}^0_{(s,-2)}\right)\Bigg)
\nonumber\\
&\
    -\frac{1}{2}\sum_s\Bigg(L_{0,2}^{(0)}\mid^2_2\left(L_{0,2}^{(1)}\mid^2_0(\id \wedge\Xi e^1_{(s,-3/2)})\wedge \tilde{e}^1_{(s,-1/2)}\right)\nonumber\\
    &\hspace{30mm} - (-1)^{e^1_s}L_{0,2}^{(0)}\mid^2_2\left(L_{0,2}^{(1)}\mid^2_0(\id \wedge e^1_{(s,-1/2)})\wedge \Xi \tilde{e}^1_{(s,-3/2)}\right)\Bigg)\,,
    \label{NSNS extra 1}
\end{align}
which vanish for the same reason as before.
Therefore, from Eq.~(\ref{NSNS 1}), we can determine $\bd{L}^{(2)}_{1,1}\mid^0_0$ as
\begin{align}
    2\pi_1\bd{L}_{1,1}^{(2)}\mid^0_0\ =&\ \pi_1\Big([\bd{Q},\bd{\lambda}_{1,1}^{(2)}\mid^0_0]+[\bd{L}_{1,0}^{(1)}\mid^0_0,\bd{\lambda}_{0,2}^{(1)}\mid^0_0]+[\bd{L}_{0,2}^{(1)}\mid^0_0,\bd{\lambda}_{1,0}^{(1)}\mid^0_0]
    \nonumber\\
    &\qquad  
    -[\bd{\lambda}_{0,3}^{(2)}\mid^0_0,\bd{\Theta}\mid^0]-[\bd{\lambda}_{0,3}^{(1)}\mid^2_2,\bd{\Theta}\mid^2]+[\bd{L}_{0,3}^{(1)}\mid^2_2,\bd{\Xi\theta}\mid^2]
    \Big)\,.
\end{align}
Next, we determine $\bd{\lambda}_{1,1}^{(2)}\mid^0_0$ so that $\bd{L}_{1,1}^{(2)}\mid^0_0$ is in the small Hilbert space, that is, it commutes with $\bd{\eta}$:
\begin{align}
    0\ =&\ 2\pi_1[\bd{\eta},\bd{L}_{1,1}^{(2)}\mid^0_0]
    \nonumber\\
    =&\ - \pi_1\Big([\bd{Q},[\bd{\eta},\bd{\lambda}_{1,1}^{(2)}\mid^0_0]]
    +[\bd{L}_{0,2}^{(1)}\mid^0_0,\bd{L}_{1,0}^{(0)}\mid^0_0]
    +[\bd{L}_{0,2}^{(0)}\mid^0_0,\bd{L}_{1,0}^{(1)}\mid^0_0]
    \nonumber\\
    &\qquad \qquad
    +[\bd{L}_{0,3}^{(1)}\mid^0_0,\bd{\Theta}\mid^0]
    +[\bd{L}_{0,3}^{(0)}\mid^2_2,\bd{\Theta}\mid^2]
    +[\bd{L}_{0,3}^{(1)}\mid^2_2,\bd{\theta}\mid^2]
    \Big)\,.
\end{align}
If we set $[\bd{\eta},\bd{\lambda}_{1,1}^{(2)}\mid^0_0]\ =\ \bd{L}_{1,1}^{(1)}\mid^0_0$\,,
it leads to the relation in the next step:
\begin{align}
&\ [\bd{Q},\bd{L}_{1,1}^{(1)}\mid^0_0]
    +[\bd{L}_{0,2}^{(1)}\mid^0_0,\bd{L}_{1,0}^{(0)}\mid^0_0]
    +[\bd{L}_{0,2}^{(0)}\mid^0_0,\bd{L}_{1,0}^{(1)}\mid^0_0]
    \nonumber\\
&\qquad  
    +[\bd{L}_{0,3}^{(1)}\mid^0_0,\bd{\Theta}\mid^0]
    +[\bd{L}_{0,3}^{(0)}\mid^2_2,\bd{\Theta}\mid^2]
    +[\bd{L}_{0,3}^{(1)}\mid^2_2,\bd{\theta}\mid^2]\ =\ 0\,.
    \label{NSNS 2}
\end{align}
We now repeat the same steps. Using
\begin{align}
    \bd{L}_{0,3}^{(1)}\mid^0_0\ =\ [\bd{Q},\bd{\lambda}_{0,3}^{(1)}\mid^0_0]
    +[\bd{L}_{0,2}^{(0)}\mid^0_0,\bd{\lambda}_{0,2}^{(1)}\mid^0_0]\,, 
\end{align}
the relation given in (\ref{NSNS 2}) can be written as
\begin{align}
    0\ =&\ \pi_1\bigg([\bd{Q},\Big(\bd{L}_{1,1}^{(1)}\mid^0_0
- [\bd{L}_{1,0}^{(0)}\mid^0_0,\bd{\lambda}_{0,2}^{(1)}\mid^0_0]
-[\bd{L}_{0,2}^{(0)}\mid^0_0,\bd{\lambda}_{1,0}^{(1)}\mid^0_0]
\nonumber\\
&\ 
+[\bd{\lambda}_{0,3}^{(1)}\mid^0_0,\bd{\Theta}\mid^0]
-[\bd{L}_{0,3}^{(0)}\mid^2_2,\bd{\Xi\theta}\mid^2]
+[\bd{\lambda}_{0,3}^{(1)}\mid^2_2,\bd{\theta}\mid^2]
    \Big)] \bigg) + \mathcal{E}\,.
\end{align}
The relations
\begin{align}
&\     \sum_s\Bigg(\Xi L_{0,2}^{(0)}\mid^2_0(\id \wedge e^1_{(s,-3/2)})\wedge \tilde{e}^1_{(s,-1/2)}\ 
\nonumber\\    
    &\hspace{30mm}  +(-1)^{\tilde{e}^1_s}L_{0,2}^{(0)}\mid^2_0(\id\wedge e^1_{(s,1/2)})\wedge\Xi \tilde{e}^1_{(s,-5/2)}\Bigg)\ =\ 0\,,\\
&\ \sum_s\Bigg(\Xi L_{0,2}^{(0)}\mid^2_0(\id \wedge e^1_{(s,-1/2)})\wedge \tilde{e}^1_{(s,-3/2)}
\nonumber\\ 
    &\hspace{30mm} +(-1)^{\tilde{e}^1_s}L_{0,2}^{(0)}\mid^2_0(\id\wedge e^1_{(s,-1/2)})\wedge\Xi \tilde{e}^1_{(s,-3/2)}\Bigg)\ =\ 0\,,
\end{align}
following from cyclicity:\footnote{
This cyclicity relation is slightly different from (\ref{q het cyc 1}). This is because $L_{0,2}^{(0)}\mid^0_0$ has a deficit of one picture.}
\begin{align}
    \sum_s\Big( L_{0,2}^{(0)}\mid^0_0(\id\wedge e^0_{(s,-1)})\wedge \tilde{e}^0_{(s,-1)}
    + L_{0,2}^{(0)}\mid^0_0(\id\wedge e^0_{(s,0)})\wedge \tilde{e}^0_{(s,-2)}\Big)\ =\ 0\,,
\end{align}
allow us to rewrite the extra contributions as
\begin{align}
\mathcal{E}  =&\     -\frac{1}{2}\sum_s\Bigg( L_{0,2}^{(0)}\mid^2_2\Big(L_{0,2}^{(0)}\mid^2_0(\id\wedge\Xi e^1_{(s,-3/2)})
    \wedge \tilde{e}^1_{(s,-1/2)}\Big)
    \nonumber\\
&\hspace{30mm}
- (-1)^{e^1_s}L_{0,2}^{(0)}\mid^2_2
\Big(L_{0,2}^{(0)}\mid^2_0(\id\wedge e^1_{(s,1/2)})\wedge\Xi \tilde{e}^1_{(s,-5/2)}\Big)\Bigg)\,.
\label{NSNS extra 2}
\end{align}
This has a form similar to (\ref{NSNS extra 1}) but the picture number assignment is slightly different.
However, since the contribution of the loop is independent of the picture number of the states propagating in the loop \cite{Verlinde:1987sd,Alvarez-Gaume:1988lei,Lechtenfeld:1989wu}, we can always shift the picture number appropriately, and this contribution therefore vanishes 
as well.

Now, we can determine $\bd{L}_{1,1}^{(1)}\mid^0_0$ as
\begin{align}
    \pi_1\bd{L}_{1,1}^{(1)}\mid^0_0\ =&\ \pi_1\Big([\bd{Q},\bd{\lambda}_{1,1}^{(1)}\mid^0_0]
    +[\bd{L}_{1,0}^{(0)}\mid^0_0,\bd{\lambda}_{0,2}^{(1)}\mid^0_0]
    +[\bd{L}_{0,2}^{(0)}\mid^0_0,\bd{\lambda}_{1,0}^{(1)}\mid^0_0]
    \nonumber\\
    &\
    -[\bd{\lambda}_{0,3}^{(1)}\mid^0_0,\bd{\Theta}\mid^0]
    +[\bd{L}_{0,3}^{(0)}\mid^2_2,\bd{\Xi\theta}\mid^2]
    -[\bd{\lambda}_{0,3}^{(1)}\mid^2_2,\bd{\theta}\mid^2]
    \Big)\,.
\end{align}
If we set $[\bd{\eta},\bd{\lambda}_{1,1}^{(1)}\mid^0_0]=2\bd{L}_{1,1}^{(0)}\mid^0_0$\,, the requirement
$[\bd{\eta},\bd{L}_{1,1}^{(1)}\mid^0_0]=0$ leads to the relation without PCO insertion:
%
\begin{align}
&\ [\bd{Q},\bd{L}_{1,1}^{(0)}]
+ [\bd{L}_{0,2}^{(0)}\mid^0_0,\bd{L}_{1,0}^{(0)}\mid^0_0]
+ [\bd{L}_{0,3}^{(0)}\mid^0_0,\bd{\theta}\mid^0]
+ [\bd{L}_{0,3}^{(0)}\mid^2_2,\bd{\theta}\mid^2]\ =\ 0\,.
\end{align}
Then, the construction is complete since we can obtain $\bd{L}_{1,1}^{(0)}\mid^0_0$ in the same way
as in bosonic closed string field theory.\footnote{
Note that the forms of $\bd{L}$ and $\bd{\lambda}$ at intermediate stages are not unique; there is always an ambiguity up to $Q$-exact or $\eta$-exact terms.
}

Before closing this subsection, we should comment on how to solve equations of the form $[\bd{\eta}, \bd{A}_n]=\bd{B}_n$, which arise at each step in determining the products. The simplest solution that preserves cyclicity is to set
\begin{align}
A_n\ =\ \xi\circ B_n\ =\ \frac{1}{n+1}\Big(\xi B_n+(-1)^{B}B_n(\xi\wedge\id)\Big)\,.
\end{align}
We implicitly adopt this choice throughout this paper. However, if this choice conflicts with the potential spurious pole problem, then we need to modify it. We should note that there is room for such a modification.

\subsection{General construction}

In order to systematically construct quantum heterotic string products, we extend the prescription for classical products given in \cite{Kunitomo:2019glq}.
%
In the classical case, we first construct products acting on the symmetric tensor algebra of the large Hilbert space $\mathcal{S}\mathcal{H}_l$. Then, by transforming them so that they commute with $\eta$, 
we obtain products closed in the symmetric tensor algebra of the small Hilbert space $\mathcal{S}\mathcal{H}^{\textrm{res}}$. 
We now investigate whether the same procedure can be applied to the quantum case.

In the quantum case, the degree-odd coderivation on the large Hilbert space that we should consider is
\begin{align}
    \bd{B}\ =\ \sum_{g=0}^\infty \sum_{n,r=0}^\infty \hbar^g \bd{B}_{g,n}^{(2g+n-r-1)}|^{2r},
    \label{def B}
\end{align}
which acts on the symmetric tensor algebra  $\mathcal{S}(\mathcal{H}_{l\,\textrm{NS}}^{(-1)}\oplus \mathcal{H}_{l\,\textrm{R}}^{(-1/2)})$ with
\begin{align}
    \mathcal{H}_{l\,\textrm{NS}}^{(-1)}\ =&\ \big\{ \Psi=\Psi_0+\xi\Psi_1\mid \Psi_0\in \mathcal{H}_{\textrm{NS}}^{(-1)} ,\ \Psi_1\in \mathcal{H}_{\textrm{NS}}^{(-2)} \big\}\,,\\    \mathcal{H}_{l\,\textrm{R}}^{(-1/2)}\ =&\ \big\{ \Psi=\Psi_0+\xi\Psi_1\mid \Psi_0\in \overline{\mathcal{H}}_{\textrm{R}}^{(-1/2)} ,\ \Psi_1\in \overline{\mathcal{H}}_{\textrm{R}}^{(-3/2)} \big\}\,.
\end{align}
On the other hand, we still need to use $\bd{\theta}$ for the reflector without extending to the large Hilbert space. The reason is that, as explained in the previous section, using the reflector of the large Hilbert space to generate a handle yields zero. Using the reflectors of the large Hilbert space, $|r^{\textrm{NS}}(1,2)\rangle$ and $|r^{\textrm{R}}(1,2)\rangle$, two components of $\theta$, namely $\theta\mid^0$ and $\theta\mid^2$\,, are obtained by symmetrizing the projected reflectors\footnote{
The necessity of this projection in the large Hilbert space has been addressed in \cite{Carow-Watamura:1989vmd,Carow-Watamura:1989cxv,DiVecchia:1988tc}.}, which are given by
\begin{align}
 (P_{-1}\otimes \eta\ )|r^{\textrm{NS}}(1,2)\rangle\quad \textrm{and}\quad
  (P_{-3/2}\otimes \eta\ )|r^{\textrm{R}}(1,2)\rangle\,,
\end{align}
respectively. The $\eta$ in front of the reflector provides the eta-zero mode required by the index theorem in each loop. 

Then, a straightforward extension of the classical construction would begin by requiring $\bd{B}$ to satisfy 
\begin{align}
[\bd{Q}-\bd{\eta}+\bd{B}+\hbar\bd{\theta},\bd{Q}-\bd{\eta}+\bd{B}+\hbar\bd{\theta}]\ =\ 0\,.
\label{quantum B}
\end{align}
or equivalently,
\begin{subequations}
    \begin{align}
&\ \pi_1[\bd{Q}-\bd{\eta}+\bd{B}+\hbar\bd{\theta},\bd{Q}-\bd{\eta}+\bd{B}+\hbar\bd{\theta}]\ =\ 0\,,
\label{eq pi1}\\
&\ \pi_2[\bd{Q}-\bd{\eta}+\bd{B}+\hbar\bd{\theta},\bd{Q}-\bd{\eta}+\bd{B}+\hbar\bd{\theta}]\ =\ 0\,.
\label{eq pi2}
\end{align}
\end{subequations}
However, the coexistence of $\bd{B}$ (extended to the large Hilbert space) and $\bd{\theta}$ (the reflector of the small Hilbert space) causes a difficulty. Specifically, the left-hand side of (\ref{eq pi2}) becomes
\begin{align}
\pi_2[\bd{Q}-\bd{\eta}+\bd{B}\, +\, & \hbar\bd{\theta},\ \bd{Q}-\bd{\eta}+\bd{B}+\hbar\bd{\theta}]
\nonumber\\ 
 = \sum_{g,n,r=0}^{\infty}\hbar^{g+1}&\sum_s\Big(B_{g,n+1}^{(2g+n-r)}\mid^{2r}(\id^{\wedge n}\wedge e^0_{(s,-1)})\wedge \tilde{e}^0_{(s,-1)}
\nonumber\\
&+\ \frac{1}{2}B_{g,n+1}^{(2g+n-r)}\mid^{2r}(\id^{\wedge n}\wedge e^1_{(s,-3/2)})\wedge\tilde{e}^1_{(s,-1/2)}
\nonumber\\
&+\ \frac{1}{2}B_{g,n+1}^{(2g+n-r)}\mid^{2r}(\id^{\wedge n}\wedge e^1_{(s,-1/2)})\wedge \tilde{e}^1_{(s,-3/2)}\Big)\,.
\end{align}
This does not vanish even if $\bd{B}$ is cyclic in the large Hilbert space, that is, cyclic with respect to $\omega_l$. Instead, an odd coderivation $B$ that is cyclic in the large Hilbert space satisfies
\begin{align}
& \sum_{s}\Big(P_{q}\pi^aB_{g,n+1}(\id^{\wedge n}\wedge e^b_{(s,-p-2)})\wedge \tilde{e}^b_{(s,p)}
+ P_{p}\pi^bB_{g,n+1}(\id^{\wedge n}\wedge e^a_{(s,-q-2)})\wedge \tilde{e}^{a}_{(s,q)}
\nonumber\\
  =&\
\frac{1}{2} \sum_{s}\Big(P_{p}\pi^b[\eta,B_{g,n+1}](\id^{\wedge n}\wedge\Xi e^a_{(s,-q-2)})\wedge \tilde{e}^a_{(s,q)}
\nonumber\\
&\hspace{35mm}
 -(-1)^{e^a_s}P_{p}\pi^b[\eta, B_{g,n+1}](\id^{\wedge n}\wedge e^a_{(s,-q-1)})\wedge\Xi \tilde{e}^a_{(s,q-1)}\Big) 
\nonumber\\
&\ - \frac{1}{2}\sum_{s}\Big(P_{p}\pi^bB_{g,n+1}(\id^{\wedge n-1}\wedge\eta\wedge\Xi e^a_{(s,-q-2)})\wedge \tilde{e}^a_{(s,q)}
\nonumber\\
&\hspace{35mm}
-(-1)^{e^a_s}P_{p}\pi^bB_{g,n+1}(\id^{\wedge n-1}\wedge\eta\wedge e^a_{(s,-q-1)})\wedge\Xi \tilde{e}^a_{(s,q-1)}\Big)
\nonumber\\
&\ + (a,q)\leftrightarrow (b,p)\,,\label{general cyc}
\end{align}
as shown in Appendix~\ref{sec:cyclicity}.
Using Eq.~(\ref{general cyc}), we have
\begin{align}
&\ \pi_2[\bd{Q}-\bd{\eta}+\bd{B}\, \,+\hbar\bd{\theta},\bd{Q}-\bd{\eta}+\bd{B}+\hbar\bd{\theta}]\nonumber\\ 
&\qquad   =\ \frac{1}{2}\sum_{g,n,r=0}^{\infty}\hbar^{g+1}\sum_{a,b=0,1}\sum_s\Big(P_{p_a}\pi^a[\eta, B_{g,n+1}^{(2g+n-r)}\mid^{2r}](\id^{\wedge n}\wedge \Xi e^{b}_{(s,-p_b-2)})\wedge \tilde{e}^b_{(s,p_b)}
 \nonumber\\
 &\hspace{50mm} -(-1)^{e^b_s}P_{p_a}\pi^a[\eta, B_{g,n+1}^{(2g+n-r)}\mid^{2r}_{2r}](\id^{\wedge n}\wedge e^b_{(s,-p_b-1)})\wedge \Xi \tilde{e}^b_{(s,p_b-1)} \Big)
 \nonumber\\
&\qquad\  - \frac{1}{2}\sum_{\substack{g,r=0\\n=1}}^{\infty}\hbar^{g+1}\sum_{a,b=0,1}\sum_s\Big(P_{p_a}\pi^a B_{g,n+1}^{(2g+n-r)}\mid^{2r}(\id^{\wedge n-1}\wedge\eta\wedge \Xi e^{b}_{(s,-p_b-2)})\wedge \tilde{e}^b_{(s,p_b)}
 \nonumber\\
 &\hspace{50mm} -(-1)^{e^b_s}P_{p_a}\pi^a B_{g,n+1}^{(2g+n-r)}\mid^{2r}_{2r}(\id^{\wedge n-1}\wedge\eta\wedge e^b_{(s,-p_b-1)})\wedge \Xi \tilde{e}^b_{(s,p_b-1)} \Big)
 \nonumber\\
 &\qquad\ -
\frac{1}{2} \sum_{g,n,r=0}^{\infty}\hbar^{g+1}\sum_s\Big(
B_{g,n+1}^{(2g+n-r)}\mid^{2r}(\id^{\wedge n}\wedge e^1_{(s,-1/2)})\wedge \tilde{e}^1_{(s,-3/2)}
\nonumber\\
&\hspace{50mm}
- B_{g,n+1}^{(2g+n-r)}\mid^{2r}(\id^{\wedge n}\wedge e^1_{(s,-3/2)})\wedge\tilde{e}^1_{(s,-1/2)}
\Big)\,.
\end{align}
where $p_a=-1$ for $a=0$ and $p_a=-3/2$ for $a=1$\,.
However, the right-hand side does not contribute to the quantum $L_\infty$ relations 
since it vanishes in the loop (when both outputs are inputs to other products). 
This is because of the cancellation mechanism discussed in the previous subsection
and the fact that the loop (in the small Hilbert space) does not depend on the picture number of the states propagating around it. While the state
\begin{align}
 &  (\id \otimes P_{q})\sum_{s, p}\left(\Xi e^a_{(s,p)}\otimes \tilde{e}^a_{(s,-p-2)}-(-1)^{e^a_s}e^a_{(s,p)}\otimes\Xi \tilde{e}^a_{(s,-p-2)}\right)\ 
 \nonumber\\
&\qquad  =\  \sum_{s}\Big( \Xi e^a_{(s,-q-2)}\otimes \tilde{e}^a_{(s,q)}-(-1)^{e^a_s}e^a_{(s,-q-1)}\otimes\Xi\tilde{e}^a_{(s,q-1)}\Big)\,,
\end{align}
in the right-hand side of (\ref{general cyc})
is the reflector in the large Hilbert space projected onto a fixed picture number,
it generates a loop (with a fixed picture number) in the large Hilbert space, whose contribution
vanishes due to the absence of an $\eta$ zero mode. It can be interpreted as a cancellation between states from two isomorphic small Hilbert spaces. The $\eta$ insertion projects the state propagating through the loop onto one of the two halves, thereby ensuring consistency
with the contribution from the propagator.
Note that this cancellation mechanism is independent of the complications arising from the spurious poles. This is because spurious poles arise from the insertion of the $\eta$ zero mode needed to prevent the loop contribution from vanishing.

In this paper, we use the symbol $\approx$ to denote an equality that holds under the convention of ignoring contributions that vanish upon loop formation (that is, when two outputs in this case are connected to inputs of different products)  and disregarding distinctions between reflectors projected onto different picture numbers.\footnote{We use the term \textit{weak} to refer to relations that hold in this sense, such as weak equality, holding in the weak sense, and so on.} For example,
\begin{align}
\sum_s\left(\Xi e^1_{(s,-3/2)}\otimes \tilde{e}^1_{(s,-1/2)}-(-1)^{e^1_s}e^1_{(s,-1/2)}\otimes \Xi \tilde{e}^1_{(s,-3/2)} \right)\ \approx\ 0\,,
\label{move Xi}
\end{align}
and this relation becomes an exact equality when a loop is formed. Under this convention, Eq.~(\ref{general cyc}) can be written as
\begin{align}
    &\sum_{s}\pi^aB_{g,n+1}(\id^{\wedge n}\wedge e^b_{(s,-p-2)})\wedge \tilde{e}^b_{(s,p)}\ \approx\ 0\,,
    \label{weak cyc}
\end{align}
and $B_{g,n+1}$ can be treated as if it were cyclic (with respect to $\omega$).
Taking this into account, we relax condition (\ref{quantum B}) to 
\begin{align}
[\bd{Q}-\bd{\eta}+\bd{B}+\hbar\bd{\theta},\bd{Q}-\bd{\eta}+\bd{B}+\hbar\bd{\theta}]\ \approx\ 0.
\label{quantum L app}
\end{align}
We construct $\bd{B}$ to satisfy this condition (and to be cyclic with respect to $\omega_l$) as follows.

As in the classical case, Eq.~(\ref{eq pi1}) holds for each picture number deficit independently, and thus is equivalent to the equations
\begin{subequations} \label{q L} 
\begin{align}
     [\bd{\mathcal{D}}+\hbar\bd{\theta}\mid^0,
     \bd{\mathcal{D}}+\hbar\bd{\theta}\mid^0]\ \approx&\ 0\,,\\
    [\bd{\mathcal{C}}-\hbar\bd{\theta}\mid^2,
    \bd{\mathcal{C}}-\hbar\bd{\theta}\mid^2]\ \approx&\ 0\,,\\
    [\bd{\mathcal{D}}+\hbar\bd{\theta}\mid^0,
    \bd{\mathcal{C}}-\hbar\bd{\theta}\mid^2]\ \approx&\ 0\,,
\label{eq:DCconditions}
\end{align}
\end{subequations}
with coderivations $\bd{\mathcal{D}}$ and $\bd{\mathcal{C}}$ derived from
\begin{align}
\pi_1\bd{\mathcal{D}}\ =\ Q+\pi^0_1\bd{B}\,,\qquad
 \pi_1\bd{\mathcal{C}}\ =\ \eta-\pi^1_1\bd{B}\,.
\end{align}
We will provide a specific method for finding $\bd{B}$ satisfying these conditions shortly.

Suppose we find a coderivation $\bd{B}$ that satisfies Eq.~(\ref{q L}). Consider 
an $IBL_\infty$-transfer, 
\begin{align}\label{eq: def of mathcalF^-1}
    \bd{\mathcal{F}}^{-1}=\sum_{n=0}^{\infty}\frac{1}{n!}(\bd{F}^{-1})^{\wedge n}\circ\Delta_n,
\end{align}
with $n$-fold comultiplication $\Delta_n:\mathcal{SH}_l\rightarrow(\mathcal{SH}_l)^{\otimes n}$ and
\begin{align}
    \bd{F}^{-1}\ =&\ \sum_{k=0}^{\infty}\frac{(-\hbar)^k}{(k+1)!}\sum_{s_1,\cdots,s_k}
\Xi \tilde{e}^1_{(s_1,-p_1-2)}
\wedge\dots\wedge\Xi \tilde{e}^1_{(s_k,-p_k-2)}
\nonumber\\
&\hspace{50mm}
\wedge(\pi_1\id -\Xi\pi_1^1\bd{B})
\bd{D}(e^1_{(s_k,p_k)})\dots\bd{D}(e^1_{(s_1,p_1)}),
\label{morph F}
\end{align}
where $\bd{D}(e^1_{(s,p)})$ is a coderivation  derived from $e^1_{(s,p)}$: 
\begin{align}
    \bd{D}(e^1_{(s,p)})=\sum_{n=0}^\infty e^1_{(s,p)} \wedge \id_n\,.
\end{align}
The symbol $(\bd{F}^{-1})^{\wedge n}$ represents symmetrization with respect to all the outputs.
Since only the R-sector basis appears in this morphism and distinctions between reflectors projected onto different picture numbers are disregarded, we suppress the corresponding labels hereafter, writing $e^1_{(s,p)}$ simply as $e_s$, and similarly for the dual basis.
This $IBL_\infty$-transfer is a generalization of the classical cohomomorphism and reduces to it at $\hbar=0$\,.
Assuming $\bd{\mathcal{F}}$ is given by
\begin{align}
    \bd{\mathcal{F}}=\sum_{n=0}^\infty \frac{1}{n!}(\bd{G})^{\wedge n}\circ\Delta_n,\label{eq:def F}
\end{align}
we solve
$\bd{\mathcal{F}}\bd{\mathcal{F}}^{-1}=\bd{\mathcal{F}}^{-1}\bd{\mathcal{F}}=\idSH$\,,
and find that
\begin{align}\label{eq:def of G}
    \bd{G}=\sum_{k=0}^{\infty}\frac{\hbar^k}{(k+1)!}\sum_{s_1,\cdots,s_k}\Xi \tilde{e}_{s_1}
\wedge\dots\wedge\Xi \tilde{e}_{s_k}\wedge(\pi_1 \id +\Xi\pi_1^1\bd{B}\bd{\mathcal{F}})
\bd{D}(e_{s_k})\dots\bd{D}(e_{s_1})
\end{align}
satisfies $\bd{\mathcal{F}}\bd{\mathcal{F}}^{-1}\approx\id_{\mathcal{SH}}$\,.
Note that $\bd{\mathcal{F}}^{-1}$ is an inverse of $\bd{\mathcal{F}}$ but $\bd{G}\bd{F}\neq \id$\,.
This 
$IBL_\infty$-transfer $\bd{\mathcal{F}}^{-1}$ transforms
$\bd{\mathcal{D}}+\hbar\bd{\theta}\mid^0$ and $\bd{\mathcal{C}}-\hbar\bd{\theta}\mid^2$ into
\begin{subequations}\label{eq:DC transformation}
 \begin{align}
     \bd{\mathcal{F}}^{-1}(\bd{\mathcal{D}}+\hbar\bd{\theta}\mid^0)\bd{\mathcal{F}}\ \approx&\
\bd{Q}+\bd{\mathcal{G}}\bd{B}\bd{\mathcal{F}}+\hbar\bd{\Theta},
\label{eq:L def}\\
     \bd{\mathcal{F}}^{-1}(\bd{\mathcal{C}}-\hbar\bd{\theta}\mid^2)\bd{\mathcal{F}}\ \approx&\ \bd{\eta},
\label{eq:eta F}
 \end{align}
 \end{subequations}
where $\bd{\mathcal{G}}\bd{B}\bd{\mathcal{F}}$ denotes the coderivation derived from 
$\mathcal{G}\pi_1\bd{B}\bd{\mathcal{F}}$ and 
\begin{align}
    \Theta\ =\ \frac{1}{2}\sum_{s}e^0_{(s,-1)}\wedge \tilde{e}^0_{(s,-1)}
    + \frac{1}{2}\sum_{s}Xe^1_{(s,-3/2)}\wedge \tilde{e}^1_{(s,-1/2)}.
\end{align}
Eqs.~(\ref{q L}) transforms into
\begin{subequations}
\begin{align}
[\bd{Q}+\bd{\mathcal{G}}\bd{B}\bd{\mathcal{F}}+\hbar\bd{\Theta},
\bd{Q}+\bd{\mathcal{G}}\bd{B}\bd{\mathcal{F}}+\hbar\bd{\Theta}]\
\approx&\ 0,\label{first eq}\\
[\bd{\eta},
\bd{Q}+\bd{\mathcal{G}}\bd{B}\bd{\mathcal{F}}+\hbar\bd{\Theta}]\
\approx&\ 0.\label{second eq}
\end{align} 
\end{subequations}
The second weak equality (\ref{second eq}) is actually an exact equality due to $[\bd{\eta},\bd{\Theta}]=0$.
Therefore, if we define $\bd{L}$ as 
\begin{align}
    \bd{L}=\bd{Q}+\bd{\mathcal{G}B\mathcal{F}}, \label{L GBF}
\end{align}
we have
\begin{align}
 [\bd{\eta}, \bd{L}+\hbar\bd{\Theta}]=0,
\end{align}
which allows us to consider the order-two coderivation
$\bd{L}+\hbar\bd{\Theta}$ to be in the small Hilbert space.
The first equation (\ref{first eq}) ensures that the quantum $L_\infty$ relations
\begin{align}
\pi_1\Big( [\bd{L}+\hbar\bd{\Theta}, \bd{L}+\hbar\bd{\Theta}]\Big)\ =\ 0,
\label{q L 1}
\end{align}
but not the cyclicity relation,
\begin{align}
\pi_2\Big( [\bd{L}+\hbar\bd{\Theta}, \bd{L}+\hbar\bd{\Theta}]\Big)\ =\ 0.
\label{q L 2}
\end{align}
Although it appears to hold by construction, it requires an independent proof, which 
we leave for future work.


Now, let us consider how to construct $\bd{B}$ to satisfy (\ref{quantum L app}).
If we define the subscripted commutator $[\cdot,\cdot]_{\mathcal{O}}$ as
the commutator that inserts an operator $\mathcal{O}$ into the intermediate state,\footnote{
In Ref.~\cite{Kunitomo:2019glq}, $[\ ,\ ]_{\pi^0}$ and $[\ ,\ ]_{\pi^1}$ were denoted as
$[\ ,\ ]^1$ and $[\ ,\ ]^2$, respectively.}
\begin{align}
    \pi_1[\bd{A}_m,\bd{B}_n]_{\mathcal{O}}\ =&\ \Big(A_m(\mathcal{O}B_n\wedge\id_{m-1})
-(-1)^{AB+\mathcal{O}(A+B)}B_n(\mathcal{O}A_m\wedge\id_{n-1})\Big),
\end{align}
Eq.~(\ref{quantum L app}) is written in the form
\begin{subequations}\label{eq:B condition equiv}
\begin{align}
  [\bd{Q},\bd{B}]+\frac{1}{2}[\bd{B},\bd{B}]_{\pi^0}+\hbar[\bd{B},\bd{\theta}\mid^0]\
\approx&\ 0\,,\\
 [\bd{\eta},\bd{B}]-\frac{1}{2}[\bd{B},\bd{B}]_{\pi^1}-\hbar[\bd{B},\bd{\theta}\mid^2]\
\approx&\ 0\,.\label{eq:Csquare}
\end{align}
\end{subequations}
We provide a prescription for constructing $\bd{B}$ 
satisfying (\ref{eq:B condition equiv}) as follows.
Consider the generating function
\begin{align}
    \bd{B}(s,t)=\sum_{g=0}^\infty \sum_{m,p,r=0}^\infty \hbar^g s^m t^p 
\bd{B}_{g,m+p+r-2g+1}^{(p)}|^{2r}=\sum_{p=0}^\infty t^p \bd{B}^{(p)}(s),
\end{align}
which reduces to $\bd{B}$ when evaluated at $(s,t)=(0,1)$: $\bd{B}(0,1)=\bd{B}$.
The parameters $t$ and $s$ count the picture number and the picture number deficit, respectively\footnote{
We define $\bd{B}_{g,n}^{(p)}\mid^{2r}\equiv0$ when $n<0$.}. 
Then, we extend Eqs.~\eqref{eq:B condition equiv} to equations 
for generating function: 
\begin{subequations}\label{eq:IJ}
\begin{align}
    I(s,t)&= [\bm{Q},\bm{B}(s,t)]+\frac{1}{2}[\bm{B}(s,t),\bm{B}(s,t)]_{\os} +\hbar[\bm{B}(s,t),\bm{\theta}]_{\os}\ \approx\ 0\,,\\
    J(s,t)&= [\bm{\eta},\bm{B}(s,t)]-\frac{1}{2}[\bm{B}(s,t),\bm{B}(s,t)]_{\ot}-\hbar[\bm{B}(s,t),\bm{\theta}]_{\ot} \approx\ 0\,,\label{eq:J}
\end{align}
\end{subequations}
with $\os=\pi^0+s\pi^1$ and $\ot=t\pi^1$. The subscripted commutators involving $\bd{\theta}$ are defined by
\begin{align}
[\bd{B}(s,t), \bd{\theta}\,]_{\os}\ =&\ [\bd{B}(s,t), \bd{\theta}\mid^0] + s [\bd{B}(s,t), \bd{\theta}\mid^2\,],\\
[\bd{B}(s,t), \bd{\theta}\,]_{\ot}\ =&\ t [\bd{B}(s,t), \bd{\theta}\mid^2].
\end{align}
The relations (\ref{eq:IJ}) reduce to Eqs.~\eqref{eq:B condition equiv} when evaluated at $(s,t)=(0,1)$.

If we introduce a degree-even coderivation (referred to as the gauge product)
\begin{align}
    \bm{\lambda}(s,t)=\sum_{g=0}^\infty\sum_{m,p,r=0}^\infty \hbar^g s^m t^p 
\coder{\lambda}{g}{m+p+r-2g+2}{p+1}{2r}{}=\sum_{p=0}^\infty t^p \bd{\lambda}^{(p+1)}(s),
\end{align}
and assume that $\bd{B}(s,t)$ satisfies the differential equations
\begin{subequations}\label{eq: diff eq}
\begin{align}
    \pi_1\pdvt\bm{B}(s,t)\ 
=&\ 
\pi_1\Big([\bm{Q},\bm{\lambda}(s,t)]+[\bm{B}(s,t),\bm{\lambda}(s,t)]_{\os}-\hbar[\bm{\lambda}(s,t),\bm{\theta}]_{\os}\Big)\,,\\
   \pi_1 \pdvs\bm{B}(s,t)\
=&\
\pi_1\Big([\bm{\eta},\bm{\lambda}(s,t)]-[\bm{B}(s,t),\bm{\lambda}(s,t)]_{\ot}+\hbar[\bm{\lambda}(s,t),\bm{\theta}]_{\ot}\Big)\,,
\end{align}
\end{subequations}
then, the $t$-derivatives of $\bd{I}(s,t)$ and $\bd{J}(s,t)$ are given by
\begin{subequations}
\begin{align}
    \pdvt\bm{I}(s,t)&\approx[\bm{I}(s,t),\bm{\lambda}(s,t)]_{\os}\\
    \pdvt\bm{J}(s,t)&\approx[\bm{J}(s,t),\bm{\lambda}(s,t)]_{\os}-\pdvs\bm{I}(s,t)-[\bm{I}(s,t),\bm{\lambda}(s,t)]_{\ot}.
\end{align}
\end{subequations}
Hence, if 
\begin{subequations}\label{eq:t=0 IJ}
\begin{align}
    \bm{I}(s,0)&=[\bm{Q},\bm{B}(s,0)]
+\frac{1}{2}[\bm{B}(s,0),\bm{B}(s,0)]_{\os}+\hbar[\bm{B}(s,0),\bm{\theta}]_{\os}\ \approx\ 0,\label{eq:B0}\\
    \bm{J}(s,0)&=[\bm{\eta},\bm{B}(s,0)]\ \approx\ 0,
\end{align}
\end{subequations}
then $\bm{I}(s,t)\approx\bm{J}(s,t)\approx0$ for arbitrary $t$.

However, 
\begin{align}
    \bd{B}(s,0)=\bd{B}^{(0)}(s)=\sum_{g=0}^\infty \sum_{m,r=0}^\infty \hbar^g s^m \bd{B}^{(0)}_{g,m+r-2g+1}|^{2r},
\end{align}
does not involve PCOs, and the relation \eqref{eq:t=0 IJ} is a quantum $L_\infty$ relation independent of the way to insert PCOs. Therefore, we can construct $\bd{B}^{(0)}(s)$ satisfying \eqref{eq:t=0 IJ} by determining the triangulation of the moduli space in a similar manner to the bosonic string product. Then, using it as an initial condition, we can obtain $\bd{B}(s,t)$ satisfying \eqref{eq:IJ} by iteratively solving the differential equations \eqref{eq: diff eq}. 
%

In order to solve the differential equations (\ref{eq: diff eq}), it is convenient to rewrite them as recurrence relations:
\begin{subequations}\label{eq:general vertices}
\begin{align}
&\    (p+1)\pi_1\bd{B}^{(p+1)}(s)
\nonumber\\ 
&\qquad  =\ \pi_1\Big([\bd{Q},\bd{\lambda}^{(p+1)}(s)]
+\sum_{p'=0}^p[\bd{B}^{(p-p')}(s),\bd{\lambda}^{(p'+1)}(s)]_{\os}
    -\hbar\bd{\lambda}^{(p+1)}(s)(\bd{\theta}\mid^0+s\bd{\theta}\mid^2)\Big)\,,\label{eq:B^p+1}\\
&\    \pi_1\partial_s\bd{B}^{(p)}(s)
\nonumber\\ 
&\qquad  =\ \pi_1\Big([\bd{\eta},\bd{\lambda}^{(p+1)}(s)]
-\sum_{p'=0}^{p-1}[\bd{B}^{(p-p'-1)}(s),\bd{\lambda}^{(p'+1)}(s)]_{\pi^1}
    +\hbar\bd{\lambda}^{(p)}(s)\bd{\theta}\mid^2\Big)\,.\label{eq:lambda^p+1}
\end{align}
\end{subequations}
For $p=0$, we have
\begin{subequations}
\begin{align}
    \pi_1\bd{B}^{(1)}(s)\ =&\ \pi_1\Big([\bd{Q},\bd{\lambda}^{(1)}(s)]
+[\bd{B}^{(0)}(s),\bd{\lambda}^{(1)}(s)]_{\os}
-\hbar\bd{\lambda}^{(1)}(s)(\bd{\theta}\mid^0+s\bd{\theta}\mid^2)\Big)\,,
\label{eq:B^0 equation}\\
    \pi_1\partial_s\bd{B}^{(0)}(s)\ =&\ \pi_1[\bd{\eta},\bd{\lambda}^{(1)}(s)]\,.
\end{align}
\end{subequations}
Since $[\bd{\eta},\bd{B}^{(0)}(s)]=0$, we can consistently determine the gauge products $\pi_1\bd{\lambda}^{(1)}(s)$ as
\begin{align}
   & \pi_1\bd{\lambda}^{(1)}(s)\nonumber\\
    &\ =\sum_{g=0}^\infty \sum_{m,r=0}^\infty\hbar^g s^m\frac{m+1}{m+r-2g+3}(\xi B^{(0)}_{g,m+r-2g+2}|^{2r}
-B^{(0)}_{g,m+r-2g+2}|^{2r}(\xi\wedge\id_{m+r-2g+1}))\nonumber\\
    &\ \equiv\xi\circ\pi_1\partial_s\bd{B}^{(0)}(s),
\end{align}
respecting the cyclicity. Then, all the terms on the right-hand side of \eqref{eq:B^0 equation} 
are given, which determines $\bd{B}^{(1)}(s)$.
Next, for $p=1$, we have
\begin{subequations}
\begin{align}
&\     2\pi_1\bd{B}^{(2)}(s)\nonumber\\     
&\     =\ \pi_1\Big([\bd{Q},\bd{\lambda}^{(2)}(s)]
+[\bd{B}^{(1)}(s),\bd{\lambda}^{(1)}(s)]_{\os}+[\bd{B}^{(0)}(s),\bd{\lambda}^{(2)}(s)]_{\os}
-\hbar\bd{\lambda}^{(2)}(s)(\bd{\theta}\mid^0+s\bd{\theta}\mid^2)\Big)\,,\label{eq:B^2}\\
&\    \pi_1\partial_s\bd{B}^{(1)}(s)\nonumber\\ 
&\     =\ \pi_1\Big([\bd{\eta},\bd{\lambda}^{(2)}(s)]
-[\bd{B}^{(0)}(s),\bd{\lambda}^{(1)}(s)]_{\pi^1}+\hbar\bd{\lambda}^{(1)}(s)\bd{\theta}\mid^2\Big)\,.\label{eq:lambda^2}
\end{align}
\end{subequations}
Here, $\bd{B}^{(1)}(s)$ and $\bd{\lambda}^{(1)}(s)$ appearing in \eqref{eq:B^2} 
are already given and $\bd{\lambda}^{(2)}(s)$ can be determined as
\begin{align}
    \pi_1\bd{\lambda}^{(2)}(s)=\xi\circ\pi_1\Big(\partial_s\bd{B}^{(1)}(s)
+[\bd{B}^{(0)}(s),\bd{\lambda}^{(1)}(s)]_{\pi^1}-\hbar\bd{\lambda}^{(1)}(s)\bd{\theta}\mid^2\Big),
\end{align}
and then $\bd{B}^{(2)}(s)$ is determined by \eqref{eq:B^2}.
Repeating the procedure, we can obtain $\bd{B}^{(p)}$ for all $p$. 
When $\bd{B}^{(q)}(s)$ and $\bd{\lambda}^{(q)}(s)$ for $q\leq p$ are already given, 
$\bd{\lambda}^{(p+1)}(s)$ is given by
\begin{align}
    \pi_1\bd{\lambda}^{(p+1)}(s)\ =\ \xi\circ\pi_1\left(\partial_s\bd{B}^{(p)}(s)
+\sum_{p'=0}^{p-1}[\bd{B}^{(p-p'-1)}(s),\bd{\lambda}^{(p'+1)}(s)]_{\pi^1}
-\hbar\bd{\lambda}^{(p)}(s)\bd{\theta}\mid^2\right),
\label{eq:lambda solution}
\end{align}
and then $\bd{B}^{(p+1)}(s)$ is determined by \eqref{eq:B^p+1}.

\sectiono{One-loop correction to two-point amplitude}\label{amplitudes}

It is interesting to verify that the quantum action we constructed reproduces the first-quantized amplitudes at the loop level. In this section, we demonstrate this by focusing on the one-loop correction to two-point amplitudes as the simplest example.

We concretely calculate two types of two-point amplitudes, NS-NS and R-R, by taking
the Siegel gauge 
\begin{align}
 b_0^+\Phi\ =\ \beta_0\Phi_{R}\ =\ 0.
\end{align}
The propagators in this gauge are given by
\begin{align}
 \Pi_{\textrm{NS}}\ =\ -\frac{b_0^+b_0^-}{L_0^+}\mathcal{P},\qquad
\Pi_{\textrm{R}}\ =\ -\frac{b_0^+b_0^-X}{L_0^+}\mathcal{P},
\end{align}
with
\begin{align}
 \mathcal{P}\ \equiv\ \int^{2\pi}_0\frac{d\theta}{2\pi}e^{i\theta L_0^-},
\end{align}
and the vertices are given by
\begin{align}
 c_0^-\mathcal{G}^{-1}L_{n}\ =\ c_0^-\pi_1\bd{B\mathcal{F}}\,,
\end{align}
where  $\bd{B}$ is given by \eqref{eq:general vertices} and $\bd{\mathcal{F}}$ can be obtained using \eqref{eq:def F} and \eqref{eq:def of G}.

\subsection{One-loop correction to NS-NS amplitude}

The one-loop correction to the NS-NS amplitude is given by
\begin{align}
    &\mathcal{A}_{NS}(\Phi_{NS},\Phi'_{NS})\nonumber\\
    &=\omega\Bigl(\Phi_{NS},L_{1,1}^{(2)}|^0_0(\Phi'_{NS}) \Bigr)
    -\omega\Bigl(\Phi_{NS},L_{0,2}^{(1)}|^0_0(\frac{b_0^+}{L_0^+}L_{1,0}^{(1)}|^0_0,\Phi'_{NS}) \Bigr)\nonumber\\
    &\quad-\frac{1}{2}\sum_s\omega\Bigl(\Phi_{NS},L_{0,3}^{(2)}|^0_0(\frac{b_0^+}{L_0^+}
e^0_{s},\tilde{e}^0_{s},\Phi'_{NS})\Bigr)
    -\frac{1}{2}\sum_s\omega\Bigl(\Phi_{NS},L_{0,3}^{(1)}|^2_2(\frac{b_0^+}{L_0^+}X e^1_{s},
 \tilde{e}^1_{s},\Phi'_{NS})\Bigr)\nonumber\\
    &\quad+\frac{1}{2}\sum_s\omega\Bigl(\Phi_{NS},L_{0,2}^{(1)}|^0_0
  \bigl(\frac{b_0^+}{L_0^+}L_{0,2}^{(1)}|^0_0(\frac{b_0^+}{L_0^+}e^0_{s},\tilde{e}^0_{s}),
 \Phi'_{NS}\bigr) \Bigr)\nonumber\\
    &\quad+\frac{1}{2}\sum_s\omega\Bigl(\Phi_{NS},L_{0,2}^{(1)}|^0_0
 \bigl(\frac{b_0^+}{L_0^+}L_{0,2}^{(0)}|^2_2(\frac{b_0^+}{L_0^+}X e^1_{s},\tilde{e}^1_{s}),\Phi'_{NS}\bigr) \Bigr)\nonumber\\
    &\quad+\frac{1}{2}\sum_s\omega\Bigl(\Phi_{NS},L_{0,2}^{(1)}|^0_0\bigl(\frac{b_0^+}{L_0^+}
 \tilde{e}^0_{s},\frac{b_0^+}{L_0^+}L_{0,2}^{(1)}|^0_0(e^0_{s},\Phi'_{NS})\bigr)\Bigr)\nonumber\\
     &\quad+\frac{1}{2}\sum_s\omega\Bigl(\Phi_{NS},L_{0,2}^{(0)}|^2_2
 \bigl(\frac{b_0^+}{L_0^+}X\tilde{e}^1_{s},\frac{b_0^+}{L_0^+}L_{0,2}^{(1)}|^2_0(e^1_{s},\Phi'_{NS})\bigr)\Bigr),
\label{NS-2point}
\end{align}
where we used the relation
\begin{align}
 (\frac{b_0^+}{L_0^+}\otimes\id-\id\otimes\frac{b_0^+}{L_0^+})\sum_s e^a_s\otimes \tilde{e}^a_s\ =\ 0.
\end{align}
The amplitude can be viewed as a bilinear map from the two-fold tensor product of the physical state space into complex numbers
\begin{align}
    \bra{\mathcal{A}}:\mathcal{H}_Q\otimes\mathcal{H}_Q\longrightarrow \mathbb{C},
\end{align}
where $\mathcal{H}_Q \subset \mathcal{H}^{\textrm{res}}$ is the subspace of states annihilated by $Q$. 
 By pulling out $\Phi_{NS}$ and $\Phi'_{NS}$ to the right and using the coalgebra representation, the amplitude can also be written as
\begin{align}\label{eq:NS amp}
    \bra{\mathcal{A}_{NS}}=\bra{\omega}P_{-1}\otimes P_{-1}\pi_1\Bigl(&\coder{L}{1}{1}{2}{0}{0}
-\coder{L}{0}{2}{1}{0}{0}\bigl(\bd{\frac{b_0^+}{L_0^+}L}_{1,0}^{(1)}|^0_0\bigr)
-\coder{L}{0}{3}{2}{0}{0}\bigl(\bd{\frac{b_0^+}{L_0^+}}\bd{\Theta}\mid^0\bigr)
-\coder{L}{0}{3}{1}{2}{2}\bigl(\bd{\frac{b_0^+}{L_0^+}}\bd{\Theta}\mid^2\bigr)\nonumber\\
     &+\coder{L}{0}{2}{1}{0}{0}\bigl(\bd{\frac{b_0^+}{L_0^+}}\bd{L}_{0,2}^{(1)}|^0_0\bigr)
\bigl(\bd{\frac{b_0^+}{L_0^+}}\bd{\Theta}\mid^0\bigr)
+\coder{L}{0}{2}{1}{0}{0}\bigl(\bd{\frac{b_0^+}{L_0^+}L}_{0,2}^{(0)}|^2_2\bigr)
\bigl(\bd{\frac{b_0^+}{L_0^+}}\bd{\Theta}\mid^2\bigr)\nonumber\\
     &-\frac{1}{2}\sum_s\coder{L}{0}{2}{1}{0}{0}\bigl(\bd{\frac{b_0^+}{L_0^+}}\bd{D}(\tilde{e}^0_{s})\bigr)
\bigl(\bd{\frac{b_0^+}{L_0^+}L}_{0,2}^{(1)}|^0_0\bigr)\bd{D}(e^0_{s})\nonumber\\
     &+\frac{1}{2}\coder{L}{0}{2}{0}{2}{2}\bigl(\bd{\frac{b_0^+}{L_0^+}L}_{0,2}^{(1)}|^2_0\bigr)
\bd{(\frac{b_0^+}{L_0^+}}\bd{\Theta}\mid^2)\Bigr),
\end{align}
where $\bd{(\frac{b_0^+}{L_0^+}L)}$ and $\bd{(\frac{b_0^+}{L_0^+}\Theta)}$ 
denote the coderivations derived from 
$\frac{b_0^+}{L_0^+}L$ and $\frac{1}{2}(\frac{b_0^+}{L_0^+}\wedge \id)\Theta$\,\, respectively.
Note that 
\begin{align}
    \pi_1\coder{L}{0}{2}{1}{0}{0}\bd{\bigl(\frac{b_0^+}{L_0^+}L_{0,2}^{(1)}|^0_0\bigr)}
\bigl(\bd{\frac{b_0^+}{L_0^+}\Theta}\mid^0\bigr)\Phi_{NS}=&\frac{1}{2}\sum_sL_{0,2}^{(1)}|^0_0
\bigl(\frac{b_0^+}{L_0^+}L_{0,2}^{(1)}|^0_0(\frac{b_0^+}{L_0^+}e^0_{s}, \tilde{e}^0_{s}),\Phi_{NS}\bigr)\nonumber\\
    &+\sum_s
L_{0,2}^{(1)}|^0_0\bigl(\frac{b_0^+}{L_0^+}\tilde{e}^0_{s},\frac{b_0^+}{L_0^+}L_{0,2}^{(1)}|^0_0(e^0_{s},\Phi_{NS})\bigr).
\end{align}
From the form of $\bd{L}$ given in \eqref{L GBF} and the definition of $\bd{\Theta}$, we have
\begin{subequations}
\begin{align}
    &\pi_1\coder{L}{1}{1}{2}{0}{0}=\pi_1\Bigl(\coder{B}{1}{1}{2}{0}{0}+\coder{B}{0}{3}{1}{2}{2}\bd{(\Xi}\bd{\theta}\mid^2)
+\frac{1}{2}\coder{B}{0}{2}{0}{2}{2}(\bd{\Xi}\coder{B}{0}{2}{0}{2}{0})(\bd{\Xi}\bd{\theta}\mid^2)\Bigr),\\
    &\pi_1\coder{L}{1}{0}{1}{0}{0}=\pi_1\Bigl(\coder{B}{1}{0}{1}{0}{0}+\coder{B}{0}{2}{0}{2}{2}(\bd{\Xi}\bd{\theta}\mid^2)\Bigr),\\
   &\pi_1\coder{L}{0}{2}{1}{0}{0}=\pi_1\coder{B}{0}{2}{1}{0}{0},\\
   &\pi_1\coder{L}{0}{2}{1}{2}{0}=X\pi_1\coder{B}{0}{2}{0}{2}{0},\\
   &\pi_1\coder{L}{0}{2}{0}{2}{2}=\pi_1\coder{B}{0}{2}{0}{2}{2},\\
   &\pi_1\coder{L}{0}{3}{2}{0}{0}=\pi_1\coder{B}{0}{3}{2}{0}{0},\\
   &\pi_1\coder{L}{0}{3}{1}{2}{2}=\pi_1\Bigl(\coder{B}{0}{3}{1}{2}{2}+\coder{B}{0}{2}{0}{2}{2}(\bd{\Xi}\coder{B}{0}{2}{0}{2}{0})\Bigr),\\
   &\bd{\Theta}\mid^0=\bd{\theta}\mid^0,\\
   &\bd{\Theta}\mid^2=\bd{X}\bd{\theta}\mid^2.
\end{align}
\end{subequations}
Here, we denote the coderivations derived from 
$\frac{1}{2}\sum_s\Xi e^1_{(s,-3/2)}\wedge \tilde{e}_{(s,-1/2)}^1$ and 
$\frac{1}{2}\sum_s X e^1_{(s,-3/2)}\wedge \tilde{e}_{(s,-1/2)}^1$ as $\bd{\Xi}\bd{\theta}\mid^2$ and $\bd{X}\bd{\theta}\mid^2$, respectively.
Substituting these relations into \eqref{eq:NS amp}, we have
\begin{align}\label{eq:NS amp coder}
     \bra{\mathcal{A}_{NS}}=&\bra{\omega}P_{-1}\otimes P_{-1}\pi_1\Bigl(\coder{B}{1}{1}{2}{0}{0}
-\coder{B}{0}{2}{1}{0}{0}\bigl(\bd{\frac{b_0^+}{L_0^+}}\coder{B}{1}{0}{1}{0}{0}\bigr)
-\coder{B}{0}{3}{2}{0}{0}\bigl(\bd{\frac{b_0^+}{L_0^+}}\bd{\theta}\mid^0\bigr)\nonumber\\
     &\hspace{3cm} +\coder{B}{0}{2}{1}{0}{0}\bigl(\bd{\frac{b_0^+}{L_0^+}}\coder{B}{0}{2}{1}{0}{0}\bigr)
\bigl(\bd{\frac{b_0^+}{L_0^+}}\bd{\theta}\mid^0\bigr)\nonumber\\
     &\hspace{3cm} -\frac{1}{2}\sum_s\coder{B}{0}{2}{1}{0}{0}\bd{D}\bigl(\frac{b_0^+}{L_0^+}\tilde{e}^0_{s}\bigr)
\bigl(\bd{\frac{b_0^+}{L_0^+}}\coder{B}{0}{2}{1}{0}{0}\bigr)\bd{D}(e^0_{s})\nonumber\\
     &\hspace{3cm} -\coder{B}{0}{3}{1}{2}{2}\bigl(\bd{(\frac{b_0^+}{L_0^+}X}-\bd{\Xi})\bd{\theta}\mid^2\bigr)\nonumber\\
     &\hspace{3cm} +\coder{B}{0}{2}{1}{0}{0}\bigl(\bd{\frac{b_0^+}{L_0^+}}\coder{B}{0}{2}{0}{2}{2}\bigr)
\bigl((\bd{\frac{b_0^+}{L_0^+}X}-\bd{\Xi})\bd{\theta}\mid^2\bigr)\nonumber\\
     &\hspace{3cm} +\frac{1}{2}\coder{B}{0}{2}{0}{2}{2}\bigl((\bd{\frac{b_0^+}{L_0^+}X}-\bd{\Xi})
\coder{B}{0}{2}{0}{2}{0}\bigr)\bigl((\bd{\frac{b_0^+}{L_0^+}X}-\bd{\Xi})\bd{\theta}\mid^2\bigr)\Bigr)
\nonumber\\
\equiv& \bra{\omega}\id\otimes\pi_1\bd{S}_1.
\end{align}
The coderivations $\bigl((\bd{\frac{b_0^+}{L_0^+}X}-\bd{\Xi})\bd{B}\bigr)$ and $\bigl((\bd{\frac{b_0^+}{L_0^+}X}-\bd{\Xi})\bd{\theta}\mid^2\bigr)$ 
are those derived from $(\frac{b_0^+}{L_0^+}X-\Xi)B$ 
and $\frac{1}{2}\sum_s(\frac{b_0^+}{L_0^+}X-\Xi) e^1_{(s,-3/2)}\wedge \tilde{e}^1_{(s,-1/2)}$, 
respectively. We used the relation
\begin{align}
    (\bd{\Xi}\coder{B}{0}{2}{0}{2}{0})\bigl((\bd{\frac{b_0^+}{L_0^+}X}-\bd{\Xi})\bd{\theta}\mid^2\bigr)
\approx \bigl((\bd{\frac{b_0^+}{L_0^+}X}-\bd{\Xi})\bd{\coder{B}{0}{2}{0}{2}{0}}\bigr)\bd{(\Xi}\bd{\theta}\mid^2)\,.
\end{align}
Further, using the relations 
\begin{subequations}    
\begin{align}
   &\pi_1\coder{B}{1}{1}{2}{0}{0}=\frac{1}{2}\pi_1\bigl([\bd{Q},\coder{\lambda}{1}{1}{2}{0}{0}]
-\coder{\lambda}{0}{2}{1}{0}{0}\coder{B}{1}{0}{1}{0}{0}+\coder{B}{0}{2}{1}{0}{0}\coder{\lambda}{1}{0}{1}{0}{0}
-\coder{\lambda}{0}{3}{2}{0}{0}\bd{\theta}\mid^0\bigr),\label{eq:B10100}\\
    &\pi_1\coder{B}{1}{0}{1}{0}{0}=\pi_1\bigl([\bd{Q},\coder{\lambda}{1}{0}{1}{0}{0}]
-\coder{\lambda}{0}{2}{1}{0}{0}\bd{\theta}\mid^0\bigr),\\
    &\pi_1\coder{B}{0}{2}{1}{0}{0}=\pi_1[\bd{Q},\coder{\lambda}{0}{2}{1}{0}{0}],\\
    &\pi_1\coder{B}{0}{3}{2}{0}{0}=\frac{1}{2}\pi_1\bigl([\bd{Q},\coder{\lambda}{0}{3}{2}{0}{0}]
+[\coder{B}{0}{2}{1}{0}{0},\coder{\lambda}{0}{2}{1}{0}{0}]\bigr),\label{eq:B02100}\\
    &\pi_1\coder{B}{0}{3}{1}{2}{2}=\pi_1\bigl([\bd{Q},\coder{\lambda}{0}{3}{1}{2}{2}]
-\coder{\lambda}{0}{2}{1}{0}{0}\coder{B}{0}{2}{0}{2}{2}\bigr),
\end{align}
\end{subequations}
following from \eqref{eq:B^p+1}, we can express  
$\bd{S}_1$ as a $Q$-exact form
\begin{align}
 \pi_1\bd{S}_1 =&\ \pi_1\Biggl[\bd{Q},\ \frac{1}{2}\coder{\lambda}{1}{1}{2}{0}{0}
-\frac{1}{2}\coder{\lambda}{0}{2}{1}{0}{0}\bigl(\bd{\frac{b_0^+}{L_0^+}}\coder{B}{1}{0}{1}{0}{0}\bigr)
-\frac{1}{2}\coder{B}{0}{2}{1}{0}{0}\bigl(\bd{\frac{b_0^+}{L_0^+}}\coder{\lambda}{1}{0}{1}{0}{0}\bigr)
-\frac{1}{2}\coder{\lambda}{0}{3}{2}{0}{0}\bigl(\bd{\frac{b_0^+}{L_0^+}}\bd{\theta}\mid^0\bigr)\nonumber\\
  &\hspace{1.3cm} +\frac{1}{2}\coder{B}{0}{2}{1}{0}{0}\bigl(\bd{\frac{b_0^+}{L_0^+}}\coder{\lambda}{0}{2}{1}{0}{0}\bigr)
\bigl(\bd{\frac{b_0^+}{L_0^+}}\bd{\theta}\mid^0\bigr)
  +\frac{1}{2}\coder{\lambda}{0}{2}{1}{0}{0}\bigl(\bd{\frac{b_0^+}{L_0^+}}\coder{B}{0}{2}{1}{0}{0}
\bigr)\bigl(\bd{\frac{b_0^+}{L_0^+}}\bd{\theta}\mid^0\bigr)\nonumber\\
  &\hspace{1.3cm} -\frac{1}{4}\sum_s\coder{B}{0}{2}{1}{0}{0}\bd{D}\bigl(\frac{b_0^+}{L_0^+}\tilde{e}^0_{s}\bigr)
\bigl(\bd{\frac{b_0^+}{L_0^+}}\coder{\lambda}{0}{2}{1}{0}{0}\bigr)\bd{D}(e^0_{s})
  -\frac{1}{4}\sum_s\coder{\lambda}{0}{2}{1}{0}{0}\bd{D}\bigl(\frac{b_0^+}{L_0^+}\tilde{e}^0_{s}\bigr)
\bigl(\bd{\frac{b_0^+}{L_0^+}}\coder{B}{0}{2}{1}{0}{0}\bigr)\bd{D}(e^0_{s})\nonumber\\
  &\hspace{1.3cm} -\coder{B}{0}{3}{1}{2}{2}\bigl(\bd{\frac{b_0^+}{L_0^+}\Xi}\bd{\theta}\mid^2\bigr)
+\coder{B}{0}{2}{1}{0}{0}\bigl(\bd{\frac{b_0^+}{L_0^+}}\coder{B}{0}{2}{0}{2}{2}\bigr)
\bigl(\bd{\frac{b_0^+}{L_0^+}\Xi}\bd{\theta}\mid^2\bigr)\nonumber\\
  &\hspace{1.3cm} +\frac{1}{2}\coder{B}{0}{2}{0}{2}{2}\bigl(\bd{\frac{b_0^+}{L_0^+}\Xi}\coder{B}{0}{2}{0}{2}{0}\bigr)
\bigl((\bd{\frac{b_0^+}{L_0^+}X}-\bd{\Xi})\bd{\theta}\mid^2\bigr)\Biggr]\nonumber\\
\equiv&\ \pi_1[\bd{Q},\bd{S}_2],
\end{align}
using the relation
\begin{align}\label{eq:e1B=Be1}
    \sum_s\bd{D}\bigl(\frac{b_0^+}{L_0^+}\tilde{e}^0_{s}\bigr)(\bd{\pi^0B})\bd{D}(e^0_{s})
\approx\sum_s(-1)^{e^0_s}\bd{D}(\tilde{e}^0_{s})\bigl(\bd{\frac{b_0^+}{L_0^+}\pi^0B}\bigr)\bd{D}(e^0_{s})\,.
\end{align}
Then, we can rewrite the amplitude as
\begin{align}\label{eq:symplectic form relations}
\bra{\mathcal{A}_{NS}} =& \bra{\omega}P_{-1}\otimes P_{-1}\pi_1[\bd{Q},\bd{S}_2]\ =\ \bra{\omega_l}\xi P_{-1} \otimes P_{-1}\pi_1[\bd{Q},\bd{S}_2]\nonumber\\
=&\ \bra{\omega_l}X_0P_{-1} \otimes P_{-1}\pi_1\bd{S}_2\
=\ \bra{\omega_l}\xi X_0\otimes [\bd{\eta},\bd{S}_2]
\nonumber\\
=&\ \bra{\omega} X_0P_{-1}\otimes P_{-2}\pi_1[\bd{\eta}, \bd{S}_2]\,,
\end{align}
except for terms vanishing when acting on states in $\mathcal{H}_Q$\,. 
Using the relations
\begin{subequations}
\begin{align}
    &\pi_1[\bd{\eta},\coder{\lambda}{1}{1}{2}{0}{0}]=\pi_1\bigl(\coder{B}{1}{1}{1}{0}{0}
-\coder{\lambda}{0}{3}{1}{2}{2}\bd{\theta}\mid^2\bigr),\\
    &\pi_1[\bd{\eta},\coder{\lambda}{1}{0}{1}{0}{0}]=\pi_1\coder{B}{1}{0}{0}{0}{0},\\
    &\pi_1[\bd{\eta},\coder{\lambda}{0}{3}{2}{0}{0}]=\pi_1\coder{B}{0}{3}{1}{0}{0},\\
    &\pi_1[\bd{\eta},\coder{\lambda}{0}{2}{1}{0}{0}]=\pi_1\coder{B}{0}{2}{0}{0}{0},\\
    &\pi_1[\bd{\eta},\coder{B}{1}{0}{1}{0}{0}]=\pi_1\coder{B}{0}{2}{0}{2}{2}\bd{\theta}\mid^2,\\
    &\pi_1[\bd{\eta},\coder{B}{0}{3}{1}{2}{2}]=\pi_1\coder{B}{0}{2}{0}{2}{2}\coder{B}{0}{2}{0}{2}{0},\\
    &\pi_1[\bd{\eta},\coder{B}{0}{2}{1}{0}{0}]=\pi_1[\bd{\eta},\coder{B}{0}{2}{0}{2}{2}]=0,
\end{align}
\end{subequations}
following from \eqref{eq:lambda^p+1}, 
we have
\begin{align}
\pi_1[\bd{\eta}, \bd{S}_2] =&\ \frac{1}{2}\pi_1\Bigl(\coder{B}{1}{1}{1}{0}{0}
-\coder{B}{0}{2}{0}{0}{0}\bigl(\bd{\frac{b_0^+}{L_0^+}}\coder{B}{1}{0}{1}{0}{0}\bigr)
-\coder{B}{0}{2}{1}{0}{0}\bigl(\bd{\frac{b_0^+}{L_0^+}}\coder{B}{1}{0}{0}{0}{0}\bigr)
-\coder{B}{0}{3}{1}{0}{0}\bigl(\bd{\frac{b_0^+}{L_0^+}}\bd{\theta}\mid^0\bigr)
\nonumber\\
&\hspace{1cm} +\coder{B}{0}{2}{1}{0}{0}\bigl(\bd{\frac{b_0^+}{L_0^+}}\coder{B}{0}{2}{0}{0}{0}\bigr)
\bigl(\bd{\frac{b_0^+}{L_0^+}}\bd{\theta}\mid^0\bigr)
+\coder{B}{0}{2}{0}{0}{0}\bigl(\bd{\frac{b_0^+}{L_0^+}}\coder{B}{0}{2}{1}{0}{0}\bigr)
\bigl(\bd{\frac{b_0^+}{L_0^+}}\bd{\theta}\mid^0\bigr)\nonumber\\
     &\hspace{1cm} -\frac{1}{2}\sum_s\coder{B}{0}{2}{1}{0}{0}\bd{D}\bigl(\frac{b_0^+}{L_0^+}\tilde{e}^0_{s}\bigr)
\bigl(\bd{\frac{b_0^+}{L_0^+}}\coder{B}{0}{2}{0}{0}{0}\bigr)\bd{D}(e^0_{s})\nonumber\\
     &\hspace{1cm} -\frac{1}{2}\sum_s\coder{B}{0}{2}{0}{0}{0}\bd{D}\bigl(\frac{b_0^+}{L_0^+}\tilde{e}^0_{s}\bigr)
\bigl(\bd{\frac{b_0^+}{L_0^+}}\coder{B}{0}{2}{1}{0}{0}\bigr)\bd{D}(e^0_{s})\nonumber\\
     &\hspace{1cm} 
-\coder{B}{0}{3}{1}{2}{2}\bigl(\bd{\frac{b_0^+}{L_0^+}}\bd{\theta}\mid^2\bigr)
+\coder{B}{0}{2}{1}{0}{0}\bigl(\bd{\frac{b_0^+}{L_0^+}}\coder{B}{0}{2}{0}{2}{2}\bigr)
\bigl(\bd{\frac{b_0^+}{L_0^+}\theta}\mid^2\bigr)
\nonumber\\
&\hspace{1cm} -\coder{B}{0}{3}{0}{2}{2}\bigl((\bd{\frac{b_0^+}{L_0^+}X-\Xi})\bd{\theta}\mid^2\bigr)
+\coder{B}{0}{2}{0}{0}{0}\bigl(\bd{\frac{b_0^+}{L_0^+}}\coder{B}{0}{2}{0}{2}{2}\bigr)
\bigl((\bd{\frac{b_0^+}{L_0^+}X}-\bd{\Xi})\bd{\theta}\mid^2\bigr)
\nonumber\\
     &\hspace{1cm} +\frac{1}{2}\coder{B}{0}{2}{0}{2}{2}\bigl(\bd{\frac{b_0^+}{L_0^+}}\coder{B}{0}{2}{0}{2}{0}\bigr)
\bigl((\bd{\frac{b_0^+}{L_0^+}X-\Xi)}\bd{\theta}\mid^2\bigr)\nonumber\\
     &\hspace{1cm} +\frac{1}{2}\coder{B}{0}{2}{0}{2}{2}\bigl((\bd{\frac{b_0^+}{L_0^+}X-\Xi})\coder{B}{0}{2}{0}{2}{0}\bigr)
\bigl(\bd{\frac{b_0^+}{L_0^+}\theta}\mid^2\bigr)\Bigr)
\nonumber\\
\equiv&\ \pi_1\bd{S}_3.
\end{align}
This $\bd{S}_3$ can also be expressed as $Q$-exact form: 
\begin{align}
 \pi_1\bd{S}_3 =&\    \pi_1\Biggl[\bd{Q},\ \frac{1}{2}\coder{\lambda}{1}{1}{1}{0}{0}-\frac{1}{2}
\coder{\lambda}{0}{2}{1}{0}{0}\bigl(\bd{\frac{b_0^+}{L_0^+}}\coder{B}{1}{0}{0}{0}{0}\bigr)
-\frac{1}{2}\coder{B}{0}{2}{0}{0}{0}\bigl(\bd{\frac{b_0^+}{L_0^+}}\coder{\lambda}{1}{0}{1}{0}{0}\bigr)
-\frac{1}{2}\coder{\lambda}{0}{3}{1}{0}{0}\bigl(\bd{\frac{b_0^+}{L_0^+}}\bd{\theta}\mid^0\bigr)\nonumber\\
    &\hspace{1cm} +\frac{1}{2}\coder{B}{0}{2}{0}{0}{0}\bigl(\bd{\frac{b_0^+}{L_0^+}}\coder{\lambda}{0}{2}{1}{0}{0}\bigr)
\bigl(\bd{\frac{b_0^+}{L_0^+}}\bd{\theta}\mid^0\bigr)+\frac{1}{2}\coder{\lambda}{0}{2}{1}{0}{0}
\bigl(\bd{\frac{b_0^+}{L_0^+}}\coder{B}{0}{2}{0}{0}{0}\bigr)\bigl(\bd{\frac{b_0^+}{L_0^+}}\bd{\theta}\mid^0\bigr)\nonumber\\
    &\hspace{1cm} -\frac{1}{4}\sum_s\coder{B}{0}{2}{0}{0}{0}\bd{D}\bigl(\frac{b_0^+}{L_0^+}\tilde{e}^0_{s}\bigr)
\bigl(\bd{\frac{b_0^+}{L_0^+}}\coder{\lambda}{0}{2}{1}{0}{0}\bigr)\bd{D}(e^0_{s})
\nonumber\\
&\hspace{1cm} -\frac{1}{4}\sum_s\coder{\lambda}{0}{2}{1}{0}{0}\bd{D}\bigl(\frac{b_0^+}{L_0^+}\tilde{e}^0_{s}\bigr)
\bigl(\bd{\frac{b_0^+}{L_0^+}}\coder{B}{0}{2}{0}{0}{0}\bigr)\bd{D}(e^0_{s}) -\coder{\lambda}{0}{3}{1}{2}{2}\bigl(\bd{\frac{b_0^+}{L_0^+}}\bd{\theta}\mid^2\bigr)
    \nonumber\\
 &\hspace{1cm}    +\coder{\lambda}{0}{2}{1}{0}{0}
\bigl(\bd{\frac{b_0^+}{L_0^+}}\coder{B}{0}{2}{0}{2}{2}\bigr)\bigl(\bd{\frac{b_0^+}{L_0^+}}\bd{\theta}\mid^2\bigr)
+\frac{1}{2}\coder{B}{0}{2}{0}{2}{2}\bigl(\bd{\frac{b_0^+}{L_0^+}}\coder{B}{0}{2}{0}{2}{0}\bigr)
\bigl(\bd{\frac{b_0^+}{L_0^+}\Xi}\bd{\theta}\mid^2\bigr)\Biggr],
\end{align}
where we used \eqref{eq:e1B=Be1} 
and also the relations
\begin{align}
    &\pi_1\coder{B}{1}{1}{1}{0}{0}=\pi_1\bigl([\bd{Q},\coder{\lambda}{1}{1}{1}{0}{0}]
-\coder{\lambda}{0}{2}{1}{0}{0}\coder{B}{1}{0}{0}{0}{0}+\coder{B}{0}{2}{0}{0}{0}\coder{\lambda}{1}{0}{1}{0}{0}
-\coder{\lambda}{0}{3}{1}{0}{0}\bd{\theta}\mid^0-\coder{\lambda}{0}{3}{1}{2}{2}\bd{\theta}\mid^2\bigr),\\
    &\pi_1\coder{B}{1}{0}{1}{0}{0}=\pi_1\bigl([\bd{Q},\coder{\lambda}{1}{0}{1}{0}{0}]
-\coder{\lambda}{0}{2}{1}{0}{0}\bd{\theta}\mid^0\bigr),\\
    &\pi_1\coder{B}{0}{3}{1}{2}{2}=\pi_1\bigl([\bd{Q},\coder{\lambda}{0}{3}{1}{2}{2}]
-\coder{\lambda}{0}{2}{1}{0}{0}\coder{B}{0}{2}{0}{2}{2}\bigr),\\
    &\pi_1\coder{B}{0}{3}{1}{0}{0}=\pi_1\bigl([\bd{Q},\coder{\lambda}{0}{3}{1}{0}{0}]
+[\coder{B}{0}{2}{0}{0}{0},\coder{\lambda}{0}{2}{1}{0}{0}]\bigr),\\
    &\pi_1\coder{B}{0}{2}{1}{0}{0}=\pi_1[\bd{Q},\coder{\lambda}{0}{2}{1}{0}{0}],\\
    &\pi_1[\bd{Q},\coder{B}{0}{2}{0}{0}{0}]=\pi_1[\bd{Q},\coder{B}{0}{2}{0}{2}{0}]
=\pi_1[\bd{Q},\coder{B}{0}{2}{0}{2}{2}]=0,\\
    &\pi_1[\bd{Q},\coder{B}{1}{0}{0}{0}{0}]=-\pi_1\coder{B}{0}{2}{0}{0}{0}\bd{\theta}\mid^0
-\pi_1\coder{B}{0}{2}{0}{2}{2}\bd{\theta}\mid^2.
\end{align}
Repeating the steps in \eqref{eq:symplectic form relations} and then explicitly computing the commutator with $\bd{\eta}$, we eventually obtain
\begin{align}\label{eq:NS amp B0}
    \bra{\mathcal{A}_{NS}}=\bra{\omega}(X_0)^2P_{-1}\otimes P_{-3}\pi_1\Bigl(&\coder{B}{1}{1}{0}{0}{0}
-\coder{B}{0}{2}{0}{0}{0}\bigl(\bd{\frac{b_0^+}{L_0^+}}\coder{B}{1}{0}{0}{0}{0}\bigr)
-\coder{B}{0}{3}{0}{0}{0}\bigl(\bd{\frac{b_0^+}{L_0^+}\theta}\mid^0\bigr)
-\coder{B}{0}{3}{0}{2}{2}\bigl(\bd{\frac{b_0^+}{L_0^+}\theta}\mid^2\bigr)\nonumber\\
    &+\coder{B}{0}{2}{0}{0}{0}\bigl(\bd{\frac{b_0^+}{L_0^+}}\coder{B}{0}{2}{0}{0}{0}\bigr)
\bigl(\bd{\frac{b_0^+}{L_0^+}}\bd{\theta}\mid^0\bigr)+\coder{B}{0}{2}{0}{0}{0}
\bigl(\bd{\frac{b_0^+}{L_0^+}}\coder{B}{0}{2}{0}{2}{2}\bigr)\bigl(\bd{\frac{b_0^+}{L_0^+}\theta}\mid^2\bigr)\nonumber\\
    &+\frac{1}{2}\sum_s\coder{B}{0}{2}{0}{0}{0}\bd{D}\bigl(\frac{b_0^+}{L_0^+}\tilde{e}^0_s\bigr)
\bigl(\bd{\frac{b_0^+}{L_0^+}}\coder{B}{0}{2}{0}{0}{0}\bigr)\bd{D}(e^0_s)\nonumber\\
    &+\frac{1}{2}\sum_s\coder{B}{0}{2}{0}{2}{2}\bd{D}\bigl(\frac{b_0^+}{L_0^+}\tilde{e}^1_s\bigr)
\bigl(\bd{\frac{b_0^+}{L_0^+}}\coder{B}{0}{2}{0}{2}{0}\bigr)\bd{D}(e^1_s)
\Bigr),
\end{align}
using the relations 
\begin{subequations}
\begin{align}
    &\pi_1[\bd{\eta},\coder{\lambda}{1}{1}{1}{0}{0}]=2\pi_1\coder{B}{1}{1}{0}{0}{0},\\
    &\pi_1[\bd{\eta},\coder{\lambda}{1}{0}{1}{0}{0}]=\pi_1\coder{B}{1}{0}{0}{0}{0},\\
    &\pi_1[\bd{\eta},\coder{\lambda}{0}{3}{1}{0}{0}]=2\pi_1\coder{B}{0}{3}{0}{0}{0},\\
    &\pi_1[\bd{\eta},\coder{\lambda}{0}{3}{1}{2}{2}]=\pi_1\coder{B}{0}{3}{0}{2}{2},\\
    &\pi_1[\bd{\eta},\coder{\lambda}{0}{2}{1}{0}{0}]=\pi_1\coder{B}{0}{2}{0}{0}{0},\\
    &\pi_1[\bd{\eta},\bd{B}^{(0)}]=0,
\end{align}
\end{subequations}
following from \eqref{eq:lambda^p+1}. 

In this way, the amplitude can finally be written as
\begin{align}
    &\mathcal{A}_{NS}(\Phi_{NS},\Phi'_{NS})\nonumber\\
    &=\omega\Bigl(\widetilde{\Phi}_{NS},B_{1,1}^{(0)}|^0_0(\Phi'_{NS}) \Bigr)
    -\omega\Bigl(\widetilde{\Phi}_{NS},B_{0,2}^{(0)}|^0_0(\frac{b_0^+}{L_0^+}B_{1,0}^{(0)}|^0_0,\Phi'_{NS}) \Bigr)\nonumber\\
    &\quad-\frac{1}{2}\sum_s
\omega\Bigl(\widetilde{\Phi}_{NS},B_{0,3}^{(0)}|^0_0(\frac{b_0^+}{L_0^+} e^0_s, \tilde{e}^0_s,\Phi'_{NS})\Bigr)
    -\frac{1}{2}\sum_s
\omega\Bigl(\widetilde{\Phi}_{NS},B_{0,3}^{(0)}|^2_2(\frac{b_0^+}{L_0^+}e^1_s,\tilde{e}^1_s,\Phi'_{NS})\Bigr)
\nonumber\\
    &\quad+\frac{1}{2}\sum_s\omega\Bigl(\widetilde{\Phi}_{NS},B_{0,2}^{(0)}|^0_0
\bigl(\frac{b_0^+}{L_0^+}B_{0,2}^{(0)}|^0_0(\frac{b_0^+}{L_0^+}e^0_s, \tilde{e}^0_s),\Phi'_{NS}\bigr) \Bigr)\nonumber\\
    &\quad+\frac{1}{2}\sum_s\omega\Bigl(\widetilde{\Phi}_{NS},B_{0,2}^{(0)}|^0_0
\bigl(\frac{b_0^+}{L_0^+}B_{0,2}^{(0)}|^2_2(\frac{b_0^+}{L_0^+}e^1_s, \tilde{e}^1_s),\Phi'_{NS}\bigr) \Bigr)\nonumber\\
    &\quad+\frac{1}{2}\sum_s\omega\Bigl(\widetilde{\Phi}_{NS},B_{0,2}^{(0)}|^0_0
\bigl(\frac{b_0^+}{L_0^+}\tilde{e}^0_s,\frac{b_0^+}{L_0^+}B_{0,2}^{(0)}|^0_0(e^0_s,\Phi'_{NS})\bigr)\Bigr)\nonumber\\
     &\quad+\frac{1}{2}\sum_s\omega\Bigl(\widetilde{\Phi}_{NS},B_{0,2}^{(0)}|^2_2
\bigl(\frac{b_0^+}{L_0^+}\tilde{e}^1_s,\frac{b_0^+}{L_0^+}B_{0,2}^{(0)}|^2_0(e^1_s,\Phi'_{NS})\bigr)\Bigr),
\end{align}
where $\widetilde{\Phi}_{NS}=(X_0)^2\Phi_{NS}$.
This final form of the amplitude is the same as that in (\ref{NS-2point}) 
obtained by the Feynman rules, except that all the PCOs are commonly placed at one of the external states $\Phi_{NS}$.
Under the assumption that $\bd{B}^{(0)}$ provides a triangulation of the moduli space,
this reproduces the amplitudes obtained in the first quantization method.

It should be noted that the prescription used here is itself unaffected by spurious poles, regardless of whether the amplitude is affected by them. 
This is because the prescription merely involves deforming the contour of the BRST charge, which is insensitive to spurious poles, and inserting $\xi$ when extending the theory to the large Hilbert space. The latter operation is also unaffected by spurious poles, since only the zero mode of $\xi$ is relevant.\footnote{The cancellation mechanism is not affected by spurious poles as already mentioned.} 
Therefore, as long as the amplitude itself is well-defined, applying this prescription reproduces the correct amplitude.

\subsection{One-loop correction to R-R amplitude}
The one-loop correction to the R-R amplitude is given by
\begin{align}
    &\mathcal{A}_{R}(\Phi_{R},\Phi'_{R})\nonumber\\
    &=\omega\Bigl(\Phi_{R},YL_{1,1}^{(2)}|^2_0(\Phi'_{R}) \Bigr)
    -\omega\Bigl(\Phi_{R},YL_{0,2}^{(1)}|^2_0(\frac{b_0^+}{L_0^+}L_{1,0}^{(1)}|^0_0,\Phi'_{R}) \Bigr)\nonumber\\
    &\quad-\frac{1}{2}\sum_s\omega\Bigl(\Phi_{R},YL_{0,3}^{(2)}|^2_0(\frac{b_0^+}{L_0^+}e^0_s, \tilde{e}^0_s,\Phi'_{R})\Bigr)
    -\frac{1}{2}\sum_s
\omega\Bigl(\Phi_{R},YL_{0,3}^{(1)}|^4_2(\frac{b_0^+}{L_0^+}Xe^1_s, \tilde{e}^1_s,\Phi'_{R})\Bigr)\nonumber\\
    &\quad+\frac{1}{2}\sum_s\omega\Bigl(\Phi_{R},YL_{0,2}^{(1)}|^2_0
\bigl(\frac{b_0^+}{L_0^+}L_{0,2}^{(1)}|^0_0(\frac{b_0^+}{L_0^+}e^0_s, \tilde{e}^0_s),\Phi'_{R}\bigr) \Bigr)\nonumber\\
    &\quad+\frac{1}{2}\sum_s\omega\Bigl(\Phi_{R},YL_{0,2}^{(1)}|^2_0
\bigl(\frac{b_0^+}{L_0^+}L_{0,2}^{(0)}|^2_2(\frac{b_0^+}{L_0^+}Xe^1_s, \tilde{e}^1_s),\Phi'_{R}\bigr) \Bigr)\nonumber\\
    &\quad+\sum_s\omega\Bigl(\Phi_{R},YL_{0,2}^{(1)}|^2_0
\bigl(\frac{b_0^+}{L_0^+}X\tilde{e}^1_s,\frac{b_0^+}{L_0^+}L_{0,2}^{(0)}|^2_2(e^1_s,\Phi'_{R})\bigr)\Bigr).
\end{align}
Pulling out $\Phi_{R}$ and $\Phi'_{R}$ to the right and using the coalgebra representation, we can express the amplitude as
\begin{align}
    \bra{\mathcal{A}_R}=\bra{\omega}P_{-1/2}\otimes YP_{-1/2}\pi_1\Bigl(&\coder{L}{1}{1}{2}{2}{0}
-\coder{L}{0}{2}{1}{2}{0}\bigl(\bd{\frac{b_0^+}{L_0^+}}\coder{L}{1}{0}{1}{0}{0}\bigr)
-\coder{L}{0}{3}{2}{2}{0}\bigl(\bd{\frac{b_0^+}{L_0^+}\Theta}\mid^0\bigr)
-\coder{L}{0}{3}{1}{4}{2}\bigl(\bd{\frac{b_0^+}{L_0^+}\Theta}\mid^2\bigr)\nonumber\\
    &+\coder{L}{0}{2}{1}{2}{0}\bigl(\bd{\frac{b_0^+}{L_0^+}}\coder{L}{0}{2}{1}{0}{0}\bigr)
\bigl(\bd{\frac{b_0^+}{L_0^+}\Theta}\mid^0\bigr)+\coder{L}{0}{2}{1}{2}{0}
\bigl(\bd{\frac{b_0^+}{L_0^+}}\coder{L}{0}{2}{0}{2}{2}\bigr)\bigl(\bd{\frac{b_0^+}{L_0^+}\Theta}\mid^2\bigr)\Bigr)\,.
\label{amp RR 1}
\end{align}
Using the form of $\bd{L}$ given in \eqref{L GBF}, 
\begin{subequations}
\begin{align}
    &\pi_1\coder{L}{1}{1}{2}{2}{0}=X\pi_1\Bigl(\coder{B}{1}{1}{1}{2}{0}+\coder{B}{0}{3}{0}{4}{2}(\bd{\Xi\theta}\mid^2)\Bigr),\\
    &\pi_1\coder{L}{1}{0}{1}{0}{0}=\pi_1\Bigl(\coder{B}{1}{0}{1}{0}{0}+\coder{B}{0}{2}{0}{2}{2}(\bd{\Xi\theta}\mid^2)\Bigr),\\
    &\pi_1\coder{L}{0}{3}{2}{2}{0}=X\pi_1\Bigl(\coder{B}{0}{3}{1}{2}{0}
+\coder{B}{0}{2}{0}{2}{0}\bigl(\bd{\Xi}\coder{B}{0}{2}{0}{2}{0}\bigr)\Bigr),\\
    &\pi_1\coder{L}{0}{3}{1}{4}{2}=X\pi_1\coder{B}{0}{3}{0}{4}{2},\\
    &\pi_1\coder{L}{0}{2}{1}{0}{0}=\pi_1\coder{B}{0}{2}{1}{0}{0},\\
    &\pi_1\coder{L}{0}{2}{1}{2}{0}=X\pi_1\coder{B}{0}{2}{0}{2}{0},\\
    &\pi_1\coder{L}{0}{2}{0}{2}{2}=\pi_1\coder{B}{0}{2}{0}{2}{2},
\end{align}
\end{subequations}
the amplitude (\ref{amp RR 1}) can be written as
\begin{align}\label{eq: R amp B}
     \bra{\mathcal{A}_R}=&\ \bra{\omega}P_{-1/2}\otimes P_{-3/2}\pi_1\Bigl(\coder{B}{1}{1}{1}{2}{0}-\coder{B}{0}{2}{0}{2}{0}
\bigl(\bd{\frac{b_0^+}{L_0^+}}\coder{B}{1}{0}{1}{0}{0}\bigr)-\coder{B}{0}{3}{1}{2}{0}
\bigl(\bd{\frac{b_0^+}{L_0^+}\theta}\mid^0\bigr)\nonumber\\
     &\hspace{3cm} +\coder{B}{0}{2}{0}{2}{0}\bigl(\bd{\frac{b_0^+}{L_0^+}}\coder{B}{0}{2}{1}{0}{0}\bigr)
\bigl(\bd{\frac{b_0^+}{L_0^+}\theta}\mid^0\bigr)
-\coder{B}{0}{3}{0}{4}{2}\bigl(\bd{(\frac{b_0^+}{L_0^+}X-\Xi)\theta}\mid^2\bigr)\nonumber\\
     &\hspace{3cm} +\coder{B}{0}{2}{0}{2}{0}\bigl(\bd{\frac{b_0^+}{L_0^+}}\coder{B}{0}{2}{0}{2}{2}\bigr)
\bigl(\bd{(\frac{b_0^+}{L_0^+}X-\Xi)\theta}\mid^2\bigr)\Bigr)\nonumber\\
\equiv&\ \bra{\omega}P_{-1/2}\otimes P_{-3/2}\pi_2\bd{S}_1\,,
\end{align}
where we used the relation
\begin{align}
    \bra{\omega}P_{-1/2}\otimes YXP_{-3/2}=\bra{\omega}XYP_{-1/2}\otimes P_{-3/2}\,, 
\end{align}
and $XY=\id$ acting on states in the restricted space $\overline{\mathcal{H}}_{\textrm{R}}^{(-1/2)}$.
As in the NS-NS amplitude, 
we can express $\bd{S}_1$ as a $Q$-exact form
\begin{align}
 \pi_1\bd{S}_1\ =&\    \pi_1\Biggl[\bd{Q},\,\coder{\lambda}{1}{1}{1}{2}{0}
-\coder{B}{0}{2}{0}{2}{0}\bigl(\bd{\frac{b_0^+}{L_0^+}}\coder{\lambda}{1}{0}{1}{0}{0}\bigr)
-\coder{\lambda}{0}{3}{1}{2}{0}\bigl(\bd{\frac{b_0^+}{L_0^+}\theta}\mid^0\bigr)
+\coder{B}{0}{2}{0}{2}{0}\bigl(\bd{\frac{b_0^+}{L_0^+}}\coder{\lambda}{0}{2}{1}{0}{0}\bigr)
\bigl(\bd{\frac{b_0^+}{L_0^+}\theta}\mid^0\bigr)\nonumber\\
    &\hspace{1cm} -\coder{B}{0}{3}{0}{4}{2}\bigl(\bd{\frac{b_0^+}{L_0^+}\Xi\theta}\mid^2\bigr)
+\coder{B}{0}{2}{0}{2}{0}\bigl(\bd{\frac{b_0^+}{L_0^+}}\coder{B}{0}{2}{0}{2}{2}\bigr)
\bigl(\bd{\frac{b_0^+}{L_0^+}\Xi\theta}\mid^2\bigr)\Biggr],
\end{align}
by using the relations
\begin{subequations}
\begin{align}
    &\pi_1\coder{B}{1}{1}{1}{2}{0}=\pi_1\bigl([\bd{Q},\coder{\lambda}{1}{1}{1}{2}{0}]
+\coder{B}{0}{2}{0}{2}{0}\coder{\lambda}{1}{0}{1}{0}{0}-\coder{\lambda}{0}{3}{1}{2}{0}\bd{\theta}\mid^0\bigr),\\
    &\pi_1\coder{B}{1}{0}{1}{0}{0}=\pi_1\bigl([\bd{Q},\coder{\lambda}{1}{0}{1}{0}{0}]
-\coder{\lambda}{0}{2}{1}{0}{0}\bd{\theta}\mid^0\bigr),\\
    &\pi_1\coder{B}{0}{3}{1}{2}{0}=\pi_1\bigl([\bd{Q},\coder{\lambda}{0}{3}{1}{2}{0}]
+\coder{B}{0}{2}{0}{2}{0}\coder{\lambda}{0}{2}{1}{0}{0}\bigr),\\
    &\pi_1\coder{B}{0}{2}{1}{0}{0}=\pi_1[\bd{Q},\coder{\lambda}{0}{2}{1}{0}{0}],\\
    &\pi_1[\bd{Q},\coder{B}{0}{3}{0}{4}{2}]=-\coder{B}{0}{2}{0}{2}{0}\coder{B}{0}{2}{0}{2}{2},\\
    &\pi_1[\bd{Q},\coder{B}{0}{2}{0}{2}{0}]=0.
\end{align}
\end{subequations}
Applying the prescription given in \eqref{eq:symplectic form relations} as before,
and using the relations
\begin{subequations}
\begin{align}
    &\pi_1[\bd{\eta},\coder{\lambda}{1}{1}{1}{2}{0}]=\pi_1\coder{B}{1}{1}{0}{2}{0},\\
    &\pi_1[\bd{\eta},\coder{\lambda}{1}{0}{1}{0}{0}]=\pi_1\coder{B}{1}{0}{0}{0}{0},\\
    &\pi_1[\bd{\eta},\coder{\lambda}{0}{3}{1}{2}{0}]=\pi_1\coder{B}{0}{3}{0}{2}{0},\\
     &\pi_1[\bd{\eta},\coder{\lambda}{0}{2}{1}{0}{0}]=\pi_1\coder{B}{0}{2}{0}{0}{0},
\end{align}
\end{subequations}
we obtain
\begin{align}
    \bra{\mathcal{A}_R}=&\ \bra{\omega}X_0 P_{-1/2}\otimes P_{-5/2}\pi_1\Bigl(\coder{B}{1}{1}{0}{2}{0}
-\coder{B}{0}{2}{0}{2}{0}\bigl(\bd{\frac{b_0^+}{L_0^+}}\coder{B}{1}{0}{0}{0}{0}\bigr)
-\coder{B}{0}{3}{0}{2}{0}\bigl(\bd{\frac{b_0^+}{L_0^+}\theta}\mid^0\bigr)
-\coder{B}{0}{3}{0}{4}{2}\bigl(\bd{\frac{b_0^+}{L_0^+}\theta}\mid^2\bigr)\nonumber\\
    &\hspace{2.9cm}+\coder{B}{0}{2}{0}{2}{0}\bigl(\bd{\frac{b_0^+}{L_0^+}}\coder{B}{0}{2}{0}{0}{0}\bigr)
\bigl(\bd{\frac{b_0^+}{L_0^+}\theta}\mid^0\bigr)
+\coder{B}{0}{2}{0}{2}{0}\bigl(\bd{\frac{b_0^+}{L_0^+}}\coder{B}{0}{2}{0}{2}{2}\bigr)
\bigl(\bd{\frac{b_0^+}{L_0^+}\theta}\mid^2\bigr)\Bigr).
\end{align}
Then, the final form of the amplitude is given by
\begin{align}
    &\mathcal{A}_{R}(\Phi_{R},\Phi'_{R})\nonumber\\
    &=\omega\Bigl(\widetilde{\Phi}_{R},B_{1,1}^{(0)}|^2_0(\Phi'_{R}) \Bigr)
    -\omega\Bigl(\widetilde{\Phi}_{R},B_{0,2}^{(0)}|^2_0(\frac{b_0^+}{L_0^+}B_{1,0}^{(0)}|^0_0,\Phi'_{R}) \Bigr)\nonumber\\
    &\quad-\frac{1}{2}\sum_s
\omega\Bigl(\widetilde{\Phi}_{R},B_{0,3}^{(0)}|^2_0(\frac{b_0^+}{L_0^+}e^0_s, \tilde{e}^0_s,\Phi'_{R})\Bigr)
    -\frac{1}{2}\sum_s
\omega\Bigl(\widetilde{\Phi}_{R},B_{0,3}^{(0)}|^4_2(\frac{b_0^+}{L_0^+}e^1_s, \tilde{e}^1_s,\Phi'_{R})\Bigr)\nonumber\\
    &\quad+\frac{1}{2}\sum_s\omega\Bigl(\widetilde{\Phi}_{R},B_{0,2}^{(0)}|^2_0
\bigl(\frac{b_0^+}{L_0^+}B_{0,2}^{(0)}|^0_0(\frac{b_0^+}{L_0^+}e^0_s, \tilde{e}^0_s),\Phi'_{R}\bigr) \Bigr)\nonumber\\
    &\quad+\frac{1}{2}\sum_s\omega\Bigl(\widetilde{\Phi}_{R},B_{0,2}^{(0)}|^2_0
\bigl(\frac{b_0^+}{L_0^+}B_{0,2}^{(0)}|^2_2(\frac{b_0^+}{L_0^+}e^1_s, \tilde{e}^1_s),\Phi'_{R}\bigr) \Bigr)\nonumber\\
    &\quad+\sum_s\omega\Bigl(\widetilde{\Phi}_{R},B_{0,2}^{(0)}|^2_0
\bigl(\frac{b_0^+}{L_0^+}\tilde{e}^1_s,\frac{b_0^+}{L_0^+}B_{0,2}^{(0)}|^2_2(e^1_s,\Phi'_{R})\bigr)\Bigr),
\end{align}
where $\widetilde{\Phi}_{R}=X_0\Phi_{R}$.
For the same reasons as for the NS-NS amplitude, we can conclude that this reproduces the amplitude in the first quantization method.

\sectiono{Discussion}\label{Discussion}

In this paper, we propose a prescription, generalizing that for the classical action governed by a classical $L_\infty$ structure, to construct a quantum action for heterotic string field theory with a quantum $L_\infty$ structure. Assuming the absence of the spurious pole problem, any quantum interaction is determined by iteratively solving differential equations and transforming the resulting solutions via an $IBL_\infty$ transfer. To verify the consistency of our prescription, we calculate the one-loop two-point function (setting aside the issue of spurious poles for the moment) and show that it agrees with the result obtained in the first-quantized formalism. 

When extending to a large Hilbert space at an intermediate step, we restrict the states propagating in loops to those in the small Hilbert space. While this obscures the cyclicity, we temporarily set this issue aside to formulate our prescription. The cyclicity of the string products constructed following our prescription requires a separate proof. 

The demonstrations presented in this paper to verify our prescription are conducted without addressing the issue of spurious poles. If spurious poles undermine the validity of our arguments, we must modify and improve the prescription. In that case, there are two potential points where the prescription could be modified. 

The first concerns the way of triangulating moduli space. In bosonic string field theory, it suffices to ensure that the Feynman rules of the string field theory reproduce the first-quantized amplitudes. However, since there is not necessarily a unique way to perform this triangulation, a suitable choice might allow us to avoid the problem of spurious poles. In this regard, a recently developed approach (which utilizes hyperbolic geometry \cite{Moosavian:2017qsp,Moosavian:2017sev,Costello:2019fuh}) provides a natural and general method for triangulation; it is hoped that this approach also circumvents the issue of spurious poles. 

Another point that can be modified is the way of integrating \eqref{eq:lambda^p+1} (along the $\eta$ direction). We have adopted the simplest approach \eqref{eq:lambda solution} (which works well in classical heterotic string field theory) but this is not unique. Our solution above is obtained by integrating $\xi$ along the boundary of the Riemann surface that describes the string product; however, in the case of quantum theory, this Riemann surface necessarily has a finite area, which allows $\xi$ to be placed at a point (or integrated along a line) within its interior. By placing $\xi$ appropriately to satisfy cyclicity, the spurious pole problem might be circumvented \cite{Erler:2017pgf}.

In any case, string field theory involves decomposing Riemann surfaces into constituent parts (propagators and vertices) and providing rules for constructing general Riemann surfaces from them. Once the distributions of PCOs for vertices have been determined, they are used to construct every Riemann surface, and the spurious pole problem must be circumvented in every case. It is difficult, in our approach, to address the spurious pole problem that arises from the global structure of Riemann surfaces, although it may be possible to discuss individual cases. This issue requires further investigation. It would be helpful to clarify the relationship between our approach and another approach \cite{Sen:2015uaa,Pius:2018pqr}, a framework in which the connections to the geometry of Riemann surfaces (and their moduli) and the path-integral formalism are more explicit. The general argument presented in \cite{Sen:2015hia,Erler:2017dgr} regarding how to circumvent the problem of spurious poles is highly illuminating.

Furthermore, a rigorous examination of quantum consistency should yield constraints on the gauge group from conditions concerning the cancellations of quantum anomalies and/or tadpoles. This requires more detailed calculation.

Extending this construction to the case of Type II superstrings also represents an important direction for future research.

\vspace{1cm}
\newpage
\noindent
{\bf \large Acknowledgments}

The early stage of the work of H.K. was partially supported in part by JSPS KAKENHI Grant Number JP18K03645.
The work of J.~T. was supported in part by JSPS KAKENHI Grant No. JP22H05115. 
Y.G. would like to thank R.~Adachi and S.~Chikazawa for helpful discussions.

\vspace{1cm}
\appendix

\sectiono{Basic properties of bases}\label{prop of basis}

We use the bra-ket notation for convenience.
Consider first the case of bosonic string field theory.
We represent a string field $\Phi\in\mathcal{H}$ by a ket $\ket{\Phi}$.
Its dual $\bra{\Phi}$ is defined by a linear map:
\begin{alignat}{3}
\bra{\Phi}\ :\ & \mathcal{H}& \qquad &\rightarrow&\qquad &\mathbb{C}  
\nonumber\\
&\text{\rotatebox[origin=c]{90}{$\in$}}& & & &\text{\rotatebox[origin=c]{90}{$\in$}} 
\nonumber\\
 &\!\!\ket{\Psi}& &\mapsto\ & &\hspace{-3mm}\braket*{\Phi}{\Psi}
\end{alignat}
with
\begin{align}
    \braket{\Phi}{\Psi}\ =\ - (-1)^{\Phi}
\omega_{\textrm{B}}(\ket{\Phi},\ket{\Psi}),\qquad\textrm{for}\ \forall \ket{\Psi} \in \mathcal{H}.
\end{align}
Since $\omega$ is degree odd by definition, the bra $\bra{\Phi}$ has opposite degree to the ket $\ket{\Phi}$:
\begin{align}
    \textrm{deg}(\bra{\Phi})\ =\ \textrm{deg}(\ket{\Phi}) + 1\qquad\mod{2}.
\end{align}
In this notation, we have
\begin{align}
    \braket{\Phi_1}{\Phi_2}\ =&\ (-1)^{(\Phi_1+1)(\Phi_2+1)}\braket{\Phi_2}{\Phi_1},\label{eq:braket sym}\\
\braket{\tilde{e}_s}{e_t}\ =&\ \delta_{s,t},
\end{align}
from Eqs.~(\ref{symp b}) and \eqref{eq: bosonic basis relation}.
The identity operator in $\mathcal{H}$ can be expressed as
\begin{align}
    \id =\ \sum_s\ket{e_s}\bra{\tilde{e}_s}\ =\ \sum_s|\tilde{e}_s\rangle\langle e_s|.\label{eq:id}
\end{align}
Then, we can prove the second equality in Eq.~(\ref{theta sym}) as
\begin{align}
 \sum_s|e_s\rangle\otimes|\tilde{e}_s\rangle\ =&\  \sum_{r,s,t}|\tilde{e}_r\rangle\langle e_r| e_s\rangle
\otimes |e_t\rangle\langle \tilde{e}_t|\tilde{e}_s\rangle
\nonumber\\
=&\ \sum_{r,s,t}(-1)^{(e_r+1)(e_s+1)}
\ket{\tilde{e}_r}\otimes\ket{e_t}\braket*{\tilde{e}_t}{\tilde{e}_s}\braket*{e_s}{e_r}
\nonumber\\
=&\ \sum_{s} |\tilde{e}_s\rangle\otimes|e_s\rangle,
\label{e tilde e}
\end{align}
where we used the fact that $\langle e_s|e_t\rangle\ne0$ if and only if 
deg($e_s$)$=$deg($e_t$)$+1$. 

We can write the symplectic form $\omega$ as
\begin{align}
 \omega_{\textrm{B}}(A,B)\ =\ \bra{\omega_{\textrm{B}}}\big(\ket{A}\otimes \ket{B}\big),\qquad
 \bra{\omega_{\textrm{B}}}\ =\  \sum_r(-1)^{e_r}\bra{e_r}\otimes\bra{\tilde{e}_r}.
\end{align}
Similar to Eq.~(\ref{e tilde e}), it is easy to show that
\begin{equation}
  \sum_r(-1)^{e_r}\bra{e_r}\otimes\bra{\tilde{e}_r}\
=\  \sum_r(-1)^{\tilde{e}_r}\bra{\tilde{e}_r}\otimes\bra{e_r},
\end{equation}
which makes $\omega_{\textrm{B}}$ (graded) anti-symmetric as in Eq.~(\ref{symp b}).
The map $\theta_{\textrm{B}}$ in Eq.~(\ref{reflector B}) is the inverse of $\omega_{\textrm{B}}$ in the sense that
\begin{align}
 \big(\id\otimes\bra{\omega_{\textrm{B}}}\big)\big(\ket{\theta_{\textrm{B}}}\otimes\id\big)\ =&\
\sum_{r,s}(-1)^{e_r}\big(\id\otimes\bra{e_r}\otimes\bra{\tilde{e}_r}\big)
\big(\ket{e_s}\otimes\ket{\tilde{e}_s}\otimes\id\big)
\nonumber\\
=&\ \sum_r\ket{e_r}\otimes\bra{\tilde{e}_r}\ =\ \id.
\end{align}

The extension to the heterotic string is straightforward.
First we consider a linear map $\bra{\Phi_{(p)}}$ defined by using $\omega$\,:
\begin{equation}
 \braket*{\Phi_{(p)}}{\Psi_{(q)}}\ =\ - (-1)^{\Phi}\omega(\ket{\Phi_{(p)}},\ket{\Psi_{(q)}}).
\label{inner small}
\end{equation} 
The string fields $\ket{\Phi_{(p)}}$ and $\ket{\Psi_{(q)}}$ are those in the small Hilbert space
with picture numbers $p$ and $q$, respectively. 
We assume that the Grassmann parity of fields does not depend on the picture number
since the PCOs are Grassmann even.
From the picture number conservation, the inner product (\ref{inner small}) can be nonzero
if and only if $p+q=-2$. We introduce the basis $\{e^a_{(s,p)}\}$ with $p\in\mathbb{Z}$ for the NS ($a=0$) sector and $p\in\mathbb{Z}+\frac{1}{2}$ for the R $(a=1)$ sector.
The dual basis $\{\tilde{e}^a_{(s,p)}\}$ is defined to satisfy
\begin{align}
 \braket*{\tilde{e}^a_{(s,p)}}{e^b_{(t,q)}}\ =\ \delta^{a,b}\delta_{s,t}\delta_{p+q+2,0}\,.
\end{align}
Then, the symplectic form $\bra{\omega}$ can be written as
\begin{align}
 \bra{\omega}\ =&\sum_{a,s,p}(-1)^{e^a_s}\bra{e^a_{(s,p)}}\otimes\bra{\tilde{e}^a_{(s,-p-2)}}
\nonumber\\
=&\sum_{a,s,p}(-1)^{\tilde{e}^a_s}\bra{\tilde{e}^a_{(s,p)}}\otimes\bra{e^a_{(s,-p-2)}}\,.
\nonumber
\end{align}
As in the bosonic case,
we can show that the symmetric property
\begin{align}
 \sum_s\ket{e^a_{(s,p)}}\otimes\ket{\tilde{e}^a_{(s,-p-2)}}\ =&\
    \sum_s\ket{\tilde{e}^a_{(s,p)}}\otimes\ket{e^a_{(s,-p-2)}}
\end{align}
holds using the completeness condition,
\begin{equation}
 \pi^aP_{p}\ =\ \sum_{s}\ket{e^a_{(s,p)}}\bra{\tilde{e}^a_{(s,-p-2)}}\,,
\end{equation}
on picture number $p$ states in the $a$ sector.
The reflector state defined by
\begin{align}
 \ket{r^a}^{(p)}\ =\ \sum_s\ket{e^a_{(s,p)}}\otimes\ket{\tilde{e}^a_{(s,-p-2)}}
\end{align}
is the inverse of $\bra{\omega}$ on the picture number $p$ states of the $a$ sector:
\begin{align}
 (\id\otimes\bra{\omega})(\ket{r^a}^{(p)}\otimes\id)\ =\
 \sum_s\ket{e^a_{(s,p)}}\bra{\tilde{e}^a_{(s,-p-2)}}\ =\ \pi^aP_{p}\,.
\end{align}
We define a state $\ket{\theta}$ by a sum of (symmetrized) reflectors:
\begin{align}
 \ket{\theta}\ =&\ \ket{r^0}^{(-1)} + \ket{r^1}^{(-3/2)}_{\textrm{sym}}\nonumber\\ 
 =&\ \frac{1}{2}\sum_s\ket{e^0_{(s,-1)}}\wedge\ket{\tilde{e}^0_{(s,-1)}}
+ \frac{1}{2}\sum_s\ket{e^1_{(s,-3/2)}}\wedge \ket{\tilde{e}^1_{(s,-1/2)}}\,.
\end{align}
which satisfies
\begin{align}
 (\id\otimes\bra{\omega})(\ket{\theta}\otimes\id)\ =&\
 \sum_s\ket{e^0_{(s,-1)}}\bra{\tilde{e}^0_{(s,-1)}}
+\frac{1}{2}\sum_s\left(\ket{e^1_{(s,-3/2)}}\bra{\tilde{e}^1_{(s,-1/2)}}
+\ket{e^1_{(s,-1/2)}}\bra{\tilde{e}^1_{(s,-3/2)}}\right)\ 
\nonumber\\
=&\
 \pi^0P_{-1}+\frac{1}{2}\pi^1\left(P_{-1/2}+P_{-3/2}\right)\,.
\end{align}

The symplectic form of the restricted small Hilbert space $\bra{\Omega}$ is written as
\begin{align}
 \bra{\Omega}\ =&\ \sum_s\Big(
(-1)^{e^0_s}\bra{e^0_{(s,-1)}}\otimes\bra{\tilde{e}^0_{(s,-1)}}
+ (-1)^{e^1_s}\bra{e^1_{(s,-1/2)}}Y\otimes\bra{\tilde{e}^1_{(s,-1/2)}}Y\Big)
\nonumber\\
=&\  \sum_s\Big(
(-1)^{\tilde{e}^0_s}\bra{\tilde{e}^0_{(s,-1)}}\otimes\bra{e^0_{(s,-1)}}
+ (-1)^{\tilde{e}^1_s}\bra{\tilde{e}^1_{(s,-1/2)}}Y\otimes\bra{e^1_{(s,-1/2)}}Y\Big)\,.
\end{align}
The basis and dual basis in the restricted space
\begin{align}
E^a_s\ =\ \{e^0_{(s,-1)}, Xe^1_{(s,-3/2)}\},\
\tilde{E}^a_s\ =\ \{\tilde{e}^0_{(s,-1)},X\tilde{e}^1_{(s,-3/2)}\}\ \in \mathcal{H}^{\textrm{res}}\
=\ \mathcal{H}_{NS}^{(-1)} + \overline{\mathcal{H}}_{R}^{(-1/2)}\,,
\end{align}
satisfy
\begin{align}
    \Omega(\tilde{E}^a_s,E^b_t)\ =\ -\Omega(E^b_t,\tilde{E}^a_s)\ =\ (-1)^{e^a_s}\delta^{a,b}\delta_{s,t}\,.
\end{align}
We can define the symplectic form $\bra{\omega_l}$ in the large Hilbert space by
\begin{align}
 \bra{\omega_l}\ =&\ \sum_{a,s,p}\Big(\langle\langle e^a_{(s,p)}\mid\otimes\langle\langle\tilde{e}^a_{(s,-p-2)}\mid\xi
+(-1)^{e^a_s}\langle\langle e^a_{(s,p)}\mid\xi\otimes\langle\langle\tilde{e}^a_{(s,-p-2)}\mid\Big)
\nonumber\\
=&\  \sum_{a,s,p}\Big(\langle\langle\tilde{e}^a_{(s,p)}\mid\otimes\langle\langle e^a_{(s,-p-2)}\mid\xi
+(-1)^{\tilde{e}^a_s}\langle\langle\tilde{e}^a_{(s,p)}\mid\xi\otimes\langle\langle e^a_{(s,-p-2)}\mid\Big)
\end{align}
where we introduced $\langle\langle \tilde{e}^a_{(s,p)}\mid$ and $\langle\langle e^a_{(s,p)}\mid$ as states that satisfy
\begin{align}
 \langle\langle\tilde{e}^a_{(s,p)}\mid\xi\ket{e^b_{(t,q)}}\ =\ \langle\langle e^b_{(t,q)}\mid\xi\ket{\tilde{e}^a_{(s,p)}}\ =\ 
\delta^{a,b}\delta_{s,t}\delta_{p+q+2,0}\,.
\label{large basis}
\end{align}
Then, we find that
\begin{align}
 \omega_l(\xi\tilde{e}^a_{(s,p)}, e^b_{(t,q)})\ =\  (-1)^{\tilde{e}^a_s}\omega_l(\tilde{e}^b_{(t,q)},\xi e^a_{(s,p)})
=&\ (-1)^{e^a_s}\delta^{a,b}\delta_{s,t}\delta_{p+q+2,0}\,.
\end{align}
The reflector state $|r_l^a\rangle$ in the large Hilbert space is defined by
\begin{align}
 |r_l^a\rangle\ =&\ - (\xi\otimes\id-\id\otimes\xi)\sum_{s,p}|e^a_{(s,p)}\rangle\otimes|\tilde{e}^a_{(s,-p-2)}\rangle
\nonumber\\
=&\ - \sum_{s,p}\Big(\xi|e^a_{(s,p)}\rangle\otimes|\tilde{e}^a_{(s,-p-2)}\rangle
- (-1)^{e^a_s}|e^a_{(s,p)}\rangle\otimes\xi|\tilde{e}^a_{(s,-p-2)}\rangle\Big),
\end{align}
which satisfies
\begin{align}
 (\id\otimes\bra{\omega_l})(\ket{r_l^a}\otimes\id)\ =\
\sum_{s,p}\Big(|\tilde{e}^a_{(s,-p-2)}\rangle\langle\langle e^a_{(s,p)}|\xi
+\xi|\tilde{e}^a_{(s,-p-2)}\rangle\langle\langle e^a_{(s,p)}|\Big)\ =\ \pi^a\,.
\end{align}
Projecting onto the fixed picture numbers $(q,-q-1)$\,, it becomes
\begin{align}
(P_{q}\otimes P_{-q-1}) |r_l^a\rangle\ =\
- \sum_{s}\Big(\xi|\tilde{e}^{a}_{(s,q-1)}\rangle\otimes|e^{a}_{(s,-q-1)}\rangle
- (-1)^{\tilde{e}^{a}_s}|\tilde{e}^{a}_{(s,q)}\rangle\otimes\xi|e^{a}_{(s,-q-2)}\rangle\Big).
\end{align}

In Appendix~\ref{sec:cyclicity}, we use the relations
\begin{align}
 \omega_l(\xi e^a_{(s,p)},B)\ =\ (-1)^{e^a_s}\langle\langle e^a_{(s,p)}\mid\xi\ket{B},\qquad
 \omega_l(e^a_{(s,p)},B)\ =\ -\langle\langle e^a_{(s,p)}\ket{B},
\end{align}
for a coderivation $B$.

From (\ref{large basis}), the completeness condition on a fixed-picture-number subspace in the NS $(a=0)$ or R $(a=1)$ sector is given by
\begin{align}
    P_q\pi^a\id\ =&\ \sum_{s}\Big(\ket{e^a_{(s,q)}}\langle\langle \tilde{e}^a_{(s,-q-2)}|\xi
    + \xi\ket{e^a_{(s,q-1)}}\langle\langle\tilde{e}^a_{(s,-q-1)}|\Big)\,,
    \label{eq:complete large}
\end{align}
from which we see that the partition function for each sector vanishes:
\begin{align}
    Z_a^{(q)}\ =&\ \textrm{Tr}_{\mathcal{H}_l^{(q)}}\Big(\mathcal{P}_{\textrm{GSO}}\, e^{\tau L+\bar{\tau}\bar{L}}\Big)\nonumber\\ 
    =&\ \sum_s\Big((-1)^{e^a_s}\langle\langle \tilde{e}^a_{(s,-q-2)}|\xi\, \mathcal{P}_{\textrm{GSO}}\, e^{\tau L+\bar{\tau}\bar{L}}\ket{e^a_{(s,q)}} 
    \nonumber\\
    &\hspace{3cm}
    +(-1)^{\tilde{e}^a_s} \langle\langle\tilde{e}^a_{(s,-q-1)}| \mathcal{P}_{\textrm{GSO}}\, e^{\tau L+\bar{\tau}\bar{L}} \xi\ket{e^a_{(s,q-1)}}  \Big)
    \nonumber\\
    =&\ 0\,.
\end{align}
This provides the simplest example of the cancellation mechanism.

From 
\begin{align}
 \bra{\omega_l}(\xi\otimes\id)\ =&\
\sum_{a,s,p}(-1)^{e^a_s}\langle\langle e^a_{(s,p)}\mid\xi\otimes\langle\langle\tilde{e}^a_{(s,-p-2)}\mid\xi\
\nonumber\\
=&\ \sum_{a,s,p}(-1)^{\tilde{e}^a_s}\langle\langle \tilde{e}^a_{(s,p)}\mid\xi\otimes\langle\langle e^a_{(s,-p-2)}\mid\xi\,,
\end{align}
we can relate $\bra{\omega_l}$ to the symplectic form of the small Hilbert space $\bra{\omega}$
by identifying $\langle\langle \tilde{e}^a_{(s,p)}\mid\xi\equiv\langle\tilde{e}^a_{(s,p)}\mid$ and
$\langle\langle e^a_{(s,p)}\mid\xi\equiv\langle e^a_{(s,p)}\mid$\,. 

\sectiono{Operator form of reflectors}\label{sec:reflectors}

We summarize the operator form of the reflector in this Appendix.
The reflector considered in this paper represents the delta-function sewing of two inputs of string products. In bra-ket notation, it is given by
\begin{align}
\ket{r(1,2)}^a_{(q)}
= \sum_{s}\ket{e^a_{(s,q)}}\otimes
\ket{\tilde{e}^a_{(s,-q-2)}},
\end{align}
and is related to $\theta$ as
\begin{align}
\theta\mid^0
= \ket{r(1,2)}^0_{(-1)}\,,\qquad
\theta\mid^2
= \frac{1}{2}\Big(
\ket{r(1,2)}^1_{(-3/2)}
+ \ket{r(1,2)}^1_{(-1/2)}
\Big)\,.
\end{align}
We can factorize it into four components
\begin{align}
\ket{r(1,2)}^a_{(q)}\ =\ \left(b_0^{-(1)}\mathcal{P}^{(1)}\otimes\mathcal{P}^{(2)}\right)\ket{r(1,2)}_{X}\otimes
\ket{r(1,2)}_{bc}\otimes\ket{r(1,2)}_{\psi}^a\otimes\ket{r(1,2)}_{\beta\gamma}^{(q)}\,,
\end{align}
with 
\begin{align}
    \mathcal{P}\ =\ \int^{2\pi}_{0}\frac{d\theta}{2\pi}e^{i\theta L^-_0}\,,
\end{align}
and the explicit form of each part is given as follows.

The $X$ and $bc$-ghost components are given by
\begin{align}
    \ket{r(1,2)}_{X}\ =&\ \exp\left(-\sum_{n>0}\frac{1}{n}\alpha^{(1)}_{-n}\cdot\alpha^{(2)}_{-n}
    - \sum_{n>0}\frac{1}{n}\tilde{\alpha}^{(1)}_{-n}\cdot\tilde{\alpha}^{(2)}_{-n}\right)\ket{k}_1\otimes\ket{-k}_2\,,\\
    \ket{r(1,2)}_{bc}\ =&\ \exp\left(\sum_{\pm}
    \left(\sum_{n\ge0}c^{\pm(1)}_{-n}b^{\pm(2)}_{-n}-\sum_{n>0}b^{\pm(1)}_{-n}c^{\pm(2)}_{-n}\right)\right)\ket{\downarrow\downarrow}_1\otimes \ket{\uparrow\uparrow}_2\,.
\end{align}
These are determined by connection conditions 
\begin{alignat}{2}
    \left(\alpha_n^{\mu(1)}+\alpha^{\mu(2)}_{-n}\right)\ket{r(1,2)}_{X}\ =&\ 0\,,&\qquad
    \left(\tilde{\alpha}_n^{\mu(1)}+\tilde{\alpha}^{\mu(2)}_{-n}\right)\ket{r(1,2)}_{X}\ =&\ 0\,,\qquad
    (n\ne0)\,,\\
    \left(b^{\pm(1)}_{n}-b^{\pm(2)}_{-n}\right)\ket{r(1,2)}_{bc}\ =&\ 0\,,&\qquad  
    \left(c^{\pm(1)}_{n}+c^{\pm(2)}_{-n}\right)\ket{r(1,2)}_{bc}\ =&\ 0\,,\qquad 
\end{alignat}
and 
\begin{align}
    \left(p^{\mu(1)}+p^{\mu(2)}\right)\ket{r(1,2)}_{X}\ =\ 0\,. 
\end{align}

The $\psi$ and $\beta\gamma$-ghost components are separately given for the NS and R parts.
Let us first consider the $\psi$ component, since the $\beta\gamma$ component is complicated when the picture number $q$ is kept general. 

The $\psi$ component for the NS sector is 
\begin{align}
    \ket{r(1,2)^\pm}_{\psi}^0\ =\ \exp\left(\pm i\sum_{r>0}\psi^{(1)}_{-r}\cdot\psi^{(2)}_{-r}\right)\ket{0}_1\otimes\ket{0}_2\,,
\end{align}
which satisfies
\begin{align}
    \left(\psi^{\mu(1)}_{r}\mp i\psi^{\mu(2)}_{-r}\right)\ket{r(1,2)^\pm}^0_{\psi}\ =\ 0\,.
\end{align}
The two choices, denoted by $\pm$, characterize the spin structure after sewing. If they are consistent with the choice of reflectors in the $\beta\gamma$ component, the two choices become identical after imposing the GSO projection.
For the R sector, the reflector for $\psi$ component is written as
\begin{align}
    \ket{r(1,2)^\pm}_{\psi}^1\ =&\ \exp\left(\pm\sum_{n>0}\psi^{(1)}_{-n}\cdot\psi^{(2)}_{-n}
    \right)\ket{0}^{\pm}_{12}\,,\\
    \ket{0}^{\pm}_{12}\ =&\ 2^{-\frac{5}{2}}\left(\sum_{\alpha,\bar{\beta}}\ket{\alpha}_1\otimes\ket{\bar{\beta}}_2({C^{-1})_{\alpha}}^{\bar{\beta}}\mp i\sum_{\bar{\alpha},\beta}\ket{\bar{\alpha}}_1\otimes\ket{\beta}_2({\bar{C}^{-1})^{\bar{\alpha}}}_{\beta}
    \right)\,,
\end{align}
following the spinor convention of \cite{Hata:1987qx}.\footnote{The gamma-matrix convention differs
slightly from that used in Ref.~\cite{Friedan:1985ge}.} As in the NS sector, it satisfies
\begin{align}
     \left(\psi^{\mu(1)}_{n}\mp i\psi^{\mu(2)}_{-n}\right)\ket{r(1,2)^\pm}^1_{\psi}\ =\ 0\,,
\end{align}
including the zero mode.

We consider the $\beta\gamma$ component for general picture number $q$\,. The oscillator ground state in picture number $q$ satisfies
\begin{alignat}{2}
    \beta_{n}\ket{q}\ =&\ 0\,, &\qquad \textrm{for}\quad n\ >&\ -q-\frac{3}{2}\,,\\
    \gamma_{n}\ket{q}\ =&\ 0\,,&\qquad \textrm{for}\quad n\ \ge&\ q+\frac{3}{2}\,. 
\end{alignat}
Since the dual basis is treated somewhat differently later, we distinguish it with a tilde.
The reflector for the $\beta\gamma$ component with picture number $q$ is given by
\begin{align}
    \ket{r(1,2)^\pm}^{(q)}_{\beta\gamma}\ =&\ \exp\left( \mp i \sum_{n>-q-3/2}\gamma^{(1)}_{-n}\beta^{(2)}_{-n} \pm i\sum_{n\ge q+3/2}\gamma^{(2)}_{-n}\beta^{(1)}_{-n}
    \right)\ket{q}_1\otimes\widetilde{\ket{-q-2}}_2\,.
\end{align}
Whether the reflector is for the NS sector or the R sector is determined by whether \(q\) is an integer or a half-integer.
The oscillator mode $n$ takes half-integer and integer values for the NS and R sectors, respectively.
The reflector satisfies
\begin{align}
    \left(\gamma_{n}^{(1)}\mp i\gamma_{-n}^{(2)}\right)\ket{r(1,2)^\pm}^{(q)}_{\beta\gamma}\ =&\ 0\,,\\
    \left(\beta_{n}^{(1)}\mp i\beta_{-n}^{(2)}\right)\ket{r(1,2)^\pm}^{(q)}_{\beta\gamma}\ =&\ 0\,,   
\end{align}
independent of the picture number (and hence of the sector).
The fact that these conditions do not depend on the picture number $q$ (in each sector) implies that the loop generated by this reflector does not depend on the picture number of the state propagating along it.

The GSO projection operator is defined by
\begin{align}
    \mathcal{P}_{\textrm{GSO}}^{\textrm{NS}(q)}\ =\ \frac{1}{2}\left(1+(-1)^{F_q}\right)\,,\qquad
        \mathcal{P}_{\textrm{GSO}}^{\textrm{R}(q)}\ =\ \frac{1}{2}\left(1+(-1)^{F_q}\Gamma_{11}\right)\,,
\end{align}
with
\begin{align}
    F_q\ =\ \sum_{n>0}\psi_{-n}\cdot\psi_{n} - \sum_{n>-q-3/2}\gamma_{-n}\beta_{n} -\sum_{n\ge q+3/2}\beta_{-n}\gamma_{n} + q + \epsilon\,,
\end{align}
where $\epsilon=0$ for the NS sector and $\epsilon=1/2$ for the R sector.
In the bosonized representation, 
\begin{align}
    \frac{1}{\sqrt{2}}\left(\pm\psi^0+\psi^9\right)\ =&\ e^{\pm iH_0}\,,\qquad 
    \frac{1}{\sqrt{2}}\left(\psi^{I}\pm i\psi^{I+4}\right)\ =\ e^{\pm iH_I}\,,\qquad (I=1,\cdots, 4)\,,\\
    \gamma\ =&\ e^{\phi}\eta\,,\qquad \beta\ =\ \partial \xi e^{-\phi}\,,
\end{align}
the GSO projection takes the same form in the NS and R sectors:
\begin{align}
\mathcal{P}_{\textrm{GSO}}\ =\ \frac{1}{2}\left(1+(-1)^{F}\right)\,,\qquad
F\ =\ \oint\frac{dz}{2\pi i}\left(\sum_{I=0}^{4}i\partial H_I-\partial\phi\right)\,.
\label{bosonized GSO}
\end{align}
We also adopt this as the GSO projection in the large Hilbert space. 

The reflector in the large Hilbert space is obtained by replacing the $\beta\gamma$ component with
\begin{align}
\ket{r(1,2)}_{\eta\xi\phi}\ =&\ \exp\left(-\sum_{n>0}\phi^{(1)}_{-n}\phi^{(2)}_{-n}
-\sum_{n\ge0}\xi^{(1)}_{-n}\eta^{(2)}_{-n}+\sum_{n>0}\eta^{(1)}_{-n}\xi^{(2)}_{-n}\right)\ket{0(1,2)}\,,\\
\ket{0(1,2)}\ =&\ \sum_{p}(-1)^p\left(e^{p\phi^{(1)}(0)}\ket{0}_1\otimes \xi_0^{(2)}e^{-(p+2)\phi^{(2)}(0)}\ket{0}_2\right)\,.
\end{align}
Here, the sum over $p$ runs over integers in the NS sector and half-integers in the R sector.
It satisfies the conditions
\begin{align}
    \left(\eta^{(1)}_{n}+\eta^{(2)}_{-n}\right)\ket{r(1,2)}_{\eta\xi\phi}\ =&\ 0\,,\qquad 
        \left(\xi^{(1)}_{n}-\xi^{(2)}_{-n}\right)\ket{r(1,2)}_{\eta\xi\phi}\ =\ 0\,,\\
            \left(\phi^{(1)}_n+\phi^{(2)}_{-n}\right)\ket{r(1,2)}_{\eta\xi\phi}\ =&\ 0\,.
            \label{con cond large}
\end{align}
Extracting $\xi_0$, it is written as
\begin{align}
\ket{r(1,2)}_{\eta\xi\phi}\ =&\    - \left(\xi_0\otimes\id - \id\otimes\xi_0\right)
\nonumber\\
&\
\times \sum_{p}\exp\left(-\sum_{n>0}\phi^{(1)}_{-n}\phi^{(2)}_{-n}
-\sum_{n>0}\xi^{(1)}_{-n}\eta^{(2)}_{-n}+\sum_{n>0}\eta^{(1)}_{-n}\xi^{(2)}_{-n}\right)\ket{p}_1\otimes\ket{-p-2}_2\,.
\label{ref large alt}
\end{align}
The picture number is counted by the operator
\begin{align}
    Q_{\textrm{p}}\ =\ \oint\frac{dz}{2\pi i}\left(\xi\eta-\partial\phi\right)(z)\,,
\end{align}
and the picture number projection satisfies
\begin{align}
    P_{q}\xi_n\ =\ \xi_nP_{q-1}\,,\qquad P_{q}\eta_n\ =\ \eta_n P_{q+1}\,.
\end{align}
The (bosonized representation of) reflector in the small Hilbert space is obtained by projecting it as 
\begin{align}
\ket{r(1,2)}_{\beta\gamma}^{(q)}
=&\  
\left(P_{q}\otimes\eta_0\right)\ket{r(1,2)}_{\eta\xi\phi}\ 
\nonumber\\
=&\ \left(P_{q}\otimes\id\right)\sum_{p}\exp\left(-\sum_{n>0}\phi^{(1)}_{-n}\phi^{(2)}_{-n}
-\sum_{n>0}\xi^{(1)}_{-n}\eta^{(2)}_{-n}+\sum_{n>0}\eta^{(1)}_{-n}\xi^{(2)}_{-n}\right)\ket{p}_1\otimes\ket{-p-2}_2\,.
\end{align}
Projecting (\ref{ref large alt}) onto a fixed picture number, we obtain
\begin{align}
    \left(P_{q}\otimes\id\right)\ket{r(1,2)}_{\eta\xi\phi}\ =&\ -(\xi_0\otimes\id)\ket{r(1,2)}^{(q-1)}_{\beta\gamma}+(\id\otimes\xi_0)\ket{r(1,2)}^{(q)}_{\beta\gamma}\,.
\end{align}
which implies (\ref{vanish})\,.  We can also replace $\xi_0$ with $\Xi$ 
since the difference lies in the small Hilbert space and vanishes under this relation.

The relation $\sum_sXe^1_{(s,-3/2)}\otimes\tilde{e}^1_{(s,-1/2)}=\sum_se^1_{(s,-1/2)}\otimes X\tilde{e}^1_{(s,-3/2)}$\,, which is needed to derive the cyclicity condition (\ref{cyc hetero}) can be written in terms of the reflector as
\begin{align}
    (X\otimes\id)\ket{r(1,2)}^1_{(-3/2)}\ =\ (\id\otimes X)\ket{r(1,2)}^1_{-1/2}\,.\label{X BPZ}
\end{align}
This relation can be derived by using the connection conditions given above and adopting the conventions
\begin{alignat}{2}
    \ket{-1/2}\ =&\ \delta(\beta_0)\ket{-3/2}\,, \qquad  \widetilde{\ket{-1/2}}\ =\ \mp i \delta(\beta_0)\widetilde{\ket{-3/2}}\,,
\end{alignat}
in $\ket{r(1,2)^{\pm}}_{\beta\gamma}^{(-3/2)}$ and $\ket{r(1,2)^{\pm}}_{\beta\gamma}^{(-1/2)}$\,.
The relation (\ref{X BPZ}) guarantees that both outputs of $\Theta\mid^2$ belong to the restricted Hilbert space.
Similarly, the inverse relation
\begin{align}
\ket{-3/2}\ =&\ \delta(\gamma_0)\ket{-1/2}\,, &\qquad 
    \widetilde{\ket{-3/2}}\ =&\ \pm i\delta(\gamma_0)\widetilde{\ket{-1/2}}\,,
\end{align}
yields 
\begin{align}
    (Y\otimes\id)\ket{r(1,2)}^1_{(-1/2)}\ =\ (\id\otimes Y)\ket{r(1,2)}^1_{-3/2}\,.    
\end{align}
Note that in deriving these relations, we have adopted the convention that $\ket{-1/2}$ is Grassmann even and $\ket{-3/2}$ is Grassmann odd. This is because when combined with $\ket{\alpha}$ in $\psi$ component, $\ket{\alpha}\otimes\ket{-1/2}$ must be Grassmann odd. The convention of Ref.~\cite{Hata:1987qx} that we follow is treats $\ket{\alpha}$ as odd.

\sectiono{Derivation of Eq.~(\ref{general cyc})}\label{sec:cyclicity}

We use the bra-ket notation. Before deriving Eq.~\eqref{general cyc}, let us show that
the relation \eqref{eq:second relation} holds if the cyclicity condition \eqref{eq:Quantum cyclic} holds. Using the completeness relation \eqref{eq:id}, we have
\begin{align}
    \sum_{s}\ket{L(\cdots,\tilde{e}_s)}\otimes \ket{e_s}\ =&\ \sum_{s,t}\ket{e_t}\braket{\tilde{e}_t}{L(\cdots,\tilde{e}_s)}\otimes\ket{e_s}
    \nonumber\\
    =&\ -\sum_{s,t}(-1)^{\tilde{e}_t}\ket{e_t}\omega_{\textrm{B}}(\tilde{e}_t,L(\cdots,\tilde{e}_s))\otimes\ket{e_s}.
\end{align}
From the cyclicity condition it further becomes
\begin{align}
    \ -\sum_{s,t}(-1)^{\tilde{e}_s\tilde{e}_t+\tilde{e}_s}\ket{e_t}\otimes\ket{e_s}\omega_{\textrm{B}}(\tilde{e}_s,L(\cdots,\tilde{e}_t))\ =&\ \sum_{s,t}(-1)^{\tilde{e}_s\tilde{e}_t}\ket{e_t}\otimes\ket{e_s}\braket{\tilde{e}_s}{L(\cdots,\tilde{e}_t)}
    \nonumber\\
    =&\ \sum_{t}(-1)^{\tilde{e}_t}\ket{e_t}\otimes\ket{L(\cdots,\tilde{e}_t)}\,.
\end{align}
Hence, if $L$ satisfies the cyclic condition \eqref{eq:Quantum cyclic}, then \eqref{eq:second relation} holds:
\begin{align}
    \sum_s L(\cdots,\tilde{e}_s)\wedge e_s\ =\ 0\,.
\end{align}

We can similarly show Eq.~\eqref{general cyc}. Using the completeness relation in the large Hilbert space \eqref{eq:complete large}, we have
\begin{align}
    \sum_s & P_{q}\pi^a\ket{B_{g,n+1}(\id^{n}\wedge e^b_{(s,-p-2)})}\otimes\ket{\tilde{e}^b_{(s,p)}}\ 
\nonumber\\
=&\
\sum_{s,t}\ket{\tilde{e}^a_{(t,q)}}\langle\langle e^a_{(t,-q-2)}\mid\xi\ket{B_{g,n+1}(\id^{n} \wedge e^b_{(s,-p-2)})}
\otimes\ket{\tilde{e}^b_{(s,p)}}
\nonumber\\
&\
+\sum_{s,t}\xi\ket{\tilde{e}^a_{(t,q-1)}}\langle\langle e^a_{(t,-q-1)}\mid B_{g,n+1}(\id^{n} 
\wedge e^b_{(s,-p-2)})\rangle \otimes\ket{\tilde{e}^b_{(s,p)}}
\nonumber\\
=&\ \sum_{s,t}(-1)^{\tilde{e}^a_{t}}\ket{\tilde{e}^a_{(t,q)}}\omega_l(\xi e^a_{(t,-q-2)}, 
B_{g,n+1}(\id^{n} \wedge e^b_{(s,-p-2)}))\otimes \ket{\tilde{e}^b_{(s,p)}}
\nonumber\\
&\
+\sum_{s,t}\xi\ket{\tilde{e}^a_{(t,q-1)}}\omega_l(e^a_{(t,-q-1)},B_{g,n+1}(\id^{n} 
\wedge e^b_{(s,-p-2)}))\otimes \ket{\tilde{e}^b_{(s,p)}}\,. 
\end{align}
From the cyclicity in the large Hilbert space, it further becomes
\begin{align}
=&\  \sum_{s,t}(-1)^{e^a_t+e^a_t\tilde{e}^b_s}
\ket{\tilde{e}^a_{(t,q)}}\otimes \ket{\tilde{e}^b_{(s,p)}}\omega_l( e^b_{(s,-p-2)}, B_{g,n+1}(\id^{n} \wedge \xi e^a_{(t,-q-2)}))
\nonumber\\
&\
+ \sum_{s,t}(-1)^{\tilde{e}^a_t\tilde{e}^b_s}
\xi\ket{\tilde{e}^a_{(t,q-1)}}\otimes \ket{\tilde{e}^b_{(s,p)}}\omega_l(e^b_{(s,-p-2)},B_{g,n+1}(\id^{n} \wedge e^a_{(t,-q-1)}))
\nonumber\\
=&\ - \sum_{s,t}(-1)^{e^a_te^b_s}
\ket{\tilde{e}^a_{(t,q)}}\otimes \ket{\tilde{e}^b_{(s,p)}}\langle\langle e^b_{(s,-p-2)}|B_{g,n+1}(\id^{n} \wedge \xi e^a_{(t,-q-2)})\rangle
\nonumber\\
&\
- \sum_{s,t}(-1)^{\tilde{e}^a_t\tilde{e}^b_s}
\xi\ket{\tilde{e}^a_{(t,q-1)}}\otimes \ket{\tilde{e}^b_{(s,p)}}\langle\langle e^b_{(s,-p-2)}|B_{g,n+1}(\id^{n} \wedge e^a_{(t,-q-1)})\rangle
\nonumber\\
=&\ - \sum_{t}(-1)^{e^a_t}
\ket{\tilde{e}^a_{(t,q)}}\otimes P_{p}\pi^b\eta\ket{B_{g,n+1}(\id^{n} \wedge \xi e^a_{(t,-q-2)})}
\nonumber\\
&\
- \sum_{t}\xi\ket{\tilde{e}^a_{(t,q-1)}}\otimes P_{p}\pi^b\eta\ket{B_{g,n+1}(\id^{n} \wedge e^a_{(t,-q-1)})}\,.
\label{eqB1}
\end{align}
Further acting $\eta$ to the right, we obtain
\begin{align}
&\    \sum_s\Big(P_q\pi^aB_{g,n+1}(\id^{\wedge n}\wedge e^b_{(s,-p-2)})\otimes \tilde{e}^b_{(s,p)}
    + (-1)^{\tilde{e}^a_s}\tilde{e}^a_{(s,q)}\otimes P_{p}\pi^b B_{g,n+1}(\id^{\wedge n}\wedge e^a_{(s,-q-2)})
    \Big)\nonumber\\
&\ = - \sum_s\Big(\xi\tilde{e}^a_{(s,q-1)}\otimes P_{p}\pi^b[\eta, B](\id^{\wedge n}\wedge e^a_{(s,-q-1)})
\nonumber\\
&\hspace{40mm}
  - (-1)^{\tilde{e}^a_s}\tilde{e}^a_{(s,q)}\otimes P_{p}\pi^b[\eta,B](\id^{\wedge n}\wedge \xi e^a_{(s,-q-2)})
\Big)\nonumber\\
&\ + \sum_s\Big(\xi\tilde{e}^a_{(s,q-1)}\otimes P_{p}\pi^b B(\eta\wedge\id^{\wedge n-1}\wedge e^a_{(s,-q-1)})
\nonumber\\
&\hspace{40mm}
- (-1)^{\tilde{e}^a_s}\tilde{e}^a_{(s,q)}\otimes P_{p}\pi^b B(\eta\wedge \id^{\wedge n-1}\wedge \xi e^a_{(s,-q-2)})\Big)\,.
\label{B1}
\end{align}
Starting from $\sum_s(-1)^{\tilde{e}^b_s}\tilde{e}^b_{(s,p)}\otimes P_q\pi^a B_{g,n+1}(\id^{\wedge n}\wedge e^b_{(s,-p-2)})$, we similarly obtain
\begin{align}
&\   \sum_s\Big( (-1)^{\tilde{e}^b_s}\tilde{e}^b_{(s,p)}\otimes P_{q}\pi^a B_{g,n+1}(\id^{\wedge n}\wedge e^b_{(s,-p-2)})
+ P_p\pi^bB_{g,n+1}(\id^{\wedge n}\wedge e^a_{(s,-q-2)})\otimes \tilde{e}^a_{(s,q)}
    \Big)\nonumber\\
&\ = - \sum_s\Big((-1)^{e^a_s} P_{p}\pi^b[\eta, B](\id^{\wedge n}\wedge e^a_{(s,-q-1)}) \otimes \xi\tilde{e}^a_{(s,q-1)}
\nonumber\\
&\hspace{40mm}
  - P_{p}\pi^b[\eta,B](\id^{\wedge n}\wedge \xi e^a_{(s,-q-2)}) \otimes \tilde{e}^a_{(s,q)}
\Big)\nonumber\\
&\ + \sum_s\Big((-1)^{e^a_s} P_{p}\pi^b B(\eta\wedge\id^{\wedge n-1}\wedge e^a_{(s,-q-1)})\otimes \xi\tilde{e}^a_{(s,q-1)}
\nonumber\\
&\hspace{40mm}
- P_{p}\pi^b B(\eta\wedge \id^{\wedge n-1}\wedge \xi e^a_{(s,-q-2)})\otimes \tilde{e}^a_{(s,q)}\Big)\,.
\label{B2}
\end{align}
Adding (\ref{B1}) and (\ref{B2}), we find
\begin{align}
& \sum_{s}\Big(P_{q}\pi^aB_{g,n+1}(\id^{\wedge n}\wedge e^b_{(s,-p-2)})\wedge \tilde{e}^b_{(s,p)}
+ P_{p}\pi^bB_{g,n+1}(\id^{\wedge n}\wedge e^a_{(s,-q-2)})\wedge\tilde{e}^{a}_{(s,q)}
\Big)
\nonumber\\
  =&\
 \sum_{s}\Big(P_{p}\pi^b[\eta,B_{g,n+1}](\id^{\wedge n}\wedge\Xi e^a_{(s,-q-2)})\wedge\tilde{e}^a_{(s,q)}
\nonumber\\
&\hspace{35mm}
 -(-1)^{e^a_s}P_{p}\pi^b[\eta, B_{g,n+1}](\id^{\wedge n}\wedge e^a_{(s,-q-1)})\wedge\Xi \tilde{e}^a_{(s,q-1)}\Big) 
\nonumber\\
&\ -\sum_{s}\Big(P_{p}\pi^bB_{g,n+1}(\id^{\wedge n-1}\wedge\eta\wedge\Xi e^a_{(s,-q-2)})\wedge\tilde{e}^a_{(s,q)}
\nonumber\\
&\hspace{35mm}
-(-1)^{e^a_s}P_{p}\pi^bB_{g,n+1}(\id^{\wedge n-1}\wedge\eta\wedge e^a_{(s,-q-1)})\wedge\Xi \tilde{e}^a_{(s,q-1)}
\Big)\,.
\end{align}
Since the left-hand side is symmetric under the simultaneous exchange of $(a,q)$ and $(b,p)$, we can symmetrize the right-hand side with respect to this exchange and finally obtain Eq.~\eqref{general cyc}.

We denote this relation as
\begin{align}
\sum_{s}\Big(P_{q}\pi^aB_{g,n+1}(\id^{\wedge n}\wedge e^b_{(s,-p-2)})\wedge \tilde{e}^b_{(s,p)}
+ P_{p}\pi^bB_{g,n+1}(\id^{\wedge n}\wedge e^a_{(s,-q-2)})\wedge\tilde{e}^{a}_{(s,q)}
\Big)\ \approx 0\,.
\end{align}
When $[\eta, B]=0$, $B$ is closed in the small Hilbert space. If we restrict the inputs within the small Hilbert space, the output is also contained within the small Hilbert space. In this case, the weak equality becomes an equality.

\sectiono{Detail calculation about $IBL_\infty$-transfer }\label{sec:IBL}

This Appendix provides the detailed calculations for deriving Eqs.~\eqref{eq:DC transformation} and also \eqref{eq:def of G}.
Since the weak equality holds regardless of the picture numbers of the bases, we omit them below and denote $\{e^0_{(s,-1)},e^1_{(s,-1/2)}\}$ and $\{\tilde{e}^0_{(s,-1)}, \tilde{e}^1_{(s,-3/2)}\}$ 
as $\{e^a_s\}$ and $\{\tilde{e}^a_{s}\}$ for simplicity.
In the calculations in this appendix, the basis $e^a_s$ always appears paired with its dual basis $\tilde{e}^a_s$, so we use the following trick to avoid the complex sign factors that would arise from swapping $ e^a_s$ or $\tilde{e}^a_s$ with other quantities.

Introduce $\epsilon^a_{s,t}$ as a quantity whose degree is equal 
to $e^a_s$ and $\epsilon^a_{s,t}=0$ for $s\neq t$.
If we define $e'^{a}_s$ and $\tilde{e}'^{a}_s$ as
$e^a_s=\epsilon^a_{s,t}e'^{a}_s$ and $\tilde{e}'^{a}_s=\tilde{e}^a_s\epsilon^a_{s,t}$, then
$e'^a_s$ and $\tilde{e}'^a_s$ can be made degree even and degree odd, respectively, regardless of $s$,
while preserving the form of their wedge product $\tilde{e}^a_s\wedge e^a_s$:
\begin{align}
\tilde{e}^a_s\wedge e^a_s=\tilde{e}^a_s\wedge \epsilon^a_{s,t} e'^a_t
=\tilde{e}^a_s\epsilon^a_{s,t}\wedge e'^a_t=\tilde{e}'^a_t\wedge e'^a_t.
\end{align}
This trick allows us to make the degree of all the $\bd{D}(e^a_s)$ and $\Xi \tilde{e}^1_s$ (the bases appear almost always in this form below) even, eliminating the need to worry about the sign factor, and allowing us to recover it whenever needed. For simplicity, we will henceforth denote $e'^a_s$ and $\tilde{e}'^a_s$ as $e^a_s$ and $\tilde{e}^a_s$, respectively.

Before going into the proofs, we provide here some useful formulas.
From the definition of $\bd{\mathcal{F}}^{-1}$ in \eqref{eq: def of mathcalF^-1},  the relations 
\begin{align}
    \bd{\mathcal{F}}^{-1}\bd{B}\ =&\ \big(\bd{\mathcal{F}}^{-1}\wedge \bd{F}^{-1}\bd{B}\big)\circ\Delta_2\label{eq:formula FB}\\
    \bd{\mathcal{F}}^{-1}\bd{\theta}\ =&\ \big(\bd{\mathcal{F}}^{-1}\wedge \bd{F}^{-1}\bd{\theta}\big)\circ\Delta_2+\frac{1}{2}\sum_{a,s}
\big(\bd{\mathcal{F}}^{-1}\wedge\bd{F}^{-1}\bd{D}(\tilde{e}^a_s)\wedge \bd{F}^{-1}\bd{D}(e^a_s)\big)\circ\Delta_3\,,\\
\bd{B}\bd{\mathcal{F}}^{-1}\ \approx&\
    \sum_{k=0}^\infty \frac{(-\hbar)^k}{k!}\sum_{s_1,\cdots,s_k} \bd{\mathcal{F}}^{-1}\wedge \Xi \tilde{e}^1_{s_k}\wedge
\dots\wedge\Xi \tilde{e}^1_{s_1}\wedge \bd{B}\bd{\mathcal{F}}^{-1}\bd{D}(e^1_{s_1})\dots\bd{D}(e^1_{s_k})\nonumber\\
\approx&\
    \sum_{k=0}^\infty \frac{(-\hbar)^k}{k!}\sum_{s_1,\cdots,s_k} \bd{\mathcal{F}}^{-1}\wedge e^1_{s_k}\wedge
\dots\wedge e^1_{s_1}\wedge \bd{B}\bd{\mathcal{F}}^{-1}\bd{D}(\Xi\tilde{e}^1_{s_1})\dots\bd{D}(\Xi\tilde{e}^1_{s_k})\,,\label{D4}\\
\bd{\theta}\bd{\mathcal{F}}^{-1}\ =&\ \bd{\mathcal{F}}^{-1}\wedge\bd{\theta}\,.
\label{eq:formula Ftheta}
\end{align}
hold for order-one coderivation $\bd{B}$ and order-two coderivation $\bd{\theta}$.
We used (weak) relations (\ref{exchange e and e-tilde}), (\ref{move Xi}), and (\ref{weak cyc})
 for deriving Eq.(\ref{D4}).\footnote{
In addition to assistance from ChatGPT, we independently proved this formula.}


\subsection{A derivation of \eqref{eq:L def}}

We show \eqref{eq:DC transformation} first.
Using the formulas \eqref{eq:formula FB}, \eqref{eq:formula Ftheta} and \eqref{D4}, 
we can show that
\begin{align}
    \bd{\mathcal{F}}^{-1}(\bd{\mathcal{D}}+\hbar\bd{\theta}\mid^0)
-(\bd{Q}+\bd{(\mathcal{G}B\mathcal{F})}+\hbar\bd{(\mathcal{G}\theta)})\bd{\mathcal{F}}^{-1}\ =\ \bd{\mathcal{F}}^{-1}\wedge \mathcal{O}
\end{align}
with
\begin{align}
  \mathcal{O}\ =&\ \bd{F}^{-1}(\bd{\mathcal{D}}+\hbar\bd{\theta}\mid^0)
+\frac{\hbar}{2}\sum_s\bd{F}^{-1}\bd{D}(\tilde{e}^0_s)\wedge\bd{F}^{-1}\bd{D}(e^0_s)-\bd{Q}\bd{F}^{-1}\nonumber\\
&\
-\sum_{n=0}^\infty\frac{(-\hbar)^n}{n!} \sum_{s_1,\cdots,s_k}\Xi\tilde{e}^1_{s_1}\wedge\dots\wedge\Xi\tilde{e}^1_{s_k}
\wedge\mathcal{G}\pi_1\bd{B\mathcal{F}}\bd{D}(e^1_{s_k})\dots\bd{D}(e^1_{s_1})-\frac{\hbar}{2}(\mathcal{G}\wedge \id)\theta\,.
\end{align}
From $\bd{\mathcal{D}}=\bd{Q+(\pi^0B)}$ and (\ref{def Theta}), $\mathcal{O}$ can be written as
\begin{align}\label{eq:inside of bracket}
 \mathcal{O}\  =&\ -[\bd{Q},\bd{F}^{-1}]+\bd{F}^{-1}(\bd{(\pi^0B)}+\hbar\bd{\theta}\mid^0)
+\frac{\hbar}{2}\sum_{s}\bd{F}^{-1}\bd{D}(\tilde{e}^0_s)\wedge\bd{F}^{-1}\bd{D}(e^0_s)\nonumber\\
    &-\sum_{n=0}^\infty \frac{(-\hbar)^n}{n!}\sum_{s_1,\cdots,s_n}\Xi\tilde{e}^1_{s_1}\wedge\dots\wedge
\Xi\tilde{e}^1_{s_n}\wedge\mathcal{G}\pi_1\bd{B\mathcal{F}}\bd{D}(e^1_{s_n})\dots\bd{D}(e^1_{s_1})
-\hbar\Theta.
\end{align}
We calculate each term in the first line and show that their sum with the second line eventually becomes zero.

The first term in the first line can be calculated as follows:
\begin{align}\label{eq:-[Q,F-1]}
    -[\bd{Q},\bd{F}^{-1}]&=-\sum_{k=0}^\infty \frac{(-\hbar)^k}{(k+1)!}\sum_{s_1,\cdots,s_k}[\bd{Q},\Xi\tilde{e}^1_{s_1}
\wedge\dots\wedge\Xi\tilde{e}^1_{s_k}\wedge(\pi_1\id-\Xi\pi_1^1\bd{B})\bd{D}(e^1_{s_k})\dots\bd{D}(e^1_{s_1})]\nonumber\\
   &=\frac{\hbar}{2}\sum_{s}X\tilde{e}^1_s\wedge e^1_s+\sum_{k=0}^\infty\frac{(-\hbar)^k}{(k+1)!}\sum_{s_1,\cdots,s_k}[\bd{Q},\Xi\tilde{e}^1_{s_1}
\wedge\dots\wedge\Xi\tilde{e}^1_{s_k}\wedge\Xi\pi_1^1\bd{B}\bd{D}(e^1_{s_k})\dots\bd{D}(e^1_{s_1})]\nonumber\\
    &=\hbar\Theta\mid^2+\sum_{k=0}^\infty\frac{k(-\hbar)^k}{(k+1)!}\sum_{s_1,\cdots,s_k}Q\Xi\tilde{e}^1_{s_1}\wedge\Xi\tilde{e}^1_{s_2}
\wedge\dots\wedge\Xi\tilde{e}^1_{s_k}\wedge\Xi\pi_1^1\bd{B}\bd{D}(e^1_{s_k})\dots\bd{D}(e^1_{s_1})\nonumber\\
    &\quad+\sum_{k=0}^\infty\frac{(-\hbar)^k}{(k+1)!}\sum_{s_1,\cdots,s_k}\Xi\tilde{e}^1_{s_1}\wedge\dots\wedge\Xi\tilde{e}^1_{s_k}
\wedge Q\Xi\pi_1^1\bd{B}\bd{D}(e^1_{s_k})\dots\bd{D}(e^1_{s_1})\nonumber\\
    &\quad-\sum_{k=0}^\infty\frac{(-\hbar)^k}{(k+1)!}\sum_{s_1,\cdots,s_k}\Xi\tilde{e}^1_{s_1}\wedge\dots\wedge\Xi\tilde{e}^1_{s_k}
\wedge\Xi\pi_1^1\bd{B}\bd{D}(e^1_{s_k})\dots\bd{D}(e^1_{s_1})\bd{Q}.
\end{align}
The last term is further calculated as
\begin{align}
     &\quad  -\sum_{k=0}^\infty\frac{(-\hbar)^k}{(k+1)!}\sum_{s_1,\cdots,s_k}\Xi\tilde{e}^1_{s_1}\wedge\dots\wedge\Xi\tilde{e}^1_{s_k}
\wedge\Xi\pi_1^1\bd{B}\bd{D}(e^1_{s_k})\dots\bd{D}(e^1_{s_1})\bd{Q}\nonumber\\
    &=-\sum_{k=0}^\infty\frac{(-\hbar)^k}{(k+1)!}\sum_{s_1,\cdots,s_k}\Xi\tilde{e}^1_{s_1}\wedge\dots\wedge\Xi\tilde{e}^1_{s_k}
\wedge\Xi\pi_1^1\bd{B}\bd{Q}\bd{D}(e^1_{s_k})\dots\bd{D}(e^1_{s_1})\nonumber\\
    &\quad+\sum_{k=0}^\infty\frac{k(-\hbar)^k}{(k+1)!}\sum_{s_1,\cdots,s_k}\Xi\tilde{e}^1_{s_1}\wedge\dots\wedge\Xi\tilde{e}^1_{s_k}
\wedge\Xi\pi_1^1\bd{B}\bd{D}(e^1_{s_k})\dots\bd{D}(Q e^1_{s_1})\nonumber\\
    &=-\sum_{k=0}^\infty\frac{(-\hbar)^k}{(k+1)!}\sum_{s_1,\cdots,s_k}\Xi\tilde{e}^1_{s_1}\wedge\dots\wedge\Xi\tilde{e}^1_{s_k}
\wedge\Xi\pi_1^1\bd{B}\bd{Q}\bd{D}(e^1_{s_k})\dots\bd{D}(e^1_{s_1})\nonumber\\
    &\quad+\sum_{k=0}^\infty\frac{k(-\hbar)^k}{(k+1)!}\sum_{s_1,\cdots,s_k}\Xi Q\tilde{e}^1_{s_1}\wedge\dots\wedge\Xi\tilde{e}^1_{s_k}
\wedge\Xi\pi_1^1\bd{B}\bd{D}(e^1_{s_k})\dots\bd{D}(e^1_{s_1}),
\end{align}
using
\begin{align}
 (Q\otimes\id +\id \otimes Q)\sum_s\tilde{e}^1_s\otimes e^1_s\ =\ 0.
\end{align}
in the second equality.
By substituting it into \eqref{eq:-[Q,F-1]}, we have
\begin{align}
     -[\bd{Q},\bd{F}^{-1}]\ 
&=\ \hbar\Theta\mid^2
+\sum_{k=0}^\infty\frac{k(-\hbar)^k}{(k+1)!}\sum_{s_1,\cdots,s_k}X\tilde{e}^1_{s_1}\wedge\Xi\tilde{e}^1_{s_2}
\wedge\dots\wedge\Xi\tilde{e}^1_{s_k}\wedge\Xi\pi_1^1\bd{B}\bd{D}(e^1_{s_k})\dots\bd{D}(e^1_{s_1})\nonumber\\
     &\ +\sum_{k=0}^\infty\frac{(-\hbar)^k}{(k+1)!}\sum_{s_1,\cdots,s_k}\Xi\tilde{e}^1_{s_1}\wedge\dots
\wedge\Xi\tilde{e}^1_{s_k}\wedge\pi_1[\bd{Q},\bd{(\Xi\pi^1B)}]\bd{D}(e^1_{s_k})\dots\bd{D}(e^1_{s_1})\nonumber\\
     &=\hbar\Theta\mid^2+\sum_{k=0}^\infty\frac{k(-\hbar)^k}{(k+1)!}\sum_{s_1,\cdots,s_k}X\tilde{e}^1_{s_1}\wedge\Xi\tilde{e}^1_{s_2}
\wedge\dots\wedge\Xi\tilde{e}^1_{s_k}\wedge\Xi\pi_1^1\bd{B}\bd{D}(e^1_{s_k})\dots\bd{D}(e^1_{s_1})\nonumber\\
     &\ +\sum_{k=0}^\infty\frac{(-\hbar)^k}{(k+1)!}\sum_{s_1,\cdots,s_k}\Xi\tilde{e}^1_{s_1}\wedge\dots\wedge\Xi\tilde{e}^1_{s_k}
\wedge X\pi_1^1\bd{B}\bd{D}(e^1_{s_k})\dots\bd{D}(e^1_{s_1})\nonumber\\
     &\ -\sum_{k=0}^\infty\frac{(-\hbar)^k}{(k+1)!}\sum_{s_1,\cdots,s_k}\Xi\tilde{e}^1_{s_1}\wedge\dots\wedge\Xi\tilde{e}^1_{s_k}
\wedge\Xi\pi_1[\bd{Q},\bd{(\pi^1B)}]\bd{D}(e^1_{s_k})\dots\bd{D}(e^1_{s_1}),
\label{QF1}
\end{align}
where, $\bd{(\Xi\pi^1B)}$ denotes the coderivation derived from $\Xi\pi^1B$.

Using the relation
\begin{align}
    \pi_1[\bd{Q},\bd{(\pi^1B)}]\ =\ -\pi_1^1\bd{B}((\bd{\pi^0B})+\hbar\bd{\theta}\mid^0)\,.
\end{align}
derived from $\pi_1[\bd{\mathcal{C}}+\hbar\bd{\theta}\mid^0,\bd{\mathcal{D}}-\hbar\bd{\theta}\mid^2]=0$\,,
we can further rewrite 
\begin{align}\label{eq:QF first term}
     -[\bd{Q},\bd{F}^{-1}]\ 
    &=\ \hbar\Theta\mid^2+X\pi_1^1\bd{B}\nonumber\\
    &\ + \sum_{k=1}^\infty\frac{(-\hbar)^{k}}{k!}\sum_{s_1,\cdots,s_k}\Xi\tilde{e}^1_{s_1}
\wedge\dots\wedge\Xi\tilde{e}^1_{s_k}\wedge X\pi_1^1\bd{B}\bd{D}(e^1_{s_k})\dots\bd{D}(e^1_{s_1})\nonumber\\
    &\ +\sum_{k=0}^\infty\frac{(-\hbar)^k}{(k+1)!}\sum_{s_1,\cdots,s_k}\Xi\tilde{e}^1_{s_1}
\wedge\dots\wedge\Xi\tilde{e}^1_{s_k}\wedge \Xi\pi_1^1\bd{B}(\bd{(\pi^0B)}
+\hbar\bd{\theta}\mid^0)\bd{D}(e^1_{s_k})\dots\bd{D}(e^1_{s_1})
\nonumber\\
&\ +\sum_{k=1}^{\infty}\frac{k(-\hbar)^{k}}{(k+1)!}\sum_{s_1,\cdots,s_k}X\tilde{e}^1_{s_1}\wedge\Xi\tilde{e}^1_{s_2}\wedge\cdots\wedge\Xi\tilde{e}^1_{s_k}\wedge\Xi\pi_1^1\bd{B}\bd{D}(e^1_{s_k})\cdots\bd{D}(e^1_{s_1})
\nonumber\\
&\ -\sum_{k=1}^{\infty}\frac{k(-\hbar)^{k}}{(k+1)!}\sum_{s_1,\cdots,s_k}\Xi\tilde{e}^1_{s_1}\wedge\Xi\tilde{e}^1_{s_2}\wedge \cdots\wedge\Xi\tilde{e}^1_{s_k}\wedge X\pi_1^1\bd{B}\bd{D}(e^1_{s_k})\cdots\bd{D}(e^1_{s_1})\,.
\end{align}
However, using (\ref{weak cyc}), the last two terms cancel in the sense of weak equality.
Therefore, we have
\begin{align}\label{eq:QF first term 2}
    -&[\bd{Q},\bd{F}^{-1}] \nonumber\\
  &\approx\ \hbar\Theta\mid^2 
 + \sum_{k=0}^\infty\frac{(-\hbar)^{k}}{k!}\sum_{s_1,\cdots,s_k}\Xi\tilde{e}^1_{s_1}
\wedge\dots\wedge\Xi\tilde{e}^1_{s_k}\wedge X\pi_1^1\bd{B}\bd{D}(e^1_{s_k})\dots\bd{D}(e^1_{s_1})\nonumber\\
    &\ +\sum_{k=0}^\infty\frac{(-\hbar)^k}{(k+1)!}\sum_{s_1,\cdots,s_k}\Xi\tilde{e}^1_{s_1}
\wedge\dots\wedge\Xi\tilde{e}^1_{s_k}\wedge \Xi\pi_1^1\bd{B}(\bd{(\pi^0B)}
+\hbar\bd{\theta}\mid^0)\bd{D}(e^1_{s_k})\dots\bd{D}(e^1_{s_1}).
\end{align}
Note that the terms that vanish only in a weak sense do not contain the $\pi_1$ component.

The second term in the first line of \eqref{eq:inside of bracket} is calculated as
\begin{align}\label{eq:QF second term}
    &\bd{F}^{-1}(\bd{(\pi^0B)}+\hbar\bd{\theta}\mid^0)\nonumber\\
    &=\pi_1^0\bd{B}-\sum_{k=0}^\infty\frac{(-\hbar)^k}{(k+1)!}\sum_{s_1,\cdots,s_k}\Xi\tilde{e}^1_{s_1}
\wedge\dots\wedge\Xi\tilde{e}^1_{s_k}\wedge \Xi\pi_1^1\bd{B}\bd{D}(e^1_{s_k})
\dots\bd{D}(e^1_{s_1})(\bd{(\pi^0B)}+\hbar\bd{\theta}\mid^0)\nonumber\\
    &=\pi_1^0\bd{B}-\sum_{k=0}^\infty\frac{(-\hbar)^k}{(k+1)!}\sum_{s_1,\cdots,s_k}\Xi\tilde{e}^1_{s_1}
\wedge\dots\wedge\Xi\tilde{e}^1_{s_k}\wedge \Xi\pi_1^1\bd{B}\bd{D}(e^1_{s_k})\dots\bd{D}(e^1_{s_1})\bd{(\pi^0B)}\nonumber\\
    &\quad+\sum_{k=0}^\infty\frac{(-\hbar)^{k+1}}{(k+1)!}\sum_{s_1,\cdots,s_k}\Xi\tilde{e}^1_{s_1}
\wedge\dots\wedge\Xi\tilde{e}^1_{s_k}\wedge \Xi\pi_1^1\bd{B}\bd{\theta}\mid^0\bd{D}(e^1_{s_k})\dots\bd{D}(e^1_{s_1})\,.
\end{align}

We can calculate the third term of the first line of \eqref{eq:inside of bracket} as
\begin{align}\label{eq:QF third term}
    &\frac{\hbar}{2}\sum_{t}\bd{F}^{-1}\bd{D}(\tilde{e}^0_{t})\wedge\bd{F}^{-1}\bd{D}(e^0_{t})\nonumber\\
    &\ =\ \frac{\hbar}{2}\sum_{t}
[\tilde{e}^0_{t}-\sum_{k=0}^\infty \frac{(-\hbar)^k}{(k+1)!}\sum_{s_1,\cdots,s_k}
\Xi\tilde{e}^1_{s_1}\wedge\dots\wedge\Xi\tilde{e}^1_{s_k}\wedge\Xi\pi_1^1\bd{B}\bd{D}(e^1_{s_k})\dots\bd{D}(e^1_{s_1})
\bd{D}(\tilde{e}^0_{t})]\nonumber\\
    &\quad\wedge[e^0_{t}-\sum_{k=0}^\infty \frac{(-\hbar)^k}{(k+1)!}\sum_{s_1,\cdots,s_k}
\Xi\tilde{e}^1_{s_1}\wedge\dots\wedge\Xi\tilde{e}^1_{s_k}
\wedge\Xi\pi_1^1\bd{B}\bd{D}(e^1_{s_k})\dots\bd{D}(e^1_{s_1})\bd{D}(e^0_{t})]\nonumber\\
    &\ =\ \hbar\Theta\mid^0
    \nonumber\\
    &\quad    + \sum_{k=0}^\infty \frac{(-\hbar)^{k+1}}{(k+1)!}\sum_{t,s_1,\cdots,s_k}
\tilde{e}^0_{t}\wedge\Xi\tilde{e}^1_{s_1}\wedge\dots\wedge\Xi\tilde{e}^1_{s_k}\wedge\Xi\pi_1^1\bd{B}\bd{D}(e^1_{s_k})\dots\bd{D}(e^1_{s_1})
\bd{D}(e^0_t)\nonumber\\
    &\quad+\frac{\hbar}{2}\sum_s\left[\sum_{k=0}^\infty \frac{(-\hbar)^k}{(k+1)!}\sum_{s_1,\cdots,s_k}
\Xi\tilde{e}^1_{s_1}\wedge\dots\wedge\Xi\tilde{e}^1_{s_k}\wedge\Xi\pi_1^1\bd{B}\bd{D}(e^1_{s_k})\dots\bd{D}(e^1_{s_1})
\bd{D}(\tilde{e}^0_s)\right]\nonumber\\
    &\qquad\wedge\left[\sum_{l=0}^\infty \frac{(-\hbar)^l}{(l+1)!}\sum_{s_1,\cdots,s_l}
\Xi\tilde{e}^1_{s_1}\wedge\dots\wedge\Xi\tilde{e}^1_{s_l}
\wedge\Xi\pi_1^1\bd{B}\bd{D}(e^1_{s_l})\dots\bd{D}(e^1_{s_1})\bd{D}(e^0_s)\right].
\end{align}
Here, from Eq.~(\ref{weak cyc}), we find
\begin{align}
    \sum_s\tilde{e}^0_s\wedge \Xi\pi_1^1\bd{B}\bd{D}(e^0_s)\approx\sum_s\Xi \tilde{e}^1_s\wedge\pi_1^0\bd{B}\bd{D}(e^1_s),
\end{align}
which enables us to rewrite \eqref{eq:QF third term} as
\begin{align}\label{eq:QF third term 2}
    &\frac{\hbar}{2}\sum_s\bd{F}^{-1}\bd{D}(\tilde{e}^0_s)\wedge\bd{F}^{-1}\bd{D}(e^0_s)\nonumber\\
    &\approx\ \hbar\Theta\mid^0 + 
    \sum_{k=1}^\infty \frac{(-\hbar)^{k}}{k!}
\sum_{s_1,\cdots,s_k}
\Xi\tilde{e}^1_{s_1}\wedge\dots\wedge\Xi\tilde{e}^1_{s_k}\wedge\pi_1^0\bd{B}\bd{D}(e^1_{s_k})
\dots\bd{D}(e^1_{s_1}) 
\nonumber\\
    &+\frac{\hbar}{2}\sum_s\left[\sum_{k=0}^\infty \frac{(-\hbar)^k}{(k+1)!}\sum_{s_1,\cdots,s_k}
\Xi\tilde{e}^1_{s_1}\wedge\dots\wedge
\Xi\tilde{e}^1_{s_k}\wedge\Xi\pi_1^1\bd{B}\bd{D}(e^1_{s_k})\dots\bd{D}(e^1_{s_1})\bd{D}(\tilde{e}^0_s)\right]\nonumber\\
    &\hspace{1.1cm}\wedge\left[\sum_{l=0}^\infty \frac{(-\hbar)^l}{(l+1)!}\sum_{s_1,\cdots,s_l}
\Xi\tilde{e}^1_{s_1}\wedge\dots\wedge\Xi\tilde{e}^1_{s_l}
\wedge\Xi\pi_1^1\bd{B}\bd{D}(e^1_{s_l})\dots\bd{D}(e^1_{s_1})\bd{D}(e^0_s)\right].
\end{align}

Summing up \eqref{eq:QF first term 2}, \eqref{eq:QF second term} and \eqref{eq:QF third term 2}, 
then $\mathcal{O}$ becomes
\begin{align}\label{eq: together }
\mathcal{O}\ \approx&\ \sum_{k=0}^\infty\frac{(-\hbar)^k}{(k+1)!}\sum_{s_1,\cdots,s_k}\Xi\tilde{e}^1_{s_1}\wedge\dots\wedge\Xi\tilde{e}^1_{s_k}\wedge 
\Xi\pi_1^1\bd{B}\bd{(\pi^0B)}\bd{D}(e^1_{s_k})\dots\bd{D}(e^1_{s_1})\nonumber\\
    &-\sum_{k=0}^\infty\frac{(-\hbar)^k}{(k+1)!}\sum_{s_1,\cdots,s_k}\Xi\tilde{e}^1_{s_1}\wedge\dots\wedge\Xi\tilde{e}^1_{s_k}\wedge 
\Xi\pi_1^1\bd{B}\bd{D}(e^1_{s_k})\dots\bd{D}(e^1_{s_1})\bd{(\pi^0B)}\nonumber\\
    &+\frac{\hbar}{2}\sum_s\left[\sum_{k=0}^\infty \frac{(-\hbar)^k}{(k+1)!}\sum_{s_1,\cdots,s_k}\Xi\tilde{e}^1_{s_1}\wedge\dots
\wedge\Xi\tilde{e}^1_{s_k}\wedge\Xi\pi_1^1\bd{B}\bd{D}(e^1_{s_k})\dots\bd{D}(e^1_{s_1})\bd{D}(\tilde{e}^0_s)\right]\nonumber\\
    &\hspace{1.1cm}\wedge\left[\sum_{l=0}^\infty \frac{(-\hbar)^l}{(l+1)!}\sum_{s_1,\cdots,s_l}
\Xi\tilde{e}^1_{s_1}\wedge\dots\wedge\Xi\tilde{e}^1_{s_l}
\wedge\Xi\pi_1^1\bd{B}\bd{D}(e^1_{s_l})\dots\bd{D}(e^1_{s_1})\bd{D}(e^0_s)\right].
\end{align}
The sum of the first two terms can further calculated as
\begin{align}
&\  \sum_{n=1}^\infty \sum_{k=0}^{n-1}\frac{(-\hbar)^n}{(n+1)!}\frac{n!}{k!(n-k)!}
\nonumber\\
   &\hspace{1cm} \times\sum_{s,s_1,\cdots,s_{n-1}}
\Xi\tilde{e}^1_{s_1}\wedge\dots \wedge\Xi\tilde{e}^1_{s_{n-1}}
\wedge\Xi\pi_1^1\bd{B}\bd{D}(e^1_{s_{n-1}})\dots\bd{D}(e^1_{s_{n-k}})
\bd{D}(\tilde{e}^0_s)\nonumber\\
 &\hspace{6.5cm} \wedge\Xi\pi_1^1\bd{B}\bd{D}(e^1_{s_{n-k-1}})\dots\bd{D}(e^1_{s_{1}})\bd{D}(e^0_s),\label{eq:intermediate}
\end{align}
using the weak relation
\begin{align}
    \sum_s\Xi\tilde{e}^1_s\wedge \Xi\pi^1_1\bd{B}\bd{(\pi^0B)}\bd{D}(e^1_s)\approx
\sum_s\Xi\pi_1^1\bd{B}\bd{D}(\tilde{e}^0_s)\wedge\Xi\pi_1^1\bd{B}\bd{D}(e^0_s),
\end{align}
follows from (\ref{weak cyc}). 
If we rewrite $n-k$ as $l+1$, \eqref{eq:intermediate} becomes
\begin{align}
 &\   -\hbar\sum_{k=0}^\infty\sum_{l=0}^\infty\frac{k+1}{k+l+2}\frac{(-\hbar)^{k+l}}{(k+1)!(l+1)!}
\nonumber\\
&\hspace{2cm} \times\sum_{s,s_1,\cdots, s_{k+l}}\Xi\tilde{e}^1_{s_1}
\wedge\dots \wedge\Xi\tilde{e}^1_{s_{k+l}}\wedge\Xi\pi_1^1\bd{B}\bd{D}(e^1_{s_{k+l}})\dots\bd{D}(e^1_{s_{l+1}})\bd{D}(\tilde{e}^0_s)\nonumber\\
     &\hspace{7.5cm}\wedge\Xi\pi_1^1\bd{B}\bd{D}(e^1_{s_{l}})\dots\bd{D}(e^1_{s_{1}})\bd{D}(e^0_s).
\end{align}
Taking into account the symmetry
\begin{align}
 \sum_s\tilde{e}^0_s\otimes e^0_s\ =\ \sum_s e^0_s\otimes\tilde{e}^0_s,
\end{align}
we can symmetrize it with respect to $k$ and $l$, and obtain
\begin{align}
&-\frac{\hbar}{2}\sum_s\left[\sum_{k=0}^\infty \frac{(-\hbar)^k}{(k+1)!}\sum_{s_1,\cdots,s_k}
\Xi\tilde{e}^1_{s_1}\wedge\dots\wedge\Xi\tilde{e}^1_{s_k}\wedge\Xi\pi_1^1\bd{B}\bd{D}(e^1_{s_k})\dots\bd{D}(e^1_{s_1})
\bd{D}(\tilde{e}^0_s)\right]\nonumber\\
    &\hspace{1.1cm}\wedge\left[\sum_{l=0}^\infty \frac{(-\hbar)^l}{(l+1)!}\sum_{s_1,\cdots,s_l}
\Xi\tilde{e}^1_{s_1}\wedge\dots\wedge\Xi\tilde{e}^1_{s_l}
\wedge\Xi\pi_1^1\bd{B}\bd{D}(e^1_{s_l})\dots\bd{D}(e^1_{s_1})\bd{D}(e^0_s)\right]\,.
\end{align}
This cancels out the last term of \eqref{eq: together }, and thus $\mathcal{O}\approx0$,
which yields
\begin{align}
    \bd{\mathcal{F}}^{-1}(\bd{\mathcal{D}}+\hbar\bd{\theta}\mid^0)
-(\bd{Q}+\bd{(\mathcal{G}B\mathcal{F})}+\hbar\bd{(\mathcal{G}\theta)})\bd{\mathcal{F}}^{-1}\approx0.
\end{align}
Multiplying $\bd{\mathcal{F}}$ from right, we obtain \eqref{eq:L def}.

\subsection{A derivation of \eqref{eq:eta F}}
 Consider
\begin{align}
\bd{\mathcal{F}}^{-1}(\bd{C}-\hbar\bd{\theta}\mid^2)-\bd{\eta}\bd{\mathcal{F}}^{-1}\ =\ \bd{\mathcal{F}}^{-1}\wedge\mathcal{O}.
\end{align}
Using \eqref{eq:formula FB} and \eqref{eq:formula Ftheta}, we have 
\begin{align}
   \mathcal{O}\ =&\ \bd{F}^{-1}\bd{C}-\hbar\bd{F}^{-1}\bd{\theta}\mid^2
-\frac{\hbar}{2}\sum_s\bd{F}^{-1}\bd{D}(\tilde{e}^1_s)\wedge\bd{F}^{-1}\bd{D}(e^1_s)-\bd{\eta}\bd{F}^{-1}
\nonumber\\
=&\     -[\bd{\eta},\bd{F}^{-1}]-\bd{F}^{-1}(\bd{\pi^1B}+\hbar\bd{\theta}\mid^2)
-\frac{\hbar}{2}\sum_s\bd{F}^{-1}\bd{D}(\tilde{e}^1_s)\wedge\bd{F}^{-1}\bd{D}(e^1_s).
\label{eq: inside of bracket}
\end{align}
As in the previous subsection, the first term can be calculated as
\begin{align}
-&[\bd{\eta},\bd{F}^{-1}]\nonumber\\ 
     &=\ \hbar\theta\mid^2 + \pi_1^1\bd{B}\nonumber\\
     &\ + \sum_{k=0}^\infty\frac{(-\hbar)^{k+1}}{(k+1)!}\sum_{s,s_1,\cdots,s_k}\Xi\tilde{e}^1_s\wedge\Xi\tilde{e}^1_{s_1}
\wedge\dots\wedge\Xi\tilde{e}^1_{s_k}\wedge\pi_1^1\bd{B}\bd{D}(e^1_{s_k})\dots\bd{D}(e^1_{s_1})\bd{D}(e^1_s)\nonumber\\     
     &\ -\sum_{k=0}^\infty\frac{(-\hbar)^k}{(k+1)!}\sum_{s_1,\cdots,s_k}\Xi\tilde{e}^1_{s_1}\wedge\dots\wedge\Xi\tilde{e}^1_{s_k}
\wedge\Xi\pi_1^1\bd{B}(\bd{(\pi^1B)}+\hbar\bd{\theta}\mid^2)\bd{D}(e^1_{s_k})\dots\bd{D}(e^1_{s_1}),
     \end{align}
using
\begin{align}
 (\id\otimes\eta+\eta\otimes\id)\sum_s\tilde{e}^1_s\otimes e^1_s\ =\ 0
\end{align}
and
     \begin{align}
         \pi_1[\bd{\eta},\bd{(\pi^1B)}]\ =\ \pi_1^1\bd{B}(\bd{(\pi^1B)}+\hbar\bd{\theta}\mid^2),\\
         \sum_s\Xi\tilde{e}^1_s\wedge\pi_1^1\bd{B}\bd{D}(e^1_s)\approx\sum_s\tilde{e}^1_s\wedge\Xi\pi_1^1\bd{B}\bd{D}(e^1_s).
     \end{align}
The second term of $\mathcal{O}$ becomes 
\begin{align}
    -&\bd{F}^{-1}(\bd{(\pi^1B)}+\hbar\bd{\theta}\mid^2)\nonumber\\
    &=-\pi_1^1\bd{B}
    +\sum_{k=0}^\infty\frac{(-\hbar)^k}{(k+1)!}\sum_{s_1,\cdots,s_k}\Xi\tilde{e}^1_{s_1}\wedge\dots\wedge\Xi\tilde{e}^1_{s_k}
\wedge\Xi\pi_1^1\bd{B}\bd{D}(e^1_{s_k})\dots\bd{D}(e^1_{s_1})\bd{(\pi^1B)}\nonumber\\
    &\quad+\sum_{k=0}^\infty\frac{(-\hbar)^{k+1}}{(k+1)!}\sum_{s_1,\cdots,s_k}\Xi\tilde{e}^1_{s_1}\wedge\dots\wedge\Xi\tilde{e}^1_{s_k}
\wedge\Xi\pi_1^1\bd{B}\bd{\theta}\mid^2\bd{D}(e^1_{s_k})\dots\bd{D}(e^1_{s_1}).
\end{align}
The third term of $\mathcal{O}$ 
can be written as
\begin{align}
    -&\frac{\hbar}{2}\sum_s\bd{F}^{-1}\bd{D}(\tilde{e}^1_s)\wedge\bd{F}^{-1}\bd{D}(e^1_s)\nonumber\\
    &=-\hbar\theta\mid^2
    -\sum_{k=0}^\infty \frac{(-\hbar)^{k+1}}{(k+1)!}\sum_{s,s_1,\cdots,s_k}
\Xi\tilde{e}^1_s\wedge\Xi\tilde{e}^1_{s_1}
\wedge\dots\wedge\Xi\tilde{e}^1_{s_k}\wedge\pi_1^1\bd{B}\bd{D}(e^1_{s_k})\dots\bd{D}(e^1_{s_1})\bd{D}(e^1_s)\nonumber\\
    &\quad-\frac{\hbar}{2}\sum_s\left[\sum_{k=0}^\infty \frac{(-\hbar)^k}{(k+1)!}\sum_{s_1,\cdots,s_k}
\Xi\tilde{e}^1_{s_1}\wedge\dots\wedge\Xi\tilde{e}^1_{s_k}
\wedge\Xi\pi_1^1\bd{B}\bd{D}(e^1_{s_k})\dots\bd{D}(e^1_{s_1})\bd{D}(\tilde{e}^1_s)\right]\nonumber\\
    &\hspace{1.5cm}\wedge\left[\sum_{l=0}^\infty \frac{(-\hbar)^l}{(l+1)!}\sum_{s_1,\cdots,s_l}
\Xi\tilde{e}^1_{s_1}\wedge\dots\wedge\Xi\tilde{e}^1_{s_l}
\wedge\Xi\pi_1^1\bd{B}\bd{D}(e^1_{s_l})\dots\bd{D}(e^1_{s_1})\bd{D}(e^1_s)\right].
\end{align}
By summing up the results, we have
\begin{align}
\mathcal{O}\ =&\ -\sum_{k=0}^\infty\frac{(-\hbar)^k}{(k+1)!}\sum_{s_1,\cdots,s_k}\Xi\tilde{e}^1_{s_1}\wedge\dots\wedge\Xi\tilde{e}^1_{s_k}
\wedge\Xi\pi_1^1\bd{B}\bd{(\pi^1B)}\bd{D}(e^1_{s_k})\dots\bd{D}(e^1_{s_1})\nonumber\\
     &\ +\sum_{k=0}^\infty\frac{(-\hbar)^k}{(k+1)!}\sum_{s_1,\cdots,s_k}\Xi\tilde{e}^1_{s_1}\wedge\dots\wedge\Xi\tilde{e}^1_{s_k}
\wedge\Xi\pi_1^1\bd{B}\bd{D}(e^1_{s_k})\dots\bd{D}(e^1_{s_1})\bd{(\pi^1B)}\nonumber\\
     &\ -\frac{\hbar}{2}\sum_s\left[\sum_{k=0}^\infty \frac{(-\hbar)^k}{(k+1)!}\sum_{s_1,\cdots,s_k}
\Xi\tilde{e}^1_{s_1}\wedge\dots\wedge\Xi\tilde{e}^1_{s_k}
\wedge\Xi\pi_1^1\bd{B}\bd{D}(e^1_{s_k})\dots\bd{D}(e^1_{s_1})\bd{D}(\tilde{e}^1_s)\right]\nonumber\\
    &\hspace{1.3cm} \wedge\left[\sum_{l=0}^\infty \frac{(-\hbar)^l}{(l+1)!}\sum_{s_1,\cdots,s_l}
\Xi\tilde{e}^1_{s_1}\wedge\dots\wedge\Xi\tilde{e}^1_{s_l}
\wedge\Xi\pi_1^1\bd{B}\bd{D}(e^1_{s_l})\dots\bd{D}(e^1_{s_1})\bd{D}(e^1_s)\right].
\end{align}
We can show that this vanishes as in a similar way to the case of previous subsection by taking into account the weak symmetry
\begin{align}
 \sum_s\tilde{e}^1_s\otimes e^1_s\ \approx\ \sum_s e^1_s\otimes\tilde{e}^1_s, 
\end{align}
and using
\begin{align}
    \sum_s\Xi\tilde{e}^1_s\wedge \Xi\pi_1^1\bd{B}\bd{(\pi^1B)}\bd{D}(e^1_s)\approx
\sum_s\Xi\pi_1^1\bd{B}\bd{D}(\tilde{e}^1_s)\wedge\Xi\pi_1^1\bd{B}\bd{D}(e^1_s).
\end{align}
Eventually, we find
\begin{align}
    \bd{\mathcal{F}}^{-1}(\bd{C}-\hbar\bd{\theta}\mid^2)-\bd{\eta}\bd{\mathcal{F}}^{-1}\ \approx\ 0\,.
\end{align}
Multiplying $\bd{\mathcal{F}}$ from right, we obtain \eqref{eq:eta F}.

\subsection{A derivation of \eqref{eq:def of G}}

Let us show that $\bd{\mathcal{F}}=\sum_{k=0}^\infty\frac{1}{k!}(\bd{G})^{\wedge k}$ with
 \begin{align}\label{eq:G determine}
     \pi_{n+1}\bd{G}=\frac{\hbar^n}{(n+1)!}\sum_{s_1,\cdots,s_n}\Xi\tilde{e}^1_{s_1}\wedge\dots\wedge\Xi\tilde{e}^1_{s_n}
\wedge(\pi_1\id+\Xi\pi_1^1\bd{B\mathcal{F}})\bd{D}(e^1_{s_n})\dots\bd{D}(e^1_{s_1}),
 \end{align}
satisfies 
$\bd{\mathcal{F}}\bd{\mathcal{F}}^{-1}\approx\idSH$
 by complete mathematical induction.
 When $n=0$\,, (\ref{eq:G determine}) gives 
 \begin{align}
     \pi_1\bd{\mathcal{F}}\bd{\mathcal{F}}^{-1}\ =&\ 
     \pi_1\bd{G}\bd{\mathcal{F}}^{-1}\ 
\nonumber\\     
     =&\ (\pi_1\id+\Xi\pi_1^1\bd{B\mathcal{F}})\bd{\mathcal{F}}^{-1}
     =\ \pi_1\id_{\mathcal{SH}}\,.
 \end{align}
In order to consider the cases of $n \ge 1$, we need some preparation.
First, we show the formula:
\begin{align}\label{eq:F rewrite}
    \pi_n\bd{\mathcal{F}}=\sum_{k=1}^n\frac{k}{n}\pi_k\bd{G}\wedge \pi_{n-k}\bd{\mathcal{F}}.
\end{align}
From the definition of $\bd{\mathcal{F}}$, we have
\begin{align}\label{eq:pinF}
    \pi_n\bd{\mathcal{F}}&=\sum_{k=0}^\infty \frac{1}{k!}\pi_n(\bd{G}^{\wedge k})\nonumber\\
    &=\sum_{l_1+2l_2+\dots+nl_n=n}\frac{1}{l_1!\dots l_n!}(\pi_1\bd{G})^{\wedge l_1}\wedge \dots \wedge (\pi_n\bd{G})^{\wedge l_n}.
\end{align}
On the other hand, the right-hand side of \eqref{eq:F rewrite} can be written as
\begin{align}\label{eq:rhs of F rewrite}
    \sum_{k=1}^n\frac{k}{n}\pi_k\bd{G}\wedge \pi_{n-k}\bd{\mathcal{F}}
=\sum_{l_1+2l_2+\dots+nl_n=n}C_{l_1,\dots ,l_n}(\pi_1\bd{G})^{\wedge l_1}\wedge \dots \wedge (\pi_n\bd{G})^{\wedge l_n},
\end{align}
with coefficient $C_{l_1,\dots ,l_n}$.
By substituting \eqref{eq:pinF} into the left-hand side of \eqref{eq:rhs of F rewrite} and comparing the coefficients of 
$(\pi_1\bd{G})^{\wedge l_1}\wedge \dots \wedge (\pi_n\bd{G})^{\wedge l_n}$ on both sides, we find that
\begin{align}
    C_{l_1,\dots ,l_n}=\sum_{k=1}^n \frac{k}{n}\frac{l_k}{l_1!\dots l_n!}=\frac{1}{l_1!\dots l_n!},
\end{align}
where, we used $l_1+2l_2+\dots+nl_n=n$ in the second equality.
Hence, \eqref{eq:F rewrite} holds.
Next, substitute \eqref{eq:F rewrite} into $\bd{\mathcal{F}}\bd{\mathcal{F}^{-1}}=\idSH$, we have
\begin{align}\label{eq:In equation}
    \pi_n\idSH&=\pi_n\bd{\mathcal{F}}\bd{\mathcal{F}}^{-1}\nonumber\\
&=\sum_{k=1}^n \frac{k}{n}(\pi_k\bd{G}\wedge \pi_{n-k}\bd{\mathcal{F}})\bd{\mathcal{F}}^{-1}\nonumber\\
    &\approx\sum_{k=1}^{n}\frac{k}{n}\sum_{m=0}^{n-k}\frac{(-\hbar)^m}{m!}
\sum_{s_1,\cdots,s_m}
\pi_k\bd{G}\bd{D}(e^1_{s_1})\dots\bd{D}(e^1_{s_m})
\bd{\mathcal{F}}^{-1}\wedge\pi_{n-k}\bd{D}(\Xi \tilde{e}^1_{s_m})\dots\bd{D}(\Xi \tilde{e}^1_{s_1})
\nonumber\\
&=\sum_{l=1}^{n}\sum_{k=1}^{l}\frac{k(-\hbar)^{l-k}}{n(l-k)!}\sum_{s_1,\cdots,s_{l-k}}\pi_k\bd{G}\bd{D}(e^1_{s_1})\cdots\bd{D}(e^1_{s_{l-k}})\bd{\mathcal{F}}^{-1}\wedge\Xi\tilde{e}^1_{s_{l-k}}\wedge\cdots\wedge\Xi\tilde{e}^1_{s_1}\wedge\pi_{n-l}\id_{\mathcal{SH}}\nonumber\\
&= \sum_{l=1}^{n}\frac{l}{n}I_l\wedge\pi_{n-l}\id_{\mathcal{SH}}\,,
\end{align}
where we define
\begin{align}
    I_l\equiv \sum_{k=1}^l \frac{k(-\hbar)^{l-k}}{l(l-k)!}\sum_{s_1,\cdots,s_{l-k}}\pi_k\bd{G}\bd{D}(e^1_{s_1})\dots\bd{D}(e^1_{s_{l-k}})
\bd{\mathcal{F}}^{-1}\wedge\Xi\tilde{e}^1_{s_{l-k}}\wedge\dots\wedge\Xi\tilde{e}^1_{s_1}\,.
\end{align}
This relation (\ref{eq:In equation}) yields $\pi_1\id_{\mathcal{SH}}=I_1$ for $n=1$, which is consistent with
\begin{align}
    I_1=\pi_1\bd{G}\bd{\mathcal{F}}^{-1}=\pi_1\bd{\mathcal{F}}\bd{\mathcal{F}}^{-1}
    =\pi_1\id_{\mathcal{SH}}\,.
\end{align}
For $n=2$, \eqref{eq:In equation} gives
\begin{align}
    \pi_2\idSH\ =&\ I_2+\frac{1}{2}\pi_1\bd{G}\bd{\mathcal{F}}^{-1}\wedge\pi_1\id_{\mathcal{SH}}\nonumber\\
    =&\
    I_2+\frac{1}{2}\pi_1\id_{\mathcal{SH}}\wedge\pi_1\id_{\mathcal{SH}}\,.
\end{align}
If we recall that 
\begin{align}
    \pi_n\id_{\mathcal{SH}}=\frac{1}{n!}\underbrace{\id\wedge\cdots\wedge\id}_n\,,
\end{align}
we find $I_2=0$\,. Then, if we assume $I_l=0$ for $2\leq l\leq n-1$, we have
\begin{align}
    \pi_n\id_{\mathcal{SH}}\ =&\ I_n + \frac{1}{n}\id\wedge\pi_{n-1}\id_{\mathcal{SH}}\,,
\end{align}
and thus, $I_n=0$\,. Therefore, we can conclude that $I_n=0$ for $\forall n\ge2$
by complete mathematical induction.

Now, let us show (\ref{eq:G determine}) for the cases of $n\ge1$\,.
Assume that 
\begin{align}\label{eq:assumption}
    \pi_{k+1}\bd{G}&=\frac{\hbar^k}{(k+1)!}\sum_{s_1,\cdots,s_k}\Xi\tilde{e}^1_{s_1}
\wedge\dots\wedge\Xi\tilde{e}^1_{s_k}\wedge(\pi_1\id+\Xi\pi_1^1\bd{B\mathcal{F})}\bd{D}(e^1_{s_k})\dots\bd{D}(e^1_{s_1})\nonumber\\
    &=\frac{\hbar^k}{(k+1)!}\sum_{s_1,\cdots,s_k}\Xi\tilde{e}^1_{s_1}\wedge\dots
\wedge\Xi\tilde{e}^1_{s_k}\wedge\pi_1\bd{G}\bd{D}(e^1_{s_k})\dots\bd{D}(e^1_{s_1}),
\end{align}
for $0<k\leq n-1$.
From the definition of $I_n$ and $I_n=0$ for $n \geq2$, we have
\begin{align}
    \pi_{n+1}\bd{G}\bd{\mathcal{F}}^{-1}\ =&\ -\sum_{k=1}^n\frac{k}{n+1}\frac{(-\hbar)^{n-k+1}}{(n-k+1)!}\nonumber\\
&\hspace{0.5cm}\times\sum_{s_1,\cdots,s_{n-k+1}}
\pi_k\bd{G}\bd{D}(e^1_{s_1})
\dots\bd{D}(e^1_{s_{n-k+1}})\bd{\mathcal{F}}^{-1}\wedge\Xi\tilde{e}^1_{s_{n-k+1}}\wedge\dots\wedge\Xi\tilde{e}^1_{s_1}.
\end{align}
From the assumption \eqref{eq:assumption}, it can be written as
\begin{align}\label{Final form of G}
&\   \pi_{n+1}\bd{G}\bd{\mathcal{F}}^{-1}\nonumber\\
&= -\frac{\hbar^n}{n+1}\left(\sum_{k=1}^n\frac{(-1)^{n-k+1}}{(n-k+1)!(k-1)!}\right)
\sum_{s_1,\cdots,s_n}
\Xi\tilde{e}^1_{s_1}\wedge\dots\wedge\Xi\tilde{e}^1_{s_n}\wedge\pi_1\bd{G}\bd{D}(e^1_{s_n})\dots\bd{D}(e^1_{s_1})
\bd{\mathcal{F}}^{-1}\nonumber\\
&=\frac{\hbar^n}{(n+1)!}\sum_{s_1,\cdots,s_n}
\Xi\tilde{e}^1_{s_1}\wedge\dots\wedge\Xi\tilde{e}^1_{s_n}
\wedge\pi_1\bd{G}\bd{D}(e^1_{s_n})\dots\bd{D}(e^1_{s_1})\bd{\mathcal{F}}^{-1}\nonumber\\
    &=\frac{\hbar^n}{(n+1)!}\sum_{s_1,\cdots,s_n}\Xi\tilde{e}^1_{s_1}\wedge\dots\wedge\Xi\tilde{e}^1_{s_n}\wedge(\pi_1\id+\Xi\pi_1^1
\bd{B\mathcal{F}})\bd{D}(e^1_{s_n})\dots\bd{D}(e^1_{s_1})\bd{\mathcal{F}}^{-1}\,,
\end{align}
where we calculated the summation as
\begin{align}
    \sum_{k=1}^n\frac{(-1)^{n-k+1}}{(n-k+1)!(k-1)!}=&\frac{1}{n!}\sum_{k=0}^{n-1}(-1)^{n-k}\mqty(n\\k)
    \nonumber\\
    =&\frac{1}{n!}\left((1-1)^n-1\right)=-\frac{1}{n!}\,.
\end{align}
We obtain \eqref{eq:G determine} by multiplying $\bd{\mathcal{F}}$ to the both side of (\ref{Final form of G}) from the right.

\end{document}